%% file: taylorImpact.tex
\documentclass[11pt]{article}
\usepackage[pdftex]{graphicx}
\usepackage[numbers]{natbib}
\usepackage{times}
\usepackage{amsmath, amssymb, amstext}
\usepackage[boxed,algonl]{algorithm2e}
\usepackage[margin=2cm]{geometry}
\usepackage[margin=20pt,small,bf]{caption}

\include{./mymacros}

\begin{document}

\title{{\normalsize Report No. C-SAFE-CD-IR-04-004}\\ \vspace{10pt}
       MPM VALIDATION: A MYRIAD OF TAYLOR IMPACT TESTS}
\author{B. Banerjee (b.banerjee.nz@gmail.com)\\
        Department of Mechanical Engineering, University of Utah, \\
        Salt Lake City, UT 84112, USA\\
        August 21, 2004}
\maketitle
\pagestyle{myheadings}
\markboth{C-SAFE-CD-IR-04-004}{C-SAFE-CD-IR-04-004}

\begin{abstract}
Taylor impacts tests were originally devised to determine the dynamic yield
strength of materials at moderate strain rates.  More recently, such
tests have been used extensively to validate numerical codes for the 
simulation of plastic deformation.  In this work, we use the material point
method to simulate a number of Taylor impact tests to compare different
Johnson-Cook, Mechanical Threshold Stress, and Steinberg-Guinan-Cochran
plasticity models and the vob Mises and Gurson-Tvergaard-Needleman yield 
conditions.  In addition to room temperature Taylor tests, high temperature
tests have been performed and compared with experimental data.
\end{abstract}

\section{INTRODUCTION}
  The Taylor impact test (\citet{Taylor48}) was originally devised as a 
  means of determining the dynamic yield strength of solids.  The test
  involves the impact of a flat-nosed cylindrical projectile on a hard
  target at normal incidence.  Taylor provided an analytical solutions 
  for the dynamic yield strength of the material
  of the projectile based on the length of the elastic region and the radius
  of the region of permanent set.  As described by \citet{Whiffin48}, that 
  use of the test was limited to relatively small deformations obtained 
  from low velocity impacts.  Though the Taylor impact test continues to be 
  used to determine yield strengths of materials at high strain rates, the
  test is limited to peak strains of around 0.6 at the center of the 
  specimen (\citet{Johnson88a}).  For higher strains and strain rates, the 
  Taylor test is currently used more as a means of validating plasticity 
  models in numerical codes for the simulation of high rate phenomena 
  such as impact and explosive deformation as suggested by 
  \citet{Zerilli87}.

  \hspace{16pt}
  In this paper, we describe our experience in validating the plasticity
  models in a parallel, multiphysics code that uses the material point method
  (\citet{Sulsky94,Sulsky95}) using Taylor impact tests for various strain
  rates and temperatures.  A number of metrics are used to compare
  simulations and experiments and suggesstions are made regarding the use
  of Taylor impacts tests for the validation of the plasticity portion of
  such codes.

  \hspace{16pt}
  The organization of this paper is as follows.  Section~\ref{sec:back}
  provides the background for the current study and describes the mutiphysics 
  code Uintah, the material point method, and the stress update algorithm,
  and various plasticity models and yield conditions.  A few validation metrics 
  are identified and their significance is discussed in 
  Section~\ref{sec:metric}.  Comparisons between experimental data and 
  simulations of Taylor impact tests using the validation metrics are described
  in Section~\ref{sec:results}.  Finally, conclusions and suggestions 
  are presented in Section~\ref{sec:conclude}.

\section{BACKGROUND} \label{sec:back}
  The goal of this work is to present some results and insights we have
  obatined during the process of validation of plasticity models used
  in the simulation of the deformation and failure of a steel container that 
  expands under the effect of gases produced by an explosively reacting high 
  energy material (PBX 9501) contained inside. The entire process is simulated 
  using the massively parallel, Common Component 
  Architecture ~\cite{Armstrong99} based, Uintah Computational 
  Framework (UCF)~\cite{Dav2000}.
  
\begin{figure}
    \centering
    \scalebox{0.3}{\includegraphics{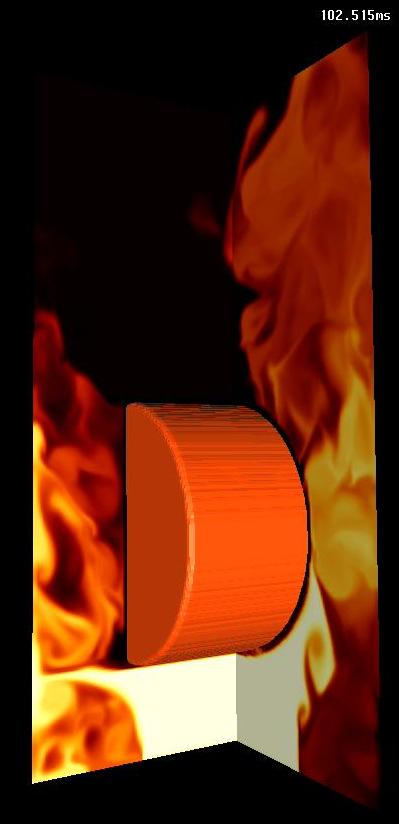}
                   \includegraphics{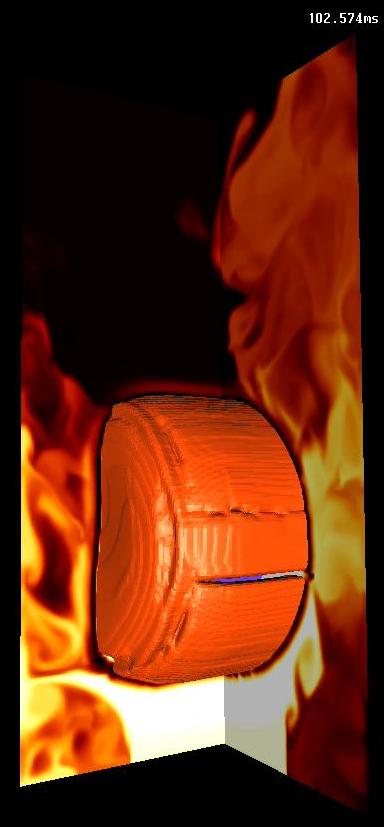}
                   \includegraphics{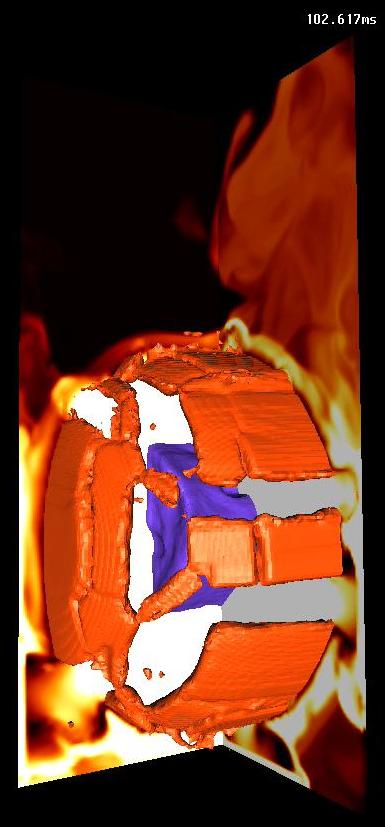}
                   \includegraphics{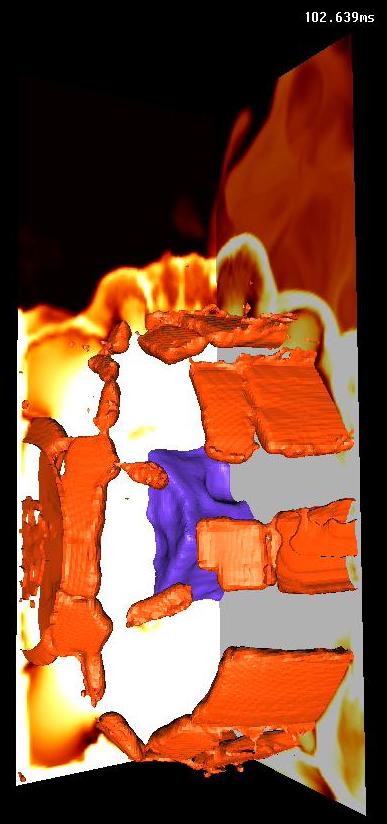}}
    \caption{\small Simulation of exploding cylinder.}
    \label{fig:poolfire}
  \end{figure}

  \hspace{16pt}
  The high energy material reacts at temperatures of 450 K and above,  This
  elevated temperature is achieved through external heating of the steel 
  container.  Experiments conducted at the University of Utah have shown that
  failure of the container can be due to ductile fracture associated with 
  void coalescence and adiabatic shear bands.  If shear bands dominate the
  steel container fragments, otherwise a few large cracks propagate along the
  cylinder and pop it open.  Figure~\ref{fig:poolfire} shows the result of 
  a simulation of a coupled fire-container-explosion using the Uintah.
  
  \hspace{16pt}
  The dynamics of the solid materials - steel and PBX 9501 - is modeled 
  using the Lagrangian Material Point Method (MPM)~\cite{Sulsky94}.  Gases 
  are generated from solid PBX 9501 using a burn model~\cite{Long02}.  
  Gas-solid interaction is accomplished using an Implicit Continuous 
  Eulerian (ICE) multi-material hydrodynamic code~\cite{Guilkey04}.  A 
  single computational grid is used for all the materials.

  \hspace{16pt}
  The constitutive response of PBX 9501 is modeled using 
  ViscoSCRAM~\cite{Bennet98}, which is a five element generalized Maxwell 
  model for the viscoelastic response coupled with statistical crack 
  mechanics.  Solid PBX 9501 is progressively converted into a gas with an 
  appropriate equation of state.  The temperature and pressure in the gas 
  increase rapidly as the reaction continues.  As a result, the steel container
  is pressurized, undergoes plastic deformation, and finally fragments.  

  \hspace{16pt}
  The main issues regarding the constitutive modeling of the steel container 
  are the selection of appropriate models for nonlinear elasticity, plasticity,
  damage, loss of material stability, and failure.  The numerical simulation 
  of the steel container involves the choice of appropriate algorithms for the
  integration of balance laws and constitutive equations, as well as the
  methodology for fracture simulation.  Models and simulation methods
  for the steel container are required to be temperature sensitive and
  valid for large distortions, large rotations, and  a range of strain 
  rates (quasistatic at the beginning of the simulation to approximately 
  $10^6$ s$^{-1}$ at fracture). 

  \hspace{16pt}
  The approach chosen for the present work is to use hypoelastic-plastic
  constitutive models that assume an additive decomposition of the rate
  of deformation tensor into elastic and plastic parts.  Hypoelastic materials
  are known not to conserve energy in a loading-unloading cycle
  unless a very small time step is used.  However, the choice of this model 
  is justified under the assumption that elastic strains are expected to be 
  small for the problem under consideration and unlikely to affect the 
  computation significantly.  

  \hspace{16pt}
  Two plasticity models for flow stress are considered 
  along with a two different yield conditions.  Explicit fracture simulation 
  is computationally expensive and prohibitive in the large simulations 
  under consideration.  The choice, therefore, has been to use damage models 
  and stability criteria for the prediction of failure (at material points) 
  and particle erosion for the simulation of fracture propagation.  

  \subsection{The Material Point Method}\label{sec:mpm}
  The Material Point Method (MPM) ~\cite{Sulsky94} is a particle method for 
  structural mechanics simulations.  In this method, the state variables 
  of the material are described on Lagrangian particles or ``material points''.
  In addition, a regular, structured Eulerian grid is used as a computational 
  scratch pad to compute spatial gradients and to solve the governing
  conservation equations.  An explicit time-stepping version of the Material 
  Point Method has been used in the simulations presented in this paper.  
  The MPM algorithm is summarized below~\cite{Sulsky95}.  

  \hspace{16pt}
  It is assumed that an particle state at the beginning of a time step 
  is known.  The mass ($m$), external force ($\Bf^{\Text}$), and 
  velocity ($\Bv$) of the particles are interpolated to the grid using 
  the relations
  \begin{equation}\label{eq:1}
    m_g = \sum_{p} S_{gp}~m_p ~,~~~~
    \Bv_g = (1/m_g)\sum_{p} S_{gp}~m_p~\Bv_p ~,~~~~
    \Bf^{\Text}_g = \sum_{p} S_{gp}~\Bf^{\Text}_p
  \end{equation}
  where the subscript ($g$) indicates a quantity at a grid node and a 
  subscript ($p$) indicates a quantity on a particle.  The symbol $\sum_p$
  indicates a summation over all particles.  The quantity ($S_{gp}$) is 
  the interpolation function of node ($g$) evaluated at the position of 
  particle ($p$).  Details of the interpolants used can be found 
  elsewhere~\cite{Bard04}.

  \hspace{16pt}
  Next, the velocity gradient at each particle is computed using the 
  grid velocities using the relation
  \begin{equation} \label{eq:2}
    \Grad{\Bv_p} = \sum_g \BG_{gp} \Bv_g
  \end{equation}
  where $\BG_{gp}$ is the gradient of the shape function of node ($g$)
  evaluated at the position of particle ($p$).  The velocity gradient at 
  each particle is used to determine the Cauchy stress ($\Bsig_p$) at the
  particle using a stress update algorithm.

  \hspace{16pt}
  The internal force at the grid nodes ($\Bf^{\Tint}_g$) is calculated 
  from the divergence of the stress using
  \begin{equation} \label{eq:3}
    \Bf^{\Tint}_g = \sum_p \BG_{gp}~\Bsig_p~V_p
  \end{equation}
  where $V_p$ is the particle volume.
  
  \hspace{16pt}
  The equation for the conservation of linear momentum is next solved on
  the grid.  This equation can be cast in the form
  \begin{equation} \label{eq:4}
    \Bm_g ~ a_g = \Bf^{\Text}_g - \Bf^{\Tint}_g
  \end{equation}
  where $\Ba_g$ is the acceleration vector at grid node ($g$).  

  \hspace{16pt}
  The velocity vector at node ($g$) is updated using an explicit (forward
  Euler) time integration, and the particle velocity and position are then
  updated using grid quantities.  The relevant equations are
  \begin{align} 
     \Bv_g(t+\Delta t) & = \Bv_g(t) + \Ba_g~\Delta t \label{eq:5} \\
     \Bv_p(t+\Delta t) & = \Bv_p(t) + \sum_g S_{gp}~\Ba_g~\Delta t ~;~~~~
     \Bx_p(t+\Delta t) = \Bx_p(t) + \sum_g S_{gp}~\Bv_g~\Delta t \label{eq:6}
  \end{align}

  \hspace{16pt}
  The above sequence of steps is repeated for each time step.  The above 
  algorithm leads to particularly simple mechanisms for handling contact.
  Details of these contact algorithms can be found elsewhere~\cite{Bard01}.

  \subsection{Plasticity and Failure Simulation}\label{sec:algo}
  A hypoelastic-plastic, semi-implicit approach~\cite{Zocher00} 
  has been used for the stress update in the simulations presented in this 
  paper.  An additive decomposition of the rate of deformation tensor into
  elastic and plastic parts has been assumed.  One advantage of this approach
  is that it can be used for both low and high strain rates.  Another advantage 
  is that many strain-rate and temperature-dependent plasticity and damage 
  models are based on the assumption of additive decomposition of strain rates,
  making their implementation straightforward.  

  \hspace{16pt}
  The stress update is performed in a co-rotational frame which is equivalent 
  to using the Green-Naghdi objective stress rate.  An incremental update of 
  the rotation tensor is used instead of a direct polar decomposition of the 
  deformation gradient.  The accuracy of model is good if elastic strains are 
  small compared to plastic strains and the material is not unloaded.
  It is also assumed that the stress tensor can be divided into a volumetric 
  and a deviatoric component. The plasticity model is used to update only the 
  deviatoric component of stress assuming isochoric behavior.  The hydrostatic 
  component of stress is updated using a solid equation of state.

  \hspace{16pt}
  Since the material in the container may unload locally after fracture, the 
  hypoelastic-plastic stress update may not work accurately under certain 
  circumstances.  An improvement would be to use a hyperelastic-plastic stress 
  update algorithm.  Also, the plasticity models are temperature dependent.
  Hence there is the issue of severe mesh dependence due to change of the
  governing equations from hyperbolic to elliptic in the softening regime
  ~\cite{Hill75,Bazant85,Tver90}.  Viscoplastic stress update models or 
  nonlocal/gradient plasticity models~\cite{Ramaswamy98,Hao00} can be used 
  to eliminate some of these effects and are currently under investigation. 
 
  \hspace{16pt}
  A particle is tagged as ``failed'' when its temperature is greater than the
  melting point of the material at the applied pressure.  An additional
  condition for failure is when the porosity of a particle increases beyond a
  critical limit.  A final condition for failure is when a bifurcation 
  condition such as the Drucker stability postulate is satisfied.  Upon failure,
  a particle is either removed from the computation by setting the stress to
  zero or is converted into a material with a different velocity field 
  which interacts with the remaining particles via contact.  Either approach
  leads to the simulation of a newly created surface.

  \hspace{16pt}
  In the parallel implementation of the stress update algorithm, sockets have 
  been added to allow for the incorporation of a variety of plasticity, damage, 
  yield, and bifurcation models without requiring any change in the stress 
  update code.  The algorithm is shown in Algorithm~\ref{algo1}.  The
  equation of state, plasticity model, yield condition, damage model, and
  the stability criterion are all polymorphic objects created using a 
  factory idiom in C++~\cite{Coplien92}.
  
  \begin{algorithm}[t]
    \caption{\small Stress Update Algorithm} \label{algo1}
    \KwData{{\bf Persistent}:Initial moduli, temperature, porosity, 
              scalar damage, equation of state, plasticity model, 
              yield condition, stability criterion, damage model\\
             {\bf Temporary}:Particle state at time $t$}
    \KwResult{Particle state at time $t+\Delta t$}

    \For{all the patches in the domain}{
      Read the particle data and initialize updated data storage\;
      \For{all the particles in the patch} {
        Compute the velocity gradient, the rate of deformation tensor 
          and the spin tensor\;
        Compute the updated left stretch tensor, rotation tensor, and
          deformation gradient\;
        Rotate the input Cauchy stress and the rate of deformation tensor
          to the material configuration\;
        Compute the current shear modulus and melting temperature\;
        Compute the pressure using the equation of state, update the
        hydrostatic stress, and compute the trial deviatoric stress\;
        Compute the flow stress using the plasticity model\;
        Evaluate the yield function\;
        \eIf{particle is elastic}{
          Rotate the stress back to laboratory coordinates\;
          Update the particle state\;
        }{
          Find derivatives of the yield function\;
          Do radial return adjustment of deviatoric stress\;
          Compute updated porosity, scalar damage, and temperature 
           increase due to plastic work\;
          Compute elastic-plastic tangent modulus and evaluate stability
          condition\;
          Rotate the stress back to laboratory coordinates\;
          Update the particle state\;
          \If{Temperature $>$ Melt Temperature or Porosity $>$ Critical Porosity
               or Unstable}{
            Tag particle as failed\;
          }
        }
      }
    }
    Convert failed particles into a material with a different velocity field;
  \end{algorithm}
      
  \subsection{Models}\label{sec:models}
  The stress in the solid is partitioned into a volumetric part and a deviatoric
  part.  Only the deviatoric part of stress is used in the plasticity 
  calculations assuming isoschoric plastic behavior.

  \hspace{16pt}
  The hydrostatic pressure ($p$) is calculated either using the bulk modulus
  ($K$) and shear modulus ($\mu$) or from a temperature-corrected Mie-Gruneisen 
  equation of state of the form~\cite{Zocher00} 
  \begin{equation}
   p = \frac{\rho_0 C_0^2 \zeta
              \left[1 + \left(1-\frac{\Gamma_0}{2}\right)\zeta\right]}
             {\left[1 - (S_{\alpha} - 1) \zeta\right]^2 + \Gamma_0 C_p T}~,~~~~
   \zeta = (\rho/\rho_0 - 1)
  \end{equation}
  where $C_0$ is the bulk speed of sound, 
  $\rho_0$ is the initial density, $\rho$ is the current density, 
  $C_p$ is the specific heat at constant volume, $T$ is the temperature, 
  $\Gamma_0$ is the Gruneisen's gamma at reference state, and $S_{\alpha}$ 
  is the linear Hugoniot slope coefficient.

  \hspace{16pt}
  Depending on the plasticity model being used, the pressure and
  temperature-dependent shear modulus ($\mu$) and the pressure-dependent 
  melt temperature ($T_m$) are calculated using the relations~\cite{Steinberg80}
  \begin{align}
    \mu & = \mu_0\left[1 + A\frac{p}{\eta^{1/3}} - B(T - 300)\right] \\
    T_m & = T_{m0} \exp\left[2a\left(1-\frac{1}{\eta}\right)\right]
              \eta^{2(\Gamma_0-a-1/3)}
  \end{align}
  where, $\mu_0$ is the shear modulus at the reference state($T$ = 300 K, 
  $p$ = 0, $\epsilon_p$ = 0), $\epsilon_p$ is the plastic strain.
  $\eta = \rho/\rho_0$ is the compression, $A = (1/\mu_0)(d\mu/dp)$, 
  $B = (1/\mu_0)(d\mu/dT)$, $T_{m0}$ is the melt temperature at 
  $\rho=\rho_0$, and $a$ is the coefficient of the first order volume 
  correction to Gruneisen's gamma.

  \hspace{16pt}
  We have explored two temperature and strain rate dependent plasticity models -
  the Johnson-Cook plasticity model~\cite{Johnson83} and the Mechanical 
  Threshold Stress (MTS) plasticity model~\cite{Follans88,Goto00a}.  The
  flow stress ($\sigma_f$) from the Johnson-Cook model is given by
  \begin{equation}
    \sigma_f = [A + B (\epsilon_p)^n][1 + C \ln(\dot{\epsilon_p^*})]
    [1 - (T^*)^m]~;~~
    \dot{\epsilon_p^{*}} = \cfrac{\dot{\epsilon_p}}{\dot{\epsilon_{p0}}}~;~~
    T^* = \cfrac{(T-T_r)}{(T_m-T_r)}
  \end{equation}
  where $\dot{\epsilon_{p0}}$ is a user defined plastic strain rate, 
  A, B, C, n, m are material constants, $T_r$ is the room temperature, and
  $T_m$ is the melt temperature.  

  \hspace{16pt}
  The flow stress for the MTS model is given by
  \begin{equation}
    \sigma_f = \sigma_a + \frac{\mu}{\mu_0} S_i \hat\sigma_i
    + \frac{\mu}{\mu_0} S_e \hat\sigma_e 
  \end{equation}
  where
  \begin{align*}
    \mu &= \mu_0 - \frac{D}{\exp\left(\frac{T_0}{T}\right) - 1} \\
    S_i &= \left[1 - \left(\frac{kT}{g_{0i}\mu b^3}
    \ln\frac{\dot\epsilon_{0i}}{\dot\epsilon}\right)^{1/qi}
    \right]^{1/pi} ~;~~
    S_e = \left[1 - \left(\frac{kT}{g_{0e}\mu b^3}
    \ln\frac{\dot\epsilon_{0e}}{\dot\epsilon}\right)^{1/qe}
    \right]^{1/pe}
  \end{align*}
  \begin{align*}
    \theta &= \theta_0 [ 1 - F(X)] + \theta_{IV} F(X) ~;~~ 
    \theta_0 = a_0 + a_1 \ln \dot\epsilon + a_2 \sqrt{\dot\epsilon} - a_3 T \\
    X & = \cfrac{\hat\sigma_e}{\hat\sigma_{es}} ~;~~ F(X) = \tanh(\alpha X) ~;~~
    \ln(\hat\sigma_{es}/\hat\sigma_{es0}) =
    \left(\frac{kT}{\mu b^3 g_{0es}}\right)
    \ln\left(\cfrac{\dot\epsilon}{\dot\epsilon_{es0}}\right)\\
    \hat\sigma_e^{(n+1)} & = \hat\sigma_e^{(n)}+\theta\Delta\epsilon
  \end{align*}
  and $\sigma_a$ is the athermal component of mechanical threshold stress,
  $\mu_0$ is the shear modulus at 0 K, $D, T_0$ are empirical constants, 
  $\hat\sigma_i$ represents the stress due to intrinsic barriers 
  to thermally activated dislocation motion and dislocation-dislocation 
  interactions, $\hat\sigma_e$ represents the stress due to 
  microstructural evolution with increasing deformation, 
  $k$ is the Boltzmann constant, $b$ is the length of the Burger's vector, 
  $g_{0[i,e]}$ are the normalized activation energies, 
  $\dot\epsilon_{0[i,e]}$ are constant strain rates,
  $q_{[i,e]}, p_{[i,e]}$ are constants, $\theta_0$ is the hardening due to 
  dislocation accumulation, $a_0, a_1, a_2, a_3, \theta_{IV}, \alpha$ are 
  constants,
  $\hat\sigma_{es}$ is the stress at zero strain hardening rate, 
  $\hat\sigma_{es0}$ is the saturation threshold stress for deformation at 0 K,
  $g_{0es}$ is a constant, and $\dot\epsilon_{es0}$ is the maximum strain rate.

  \hspace{16pt}
  We have decided to focus on ductile failure of the steel container.
  Accordingly, two yield criteria have been explored - the von Mises condition
  and the Gurson-Tvergaard-Needleman (GTN) yield 
  condition~\cite{Gurson77,Tver84} which depends on porosity.  An associated 
  flow rule is used to determine the plastic rate parameter in either case.
  The von Mises yield condition is given by
  \begin{equation}
    \Phi = \left(\frac{\sigma_{eq}}{\sigma_f}\right)^2 - 1 = 0 ~;~~~
    \sigma_{eq} = \sqrt{\frac{3}{2}\sigma^{d}:\sigma^{d}}
  \end{equation}
  where $\sigma_{eq}$ is the von Mises equivalent stress, 
  $\sigma^{d}$ is the deviatoric part of the Cauchy stress, and
  $\sigma^{f}$ is the flow stress.
  The GTN yield condition can be written as
  \begin{equation}
    \Phi = \left(\frac{\sigma_{eq}}{\sigma_f}\right)^2 +
    2 q_1 f_* \cosh \left(q_2 \frac{Tr(\sigma)}{2\sigma_f}\right) -
    (1+q_3 f_*^2) = 0
  \end{equation}
  where $q_1,q_2,q_3$ are material constants and $f_*$ is the porosity 
  (damage) function given by
  \begin{equation}
    f* = 
    \begin{cases}
      f & \text{for}~~ f \le f_c,\\ 
      f_c + k (f - f_c) & \text{for}~~ f > f_c 
    \end{cases}
  \end{equation}
  where $k$ is a constant and $f$ is the porosity (void volume fraction).  The 
  flow stress in the matrix material is computed using either of the two 
  plasticity models discussed earlier.  Note that the flow stress in the matrix 
  material also remains on the undamaged matrix yield surface and uses an 
  associated flow rule.

  \hspace{16pt}
  The evolution of porosity is calculated as the sum of the rate of growth 
  and the rate of nucleation~\cite{Ramaswamy98a}.  The rate of growth of
  porosity and the void nucleation rate are given by the following equations
  ~\cite{Chu80}
  \begin{align}
    \dot{f} &= \dot{f}_{\text{nucl}} + \dot{f}_{\text{grow}} \\
    \dot{f}_{\text{grow}} & = (1-f) \text{Tr}(\BD_p) \\
    \dot{f}_{\text{nucl}} & = \cfrac{f_n}{(s_n \sqrt{2\pi})}
            \exp\left[-\Half \cfrac{(\epsilon_p - \epsilon_n)^2}{s_n^2}\right]
            \dot{\epsilon}_p
  \end{align}
  where $\BD_p$ is the rate of plastic deformation tensor, $f_n$ is the volume 
  fraction of void nucleating particles , $\epsilon_n$ is the mean of the 
  distribution of nucleation strains, and $s_n$ is the standard 
  deviation of the distribution.

  \hspace{16pt}
  Part of the plastic work done is converted into heat and used to update the 
  temperature of a particle.  The increase in temperature ($\Delta T$) due to 
  an increment in plastic strain ($\Delta\epsilon_p$) is given by the equation
  ~\cite{Borvik01}
  \begin{equation}
    \Delta T = \cfrac{\chi\sigma_f}{\rho C_p} \Delta \epsilon_p
  \end{equation}
  where $\chi$ is the Taylor-Quinney coefficient, and $C_p$ is the specific
  heat.  A special equation for the dependence of $C_p$ upon temperature is
  also used for steel~\cite{Goto00}.
  \begin{equation}
    C_p = 10^3(0.09278 + 7.454\times 10^{-4} T + 12404.0/T^2)
  \end{equation}

  \hspace{16pt}
  Under normal conditions, the heat generated at a material point is conducted 
  away at the end of a time step using the heat equation.  If special adiabatic 
  conditions apply (such as in impact problems), the heat is accumulated at a 
  material point and is not conducted to the surrounding particles.  This 
  localized heating can be used to simulate adiabatic shear band formation.

  \hspace{16pt}
  After the stress state has been determined on the basis of the yield condition
  and the associated flow rule, a scalar damage state in each material point can
  be calculated using either of two damage models - the Johnson-Cook model
  ~\cite{Johnson85} or the Hancock-MacKenzie model~\cite{Hancock76}.  While 
  the Johnson-Cook model has an explicit dependence on temperature, 
  the Hancock-McKenzie model depends on the temperature implicitly, via the 
  stress state.  Both models depend on the strain rate to determine the 
  value of the scalar damage parameter.

  \hspace{16pt}
  The damage evolution rule for the Johnson-Cook damage model can be written as
  \begin{equation}
    \dot{D} = \cfrac{\dot{\epsilon_p}}{\epsilon_p^f} ~;~~
    \epsilon_p^f = 
      \left[D_1 + D_2 \exp \left(\cfrac{D_3}{3} \sigma^*\right)\right]
      \left[1+ D_4 \ln(\dot{\epsilon_p}^*)\right]
      \left[1+D_5 T^*\right]~;~~
    \sigma^*= \cfrac{\text{Tr}(\Bsig)}{\sigma_{eq}}~;~~
  \end{equation}
  where $D$ is the damage variable which has a value of 0 for virgin material
  and a value of 1 at fracture, $\epsilon_p^f$ is the fracture strain, 
  $D_1, D_2, D_3, D_4, D_5$ are constants, $\Bsig$ is the Cauchy stress, and
  $T^*$ is the scaled temperature as in the Johnson-Cook plasticity model.

  \hspace{16pt}
  The Hancock-MacKenzie damage evolution rule can be written as
  \begin{equation}
    \dot{D} = \cfrac{\dot{\epsilon_p}}{\epsilon_p^f} ~;~~
    \epsilon_p^f = \frac{1.65}{\exp(1.5\sigma^*)}
  \end{equation}

  \hspace{16pt}
  The determination of whether a particle has failed can be made on the 
  basis of either or all of the following conditions:
  \begin{itemize}
    \item The particle temperature exceeds the melting temperature.
    \item The TEPLA-F fracture condition~\cite{Johnson88} is satisfied.
       This condition can be written as
       \begin{equation}
         (f/f_c)^2 + (\epsilon_p/\epsilon_p^f)^2 = 1
       \end{equation}
       where $f$ is the current porosity, $f_c$ is the maximum 
       allowable porosity, $\epsilon_p$ is the current plastic strain, and
       $\epsilon_p^f$ is the plastic strain at fracture.
    \item An alternative to ad-hoc damage criteria is to use the concept of 
       bifurcation to determine whether a particle has failed or not.  Two
       stability criteria have been explored in this paper - the Drucker
       stability postulate~\cite{Drucker59} and the loss of hyperbolicity
       criterion (using the determinant of the acoustic tensor)
       \cite{Rudnicki75,Perzyna98}.  
  \end{itemize}

  \hspace{16pt}
  The simplest criterion that can be used is the Drucker stability postulate 
  \cite{Drucker59} which states that time rate of change of the rate of 
  work done by a material cannot be negative.  Therefore, the material is 
  assumed to become unstable (and a particle fails) when
  \begin{equation}
    \dot\Bsig:\BD^p \le 0
  \end{equation}

  \hspace{16pt}
  Another stability criterion that is less restrictive is the acoustic
  tensor criterion which states that the material loses stability if the 
  determinant of the acoustic tensor changes sign~\cite{Rudnicki75,Perzyna98}.  
  Determination of the acoustic tensor requires a search for a normal vector 
  around the material point and is therefore computationally expensive.  A 
  simplification of this criterion is a check which assumes that the direction 
  of instability lies in the plane of the maximum and minimum principal 
  stress~\cite{Becker02}.  In this approach, we assume that the strain is 
  localized in a band with normal $\Bn$, and the magnitude of the velocity 
  difference across the band is $\Bg$.  Then the bifurcation condition 
  leads to the relation 
  \begin{equation} 
    R_{ij} g_{j} = 0 ~;~~~
    R_{ij} = M_{ikjl} n_k n_l + M_{ilkj} n_k n_l - \sigma_{ik} n_j n_k
  \end{equation} 
  where $M_{ijkl}$ are the components of the co-rotational tangent
  modulus tensor and $\sigma_{ij}$ are the components of the co-rotational 
  stress tensor.  If $\det(R_{ij}) \le 0 $, then $g_j$ can be arbitrary and 
  there is a possibility of strain localization.  If this condition for 
  loss of hyperbolicity is met,  then a particle deforms in an unstable 
  manner and failure can be assumed to have occurred at that particle.  

\section{VALIDATION METRICS} \label{sec:metric}
  The attractiveness of the Taylor impact test arises because of the 
  simplicity and inexpensiveness of the test.  A flat-ended cylinder is 
  fired on a target at a large enough velocity and the final deformed shape 
  is measured.  The drawback of this test is that intermediate states of 
  the cylinder are difficult to measure and hence are generally not.  The
  validation metrics that we consider in this paper are based on the final
  shape of the cylinder though other metrics may be considered if
  measurements of these are made during the course of an impact test.  We
  note that the Taylor test could also be used to validate simulations of
  dynamic fracture though we do not address that issue in this paper.

  \hspace{16pt}
  There is a large literature on the systematic verification and validation
  of computational codes (see \citet{Oberkampf02,Babuska04} and references
  therein).  It has been suggested that validation metrics be developed
  that can be used to compare experimental data and simulation results.  The
  metrics discussed in this paper are intended to be a step in that direction
  but they are not intended to be complete or comprehensive.

  \hspace{16pt}
  The most common metric used in the literature is the ``calibrated eyeball''
  approach or ``view-graph norm'' (\citet{Oberkampf02}) 
  where a plot of the simulated deformed configuration is 
  superimposed on the experimental data and a subjective judgement of 
  accuracy is made.  We believe that there is value to this approach and
  present all our data in this form.  However, we also believe that more 
  quantitative descriptions of the difference between experiment and simulations
  can be obtained and present comparisons using other metrics.

  Metrics, sensitivity studies, and determination of experimental variability are essential.
  Some quantities of interest are:
  \begin{enumerate}
    \item Metrics
    \begin{enumerate}
    \item Regression between profiles
    \item Length change
    \item Diameter change
    \item Volume change
    \item Middle bulge difference
    \item Length of elastic zone
    \item plastic strain
    \item temperature
    \item time of impact
    \item energy conversion at impact
    \end{enumerate}
    \item Sensitivity studies
    \begin{enumerate}
    \item  mesh size (quantify discretization errors)
    \item  plasticity model parameters 
    \item  plasticity model
    \item  impact velocity 
    \item  temperature
    \item  length and diameter
    \end{enumerate}
    \item Variability in experimental data
    \begin{enumerate}
    \item  material
    \item  geometry
    \item  velocity 
    \item  temperature
    \item  measurement error
    \end{enumerate}
  \end{enumerate}

\section{Taylor impact simulations} \label{sec:results}
  In this section, we compare the final deformed shapes from
  simulations of Taylor impact tests with experimentally obtained data.
  In cases where images or profiles of the deformed shapes of the cylinders
  were available, these were digitized using a scanner and then imported 
  into XFig~(\citet{XFig04}).  The scanned images were overlaid with manually 
  digitized lines that were drawn as accurately as possible after expanding 
  the images to a resultion of 1024$\times$1260.  The digitized curves
  were then rotated so that the axes were aligned with the grid.  The XFig
  coordinates were then scaled to length units using cues from the digitized
  images and their axes (if any were provided).  Some small errors (1\%-2\%) are
  expected in this procedure.  However, the overall profiles of the cylinders
  are captured accurately in most cases.

  The simulations were run for 150 $\mu$s - 200 $\mu$s depending on the 
  problem.  The simulation times were chosen such that the cylinders bounced
  off the anvil and moved away for at least 20 $\mu$s.  It was observed at
  beyond this time, the deformed shape of the cylinder reamined constant
  and all elastic strains and rotations had been recovered.

\begin{figure}[htbp!]
    \begin{minipage}[t]{0.5\linewidth}
      \centering
      \scalebox{0.35}{\includegraphics{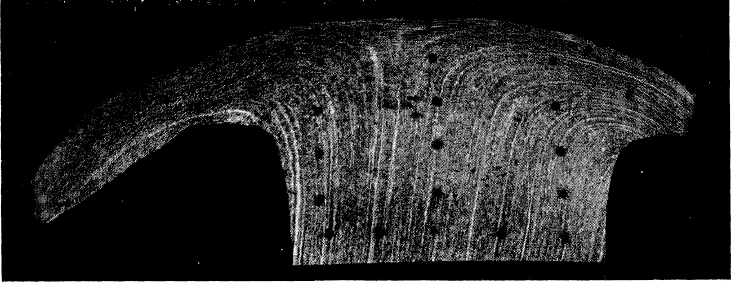}}\\
      (a) Actual profile (\citet{Carrington48}).
    \end{minipage}
    \begin{minipage}[t]{0.5\linewidth}
      \centering
      \scalebox{0.35}{\includegraphics{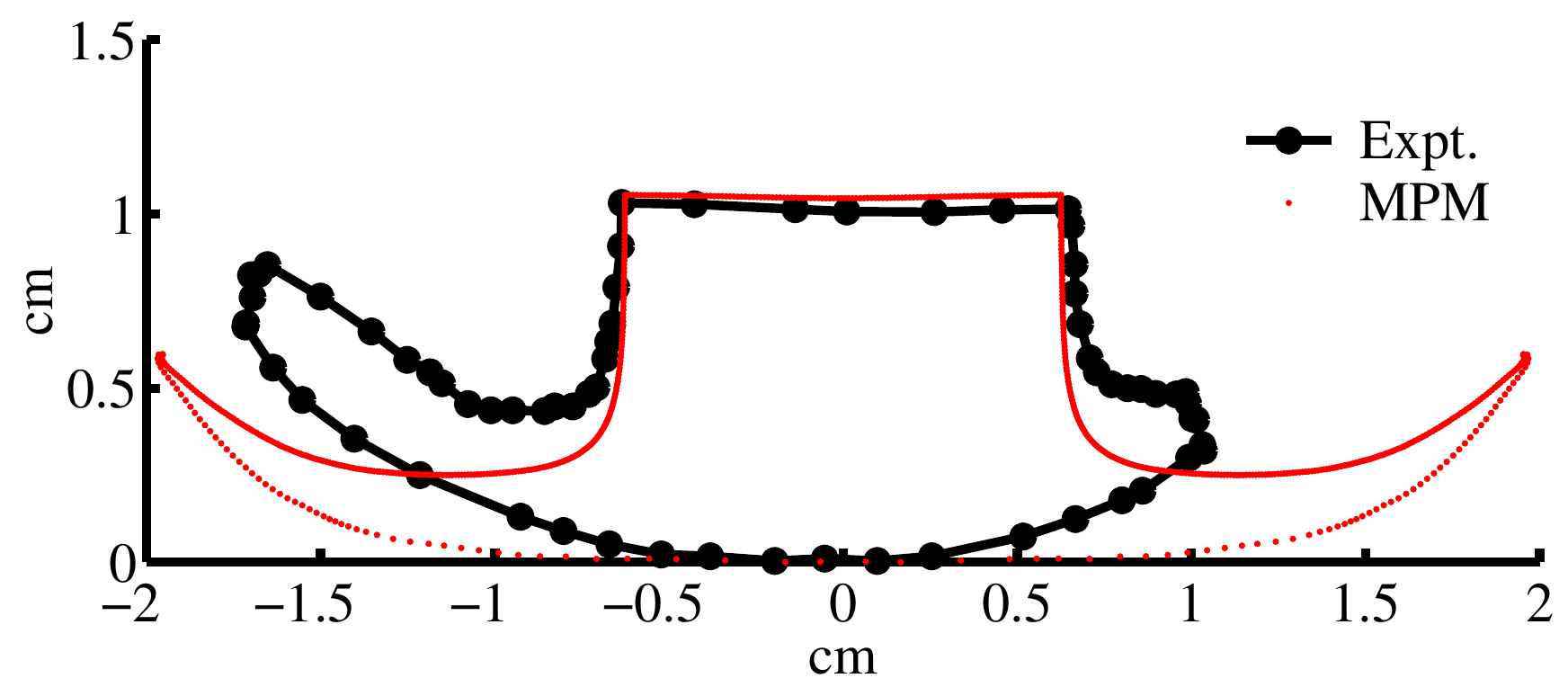}}\\
      (b) Computed vs. actual profile.
    \end{minipage}
    \vspace{12pt}
   \caption{\small Comparison of experimental vs. computed shapes. $L_0$ = 25.37 mm ,
            $D_0$ = 12.7 mm, $V_0$ = 652.3 m/s.}
   \label{fig:carr}
  \end{figure}

  \hspace{16pt}
  In the paper by \citet{Carrington48} a highly deformed mild steel
  specimen has been shown (plate 1, figure 3).  To determine if MPM could
  be used to simulate such large deformations, we ran a Taylor impact test
  on the problem geometry using the Johnson-Cook plasticity model for
  4340 steel.  The final deformed shape from Carrington's paper
  is compared with our predicted shape in Figure~\ref{fig:carr}.  The initial
  velocity is $V_0 = 2140 ft/s = 652.3 m/s$, the initial diameter of the
  cylinder is $D_0 = 0.5 in = 12.7 mm$, and the initial length of the
  cylinder is $L_0 = 0.999 in = 25.37 mm$.

  \hspace{16pt}
  In our simulation, the 4340 steel cylinder was impacted against a stiff 
  anvil using frictional contact.  The 4340 steel flows much more readily 
  than the mild steel used by \citet{Carrington48}.  The experiment also 
  shows that the tips of the ``mushroom'' have broken off.  We did not simulate
  any fracture and hence we do not see that effect.  However, the overall 
  shape of the deformed specimen suggests that our simulations can provide
  good qualitative descriptions of large deformations.  To quantify how
  well our simulations fit experimental data, we ran a series of Taylor impact
  tests on various materials and compared them against experimental data.
  Some of those results are presented in this report.

  \subsection{Taylor impact tests on copper}
  In this section we present the results from Taylor tests on copper 
  specimens for different initial temperatures and impact velocities.
  Table~\ref{tab:copper} shows the initial dimensions, velocity, and 
  temperature of the specimens (along with the type of copper used and
  the source of the data) that we have simulated and compared with 
  experimental data.
  \begin{table}[t]
    \caption{\small Initial data for copper simulations.  OFHC = oxygen free
     high conductivity.  ETP = electrolytic tough pitch.}
    \begin{tabular}{lllllll}
       \hline
       \hline
       Case & Material
            & Initial & Initial
            & Initial & Initial 
            & Source \\
            & 
            & Length & Diameter 
            & Velocity & Temperature \\
            & 
            & ($L_0$ mm) & ($D_0$ mm)
            & ($V_0$ m/s) & ($T_0$ K)\\
       \hline
       \hline
        Cu-A   & OFHC Cu & 23.47 & 7.62 & 210 & 298   & \citet{Wilkins73} \\
        Cu-B   & OFHC Cu & 25.4  & 7.62 & 130 & 298   & \citet{Johnson83} \\
        Cu-C   & OFHC Cu & 25.4  & 7.62 & 146 & 298   & \citet{Johnson83} \\
        Cu-D   & OFHC Cu & 25.4  & 7.62 & 190 & 298   & \citet{Johnson83} \\
        Cu-E   & ETP  Cu & 30    & 6.00 & 277 & 295   & \citet{Gust82} \\
        Cu-F   & ETP  Cu & 30    & 6.00 & 188 & 718   & \citet{Gust82} \\
        Cu-G   & ETP  Cu & 30    & 6.00 & 211 & 727   & \citet{Gust82} \\
        Cu-H   & ETP  Cu & 30    & 6.00 & 178 & 1235  & \citet{Gust82} \\
        Cu-I   & Annealed Cu. &   &  &  &  & \citet{Zocher00}\\
        Cu-J   & With porosity &  &  &  &  & \citet{Addessio93a}\\
       \hline
       \hline
    \end{tabular}
    \label{tab:copper}
  \end{table}

  \subsubsection{Room temperature impact of copper}
\begin{figure}[htb!]
    \begin{minipage}[t]{0.48\linewidth}
      \centering
      \scalebox{0.35}{\includegraphics{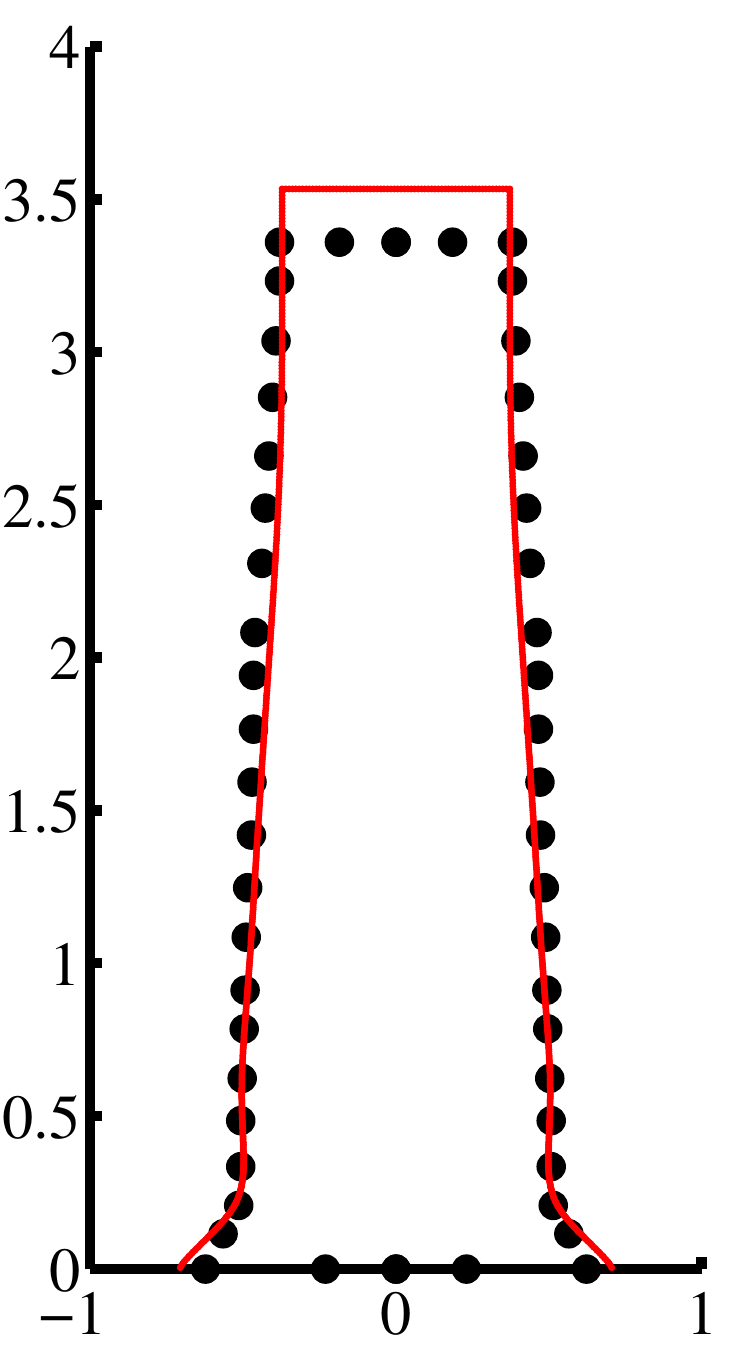}}\\
      (a) Johnson-Cook.
    \end{minipage}
    \begin{minipage}[t]{0.48\linewidth}
      \centering
      \scalebox{0.35}{\includegraphics{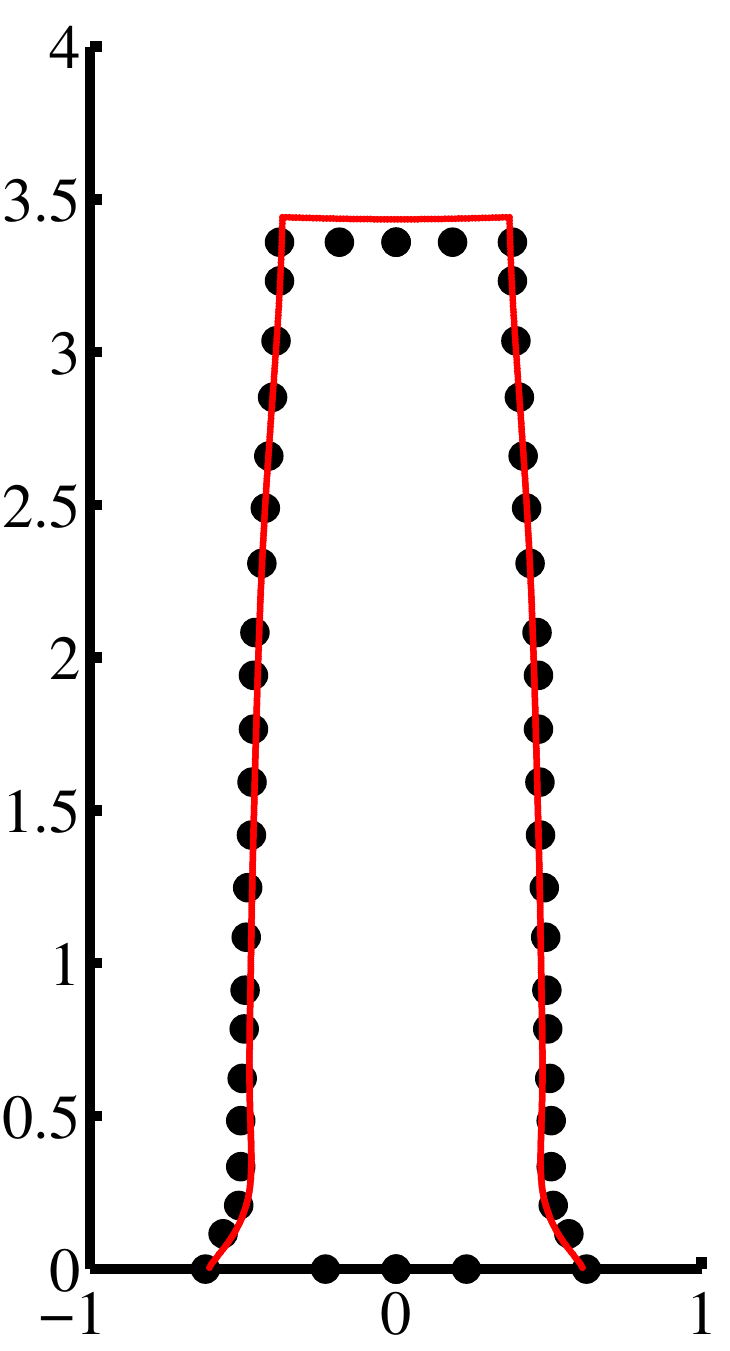}}\\
      (b) Mechanical Threshold Stress.
    \end{minipage}
    \vspace{12pt}
   \caption{\small Comparison of experimental and computed shapes of annealed copper
            cylinder Cu-I using the Johnson-Cook and Mechanical Threshold
            Stress plasticity models.  The axes are shown in cm units.}
   \label{fig:CuModelRoom}
  \end{figure}

  Comparisons between the computed and experimental profiles of annealed copper
  specimen Cu-I are shown in Figure~\ref{fig:CuModelRoom}.  The MTS model predicts the
  final length quite accurately (at this is true for other room temperature simulations
  of copper).  The profile shape is also computed accurately.  The Johnson-Cook model 
  overestimate the final length.  However, the difference is small and may be attributed to 
  material variability.

\begin{figure}[htb!]
    \begin{minipage}[t]{0.48\linewidth}
      \centering
      \scalebox{0.48}{\includegraphics{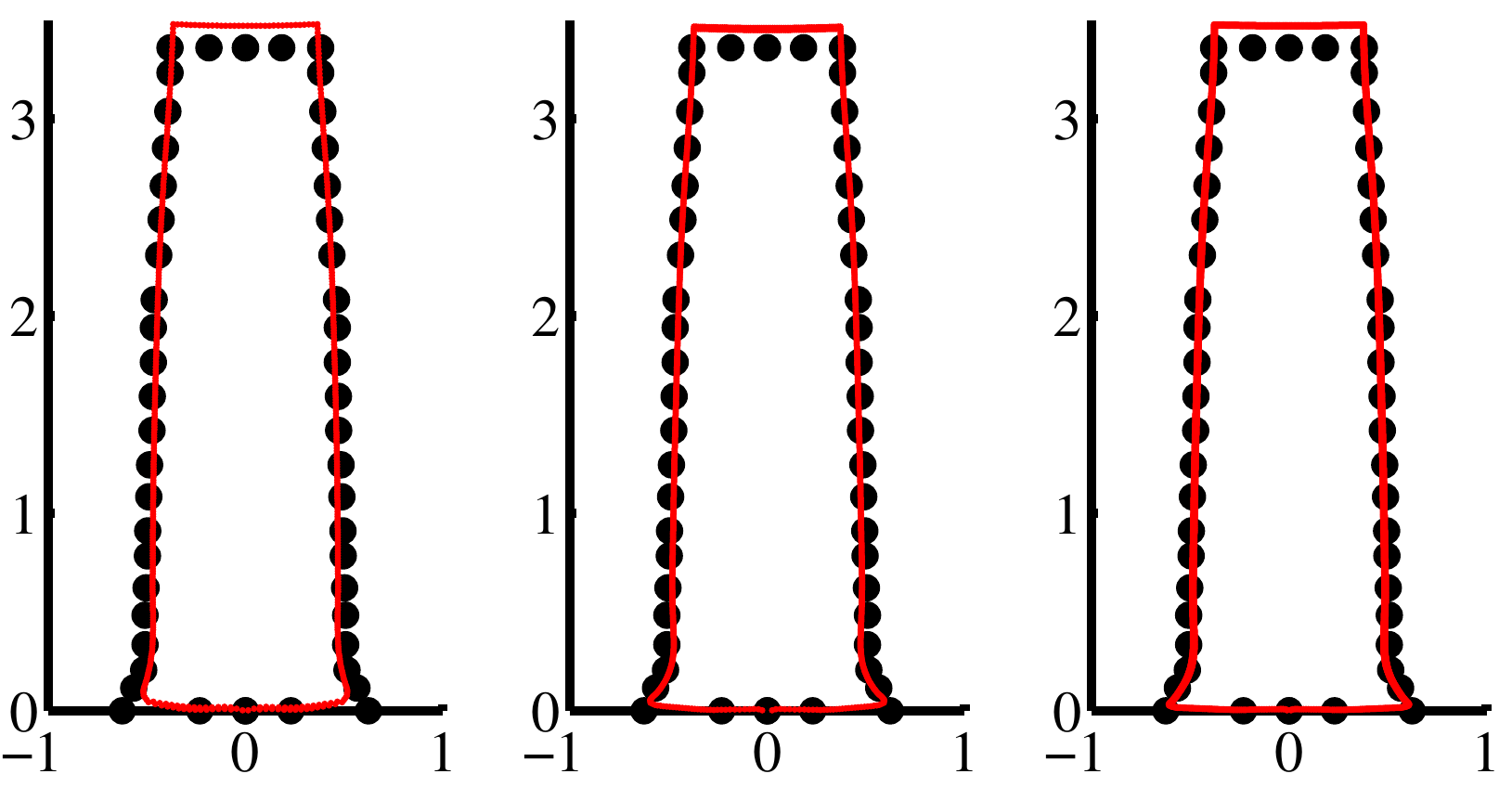}}\\
      (a) With friction.
    \end{minipage}
    \begin{minipage}[t]{0.48\linewidth}
      \centering
      \scalebox{0.48}{\includegraphics{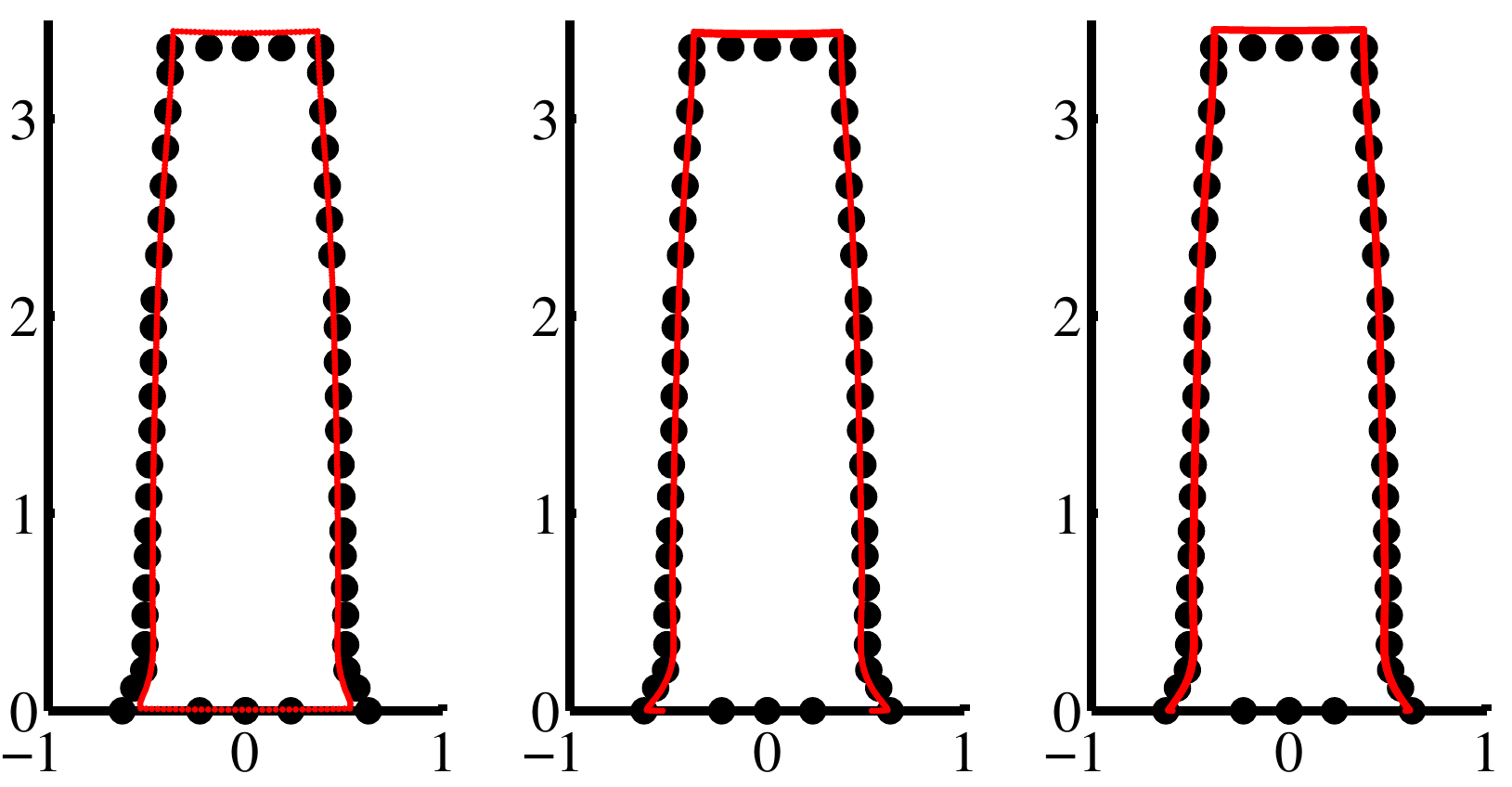}}\\
      (b) Without friction.
    \end{minipage}
    \vspace{12pt}
    \caption{\small Comparison of experimental and computed shapes of 6061T6 aluminum
             cylinders using the Johnson-Cook (JC) with increasing mesh
             refinement. The axes are in cm.}
    \label{fig:CuRefine}
  \end{figure}

  Simulations of impact case Cu-I with increasing mesh refinement are shown
  in Figure~\ref{fig:CuRefine}.  The number of MPM particles is doubled with each refinement.
  We observe that the solution does not change much as we refine the mesh.  However, this 
  is true only at low temperatures and moderate impact velocities.  Significant mesh
  dependence is observed at high temperatures where softening becomes dominant as wee will see
  in our calculations with 6061-T6 aluminum.

  \subsubsection{High temperature impact of copper}
\begin{figure}[htb!]
    \begin{minipage}[t]{0.16\linewidth}
      \centering
      \scalebox{0.30}{\includegraphics{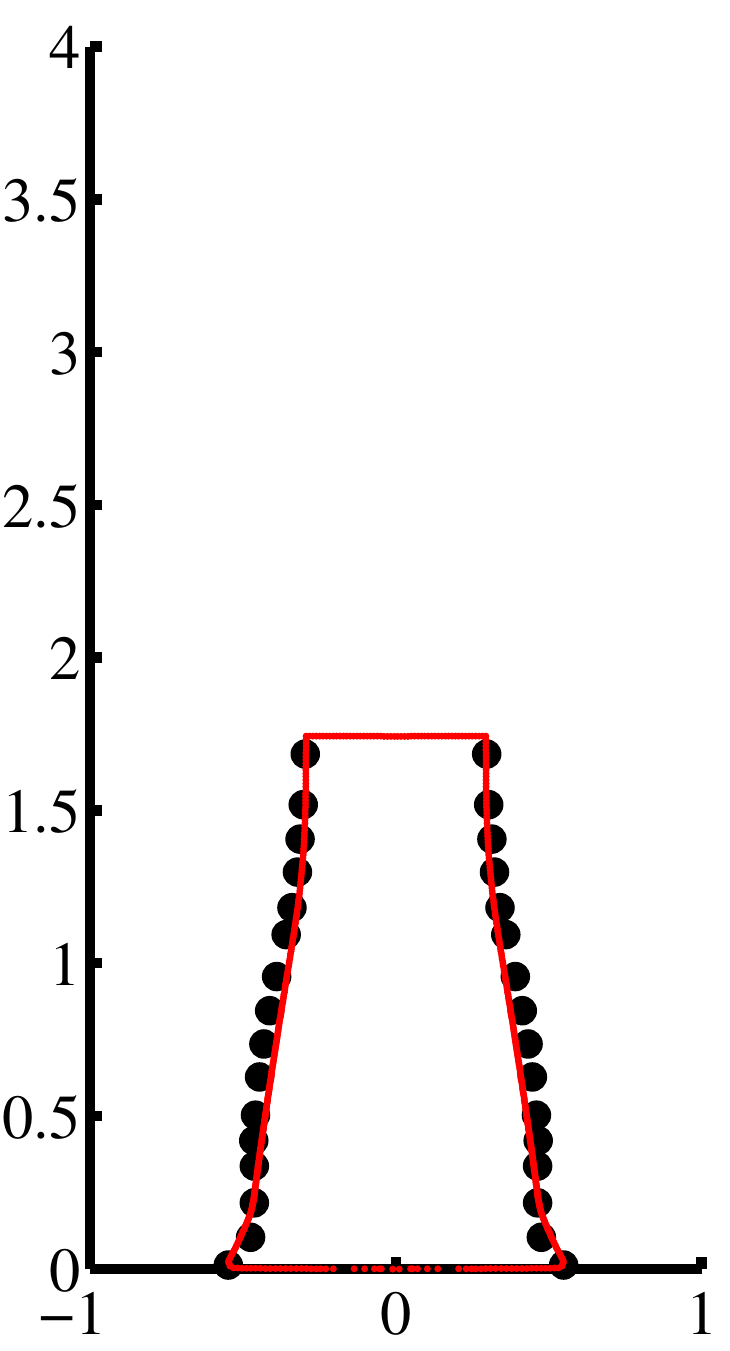}}\\
      (a) JC (Cu-F). \\
      \vspace{12pt}
      \scalebox{0.30}{\includegraphics{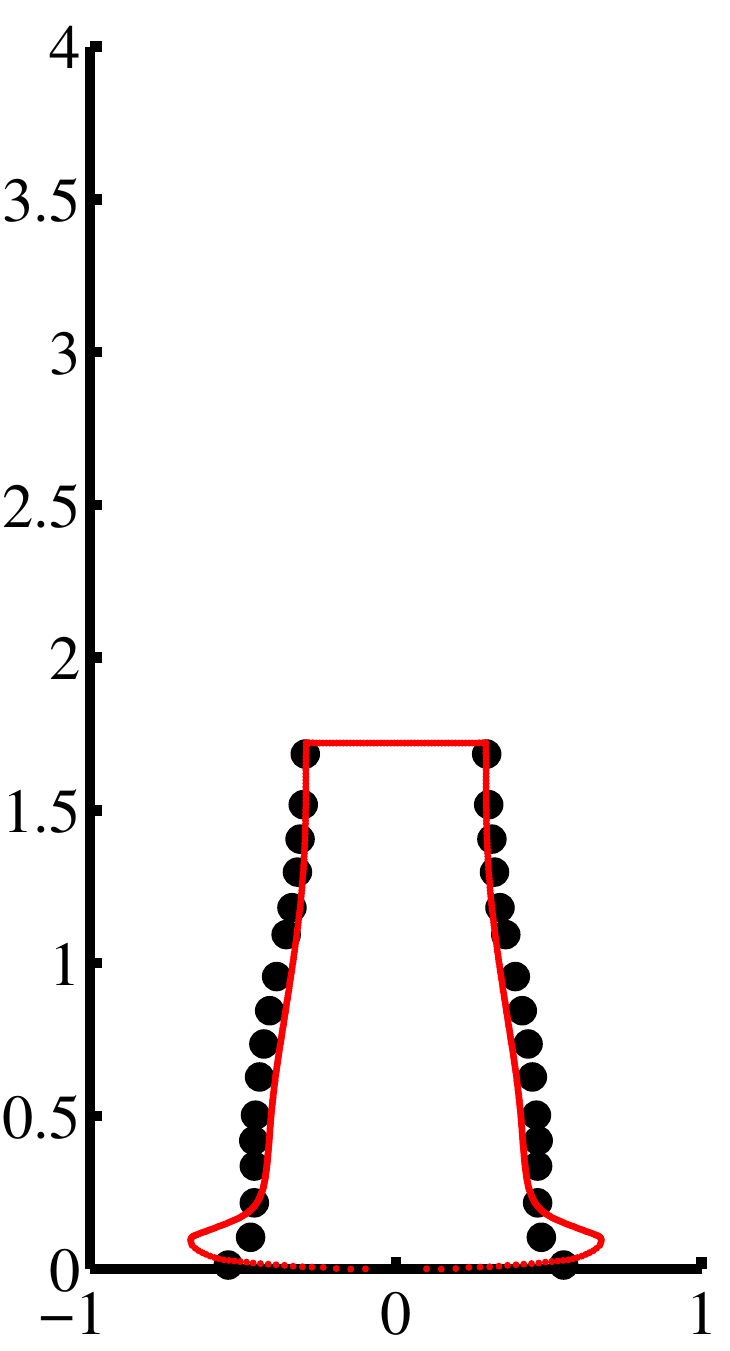}}\\
      (g) JC (Cu-F) with friction.
    \end{minipage}
    \begin{minipage}[t]{0.16\linewidth}
      \centering
      \scalebox{0.30}{\includegraphics{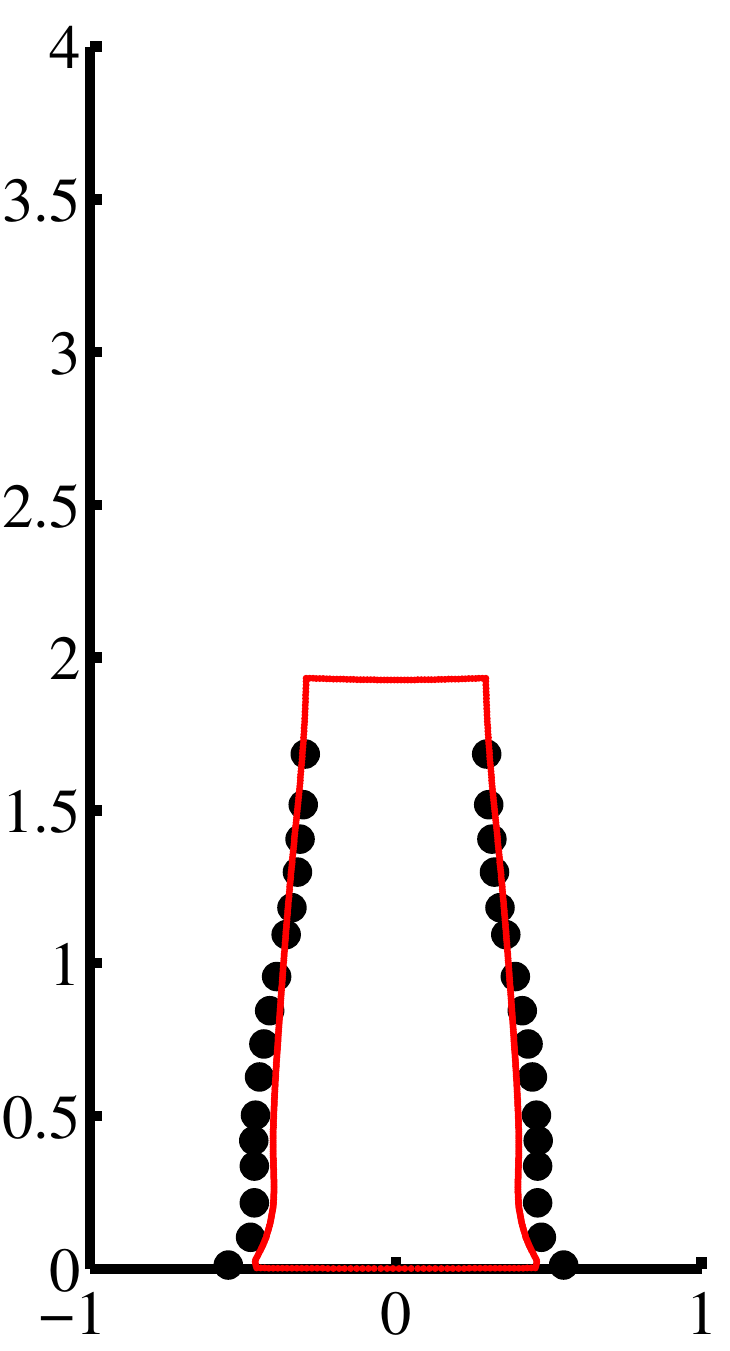}}\\
      (b) MTS (Cu-F). \\
      \vspace{12pt}
      \scalebox{0.30}{\includegraphics{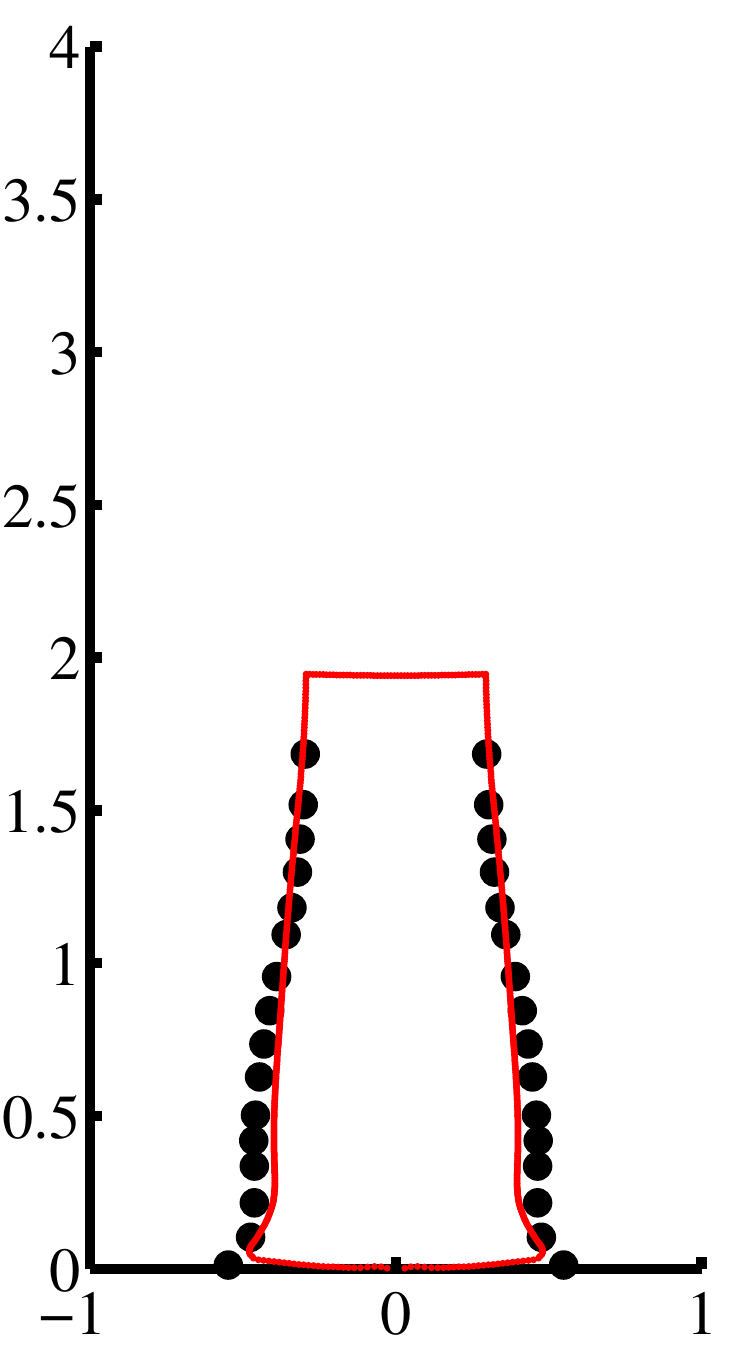}}\\
      (h) MTS (Cu-F) with friction.
    \end{minipage}
    \begin{minipage}[t]{0.16\linewidth}
      \centering
      \scalebox{0.30}{\includegraphics{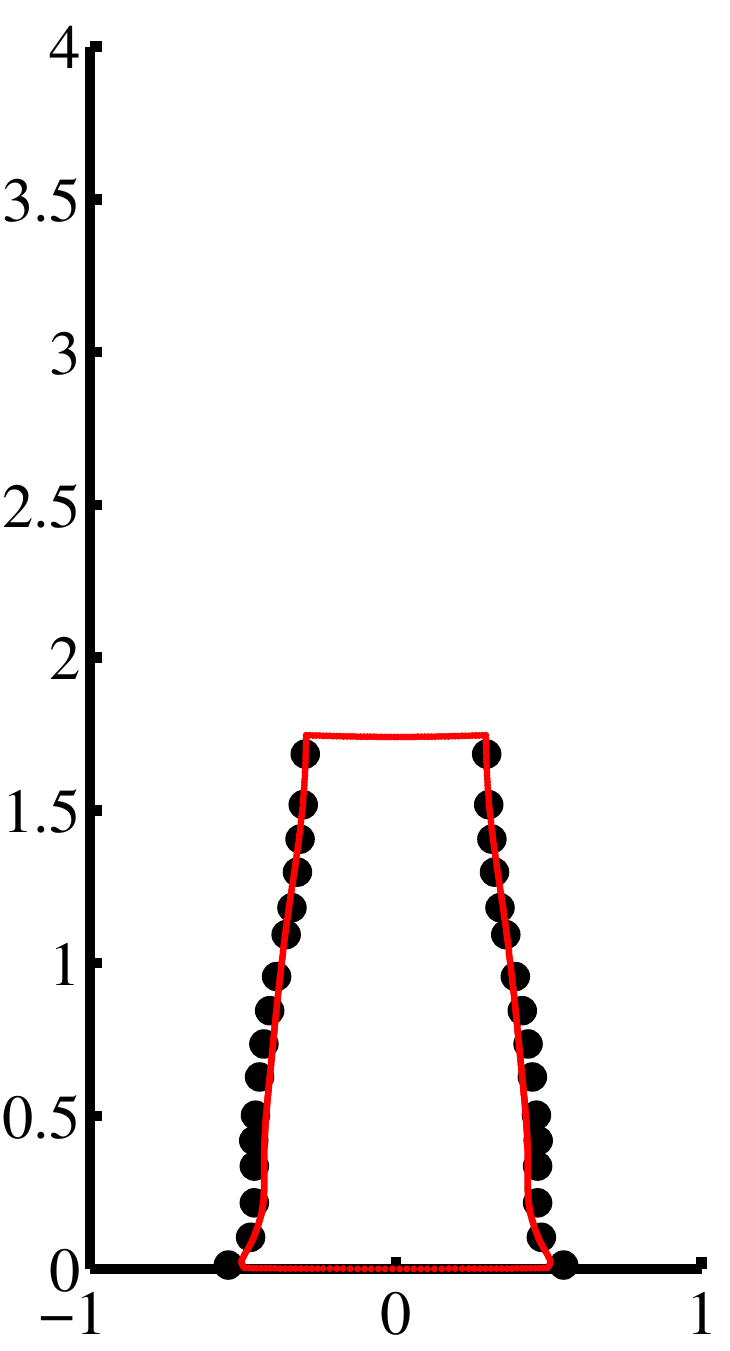}}\\
      (c) SCG (Cu-F). \\
      \vspace{12pt}
      \scalebox{0.30}{\includegraphics{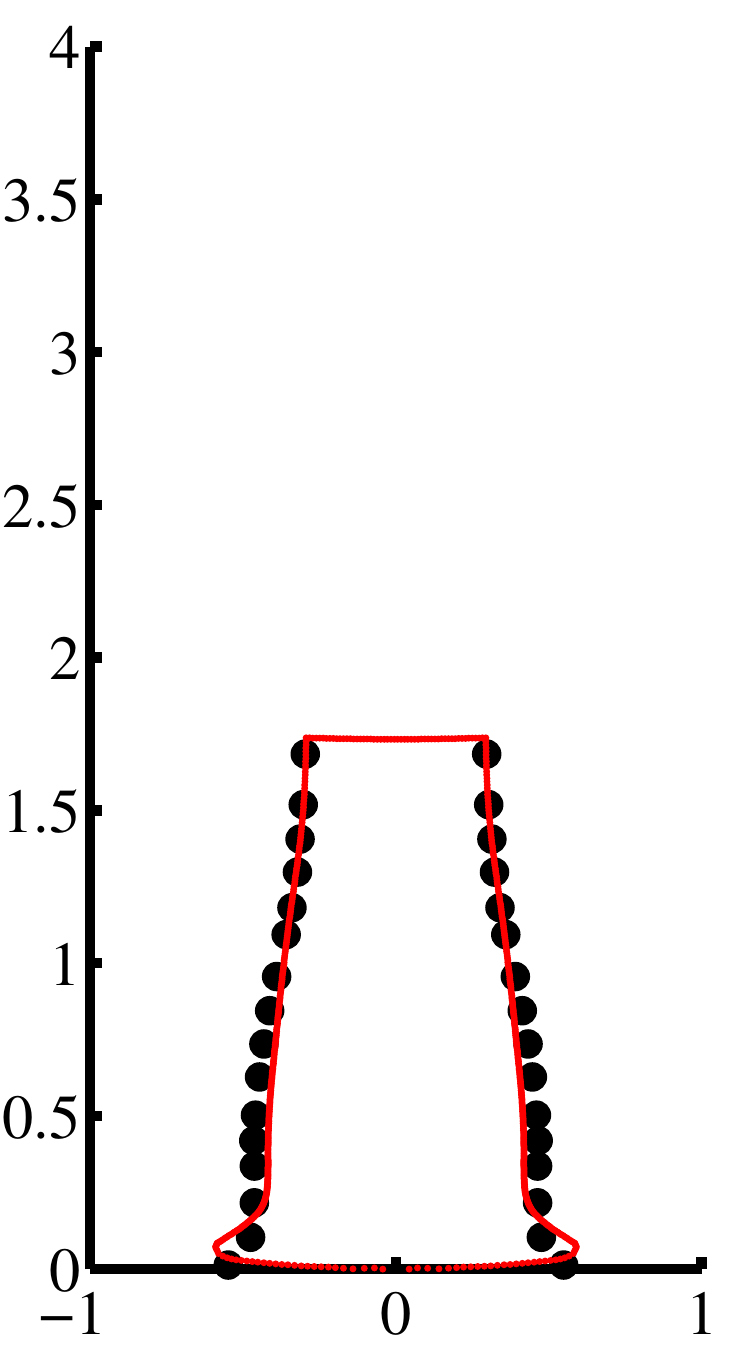}}\\
      (i) SCG (Cu-F) with friction.
    \end{minipage}
    \begin{minipage}[t]{0.16\linewidth}
      \centering
      \scalebox{0.30}{\includegraphics{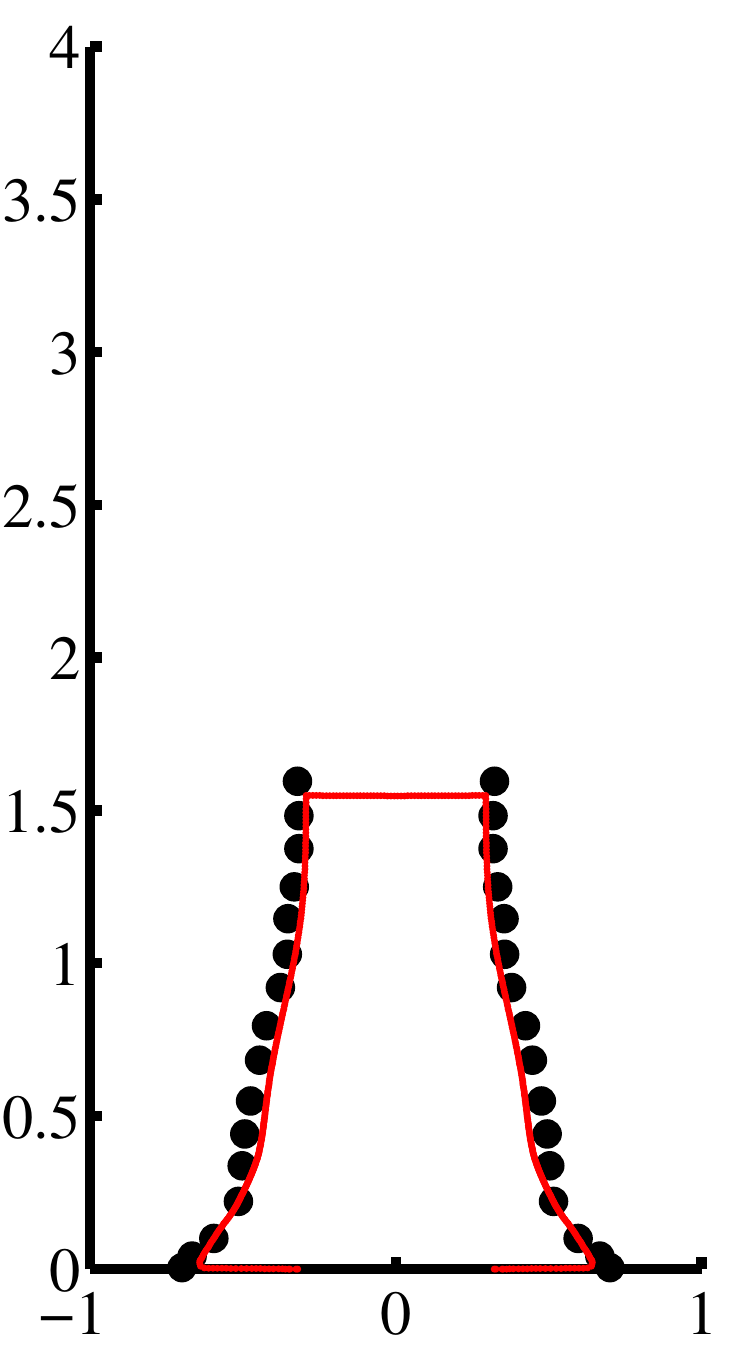}}\\
      (d) JC (Cu-G). \\
      \vspace{12pt}
      \scalebox{0.30}{\includegraphics{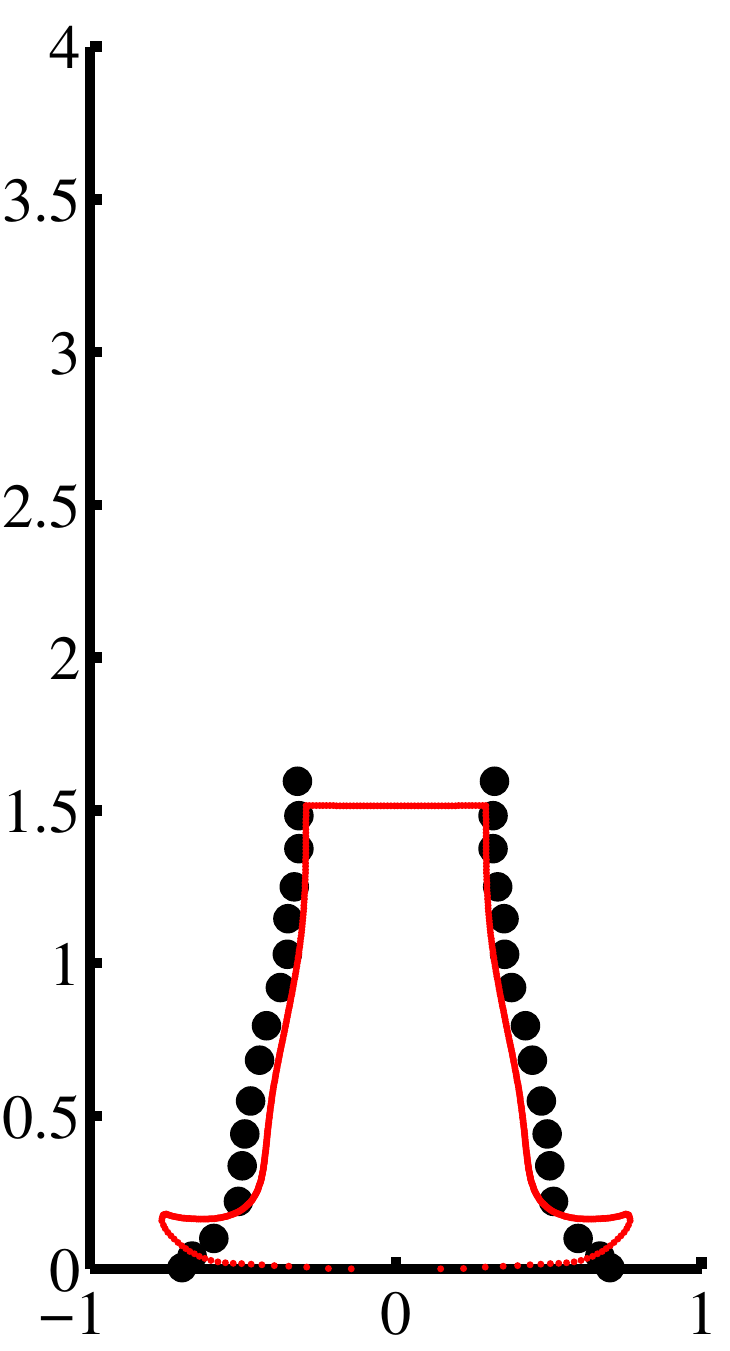}}\\
      (j) JC (Cu-G) with friction.
    \end{minipage}
    \begin{minipage}[t]{0.16\linewidth}
      \centering
      \scalebox{0.30}{\includegraphics{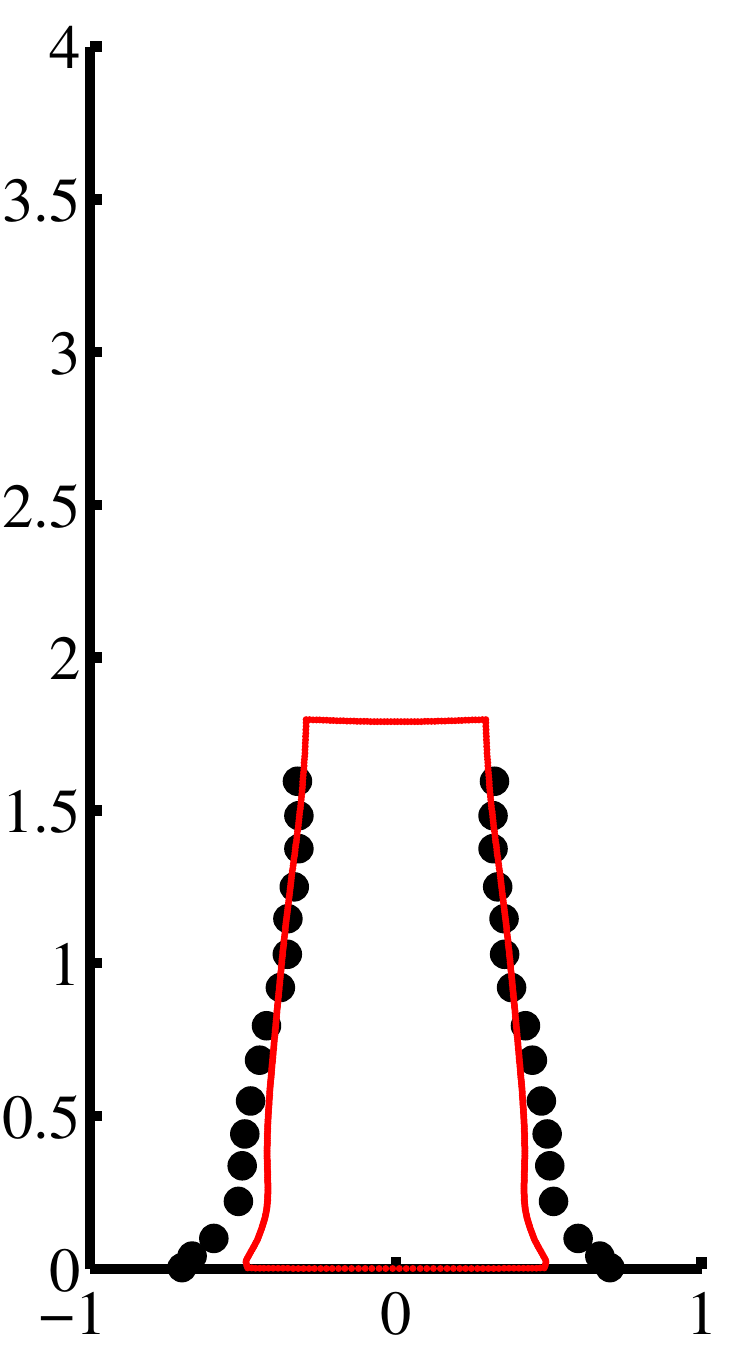}}\\
      (e) MTS (Cu-G). \\
      \vspace{12pt}
      \scalebox{0.30}{\includegraphics{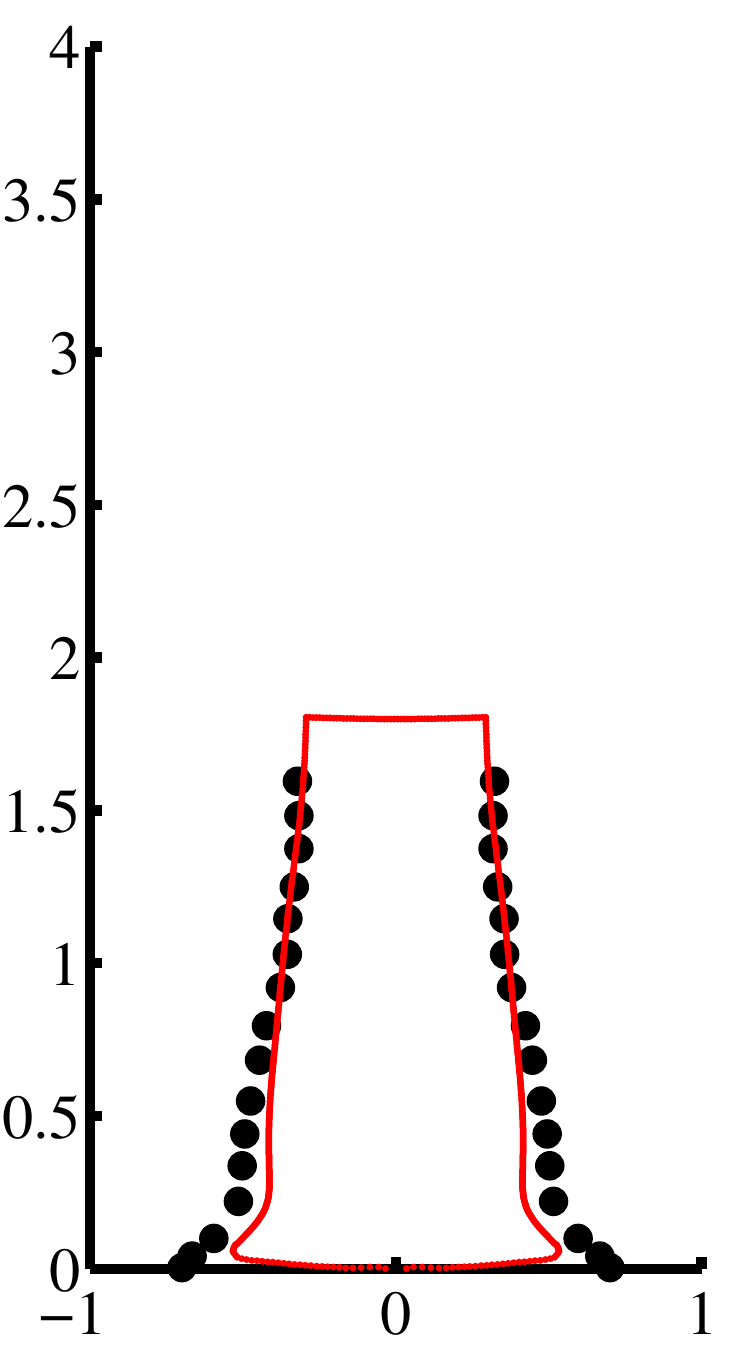}}\\
      (k) MTS (Cu-G) with friction.
    \end{minipage}
    \begin{minipage}[t]{0.16\linewidth}
      \centering
      \scalebox{0.30}{\includegraphics{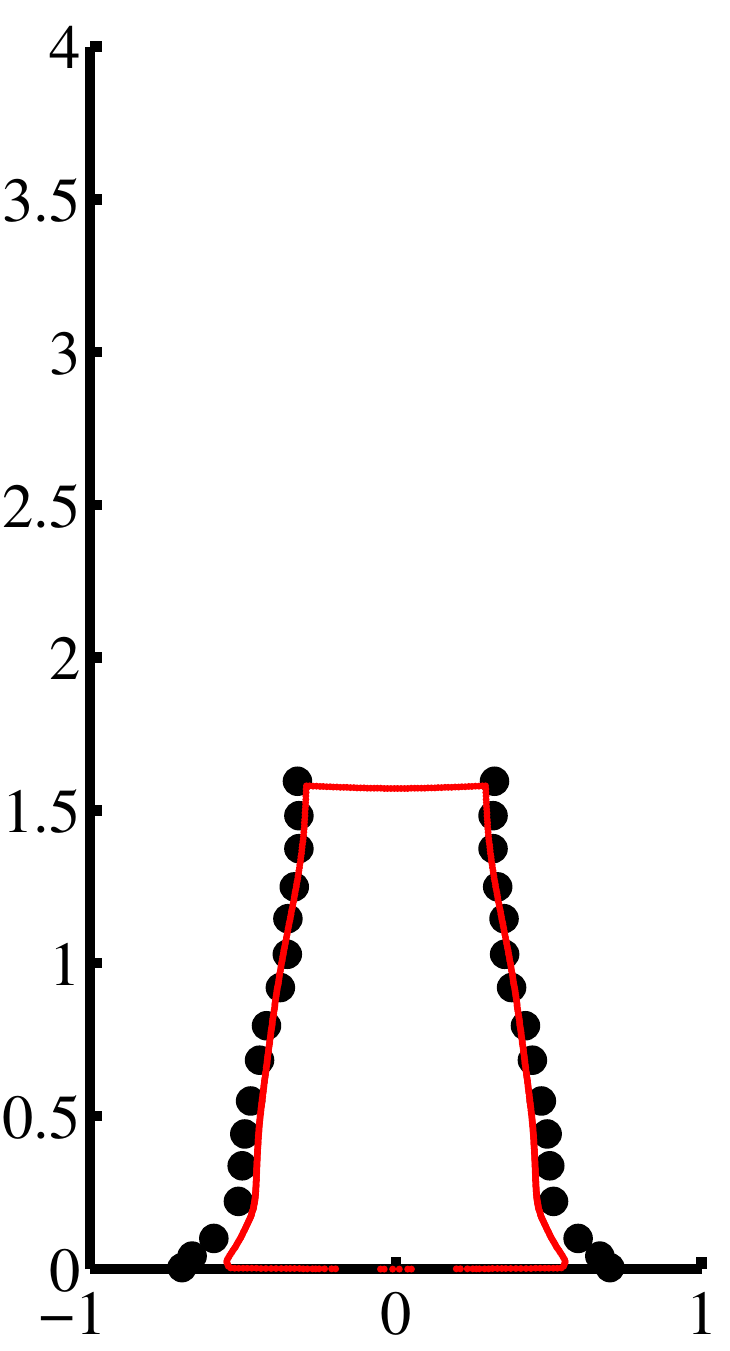}}\\
      (f) SCG (Cu-G). \\
      \vspace{12pt}
      \scalebox{0.30}{\includegraphics{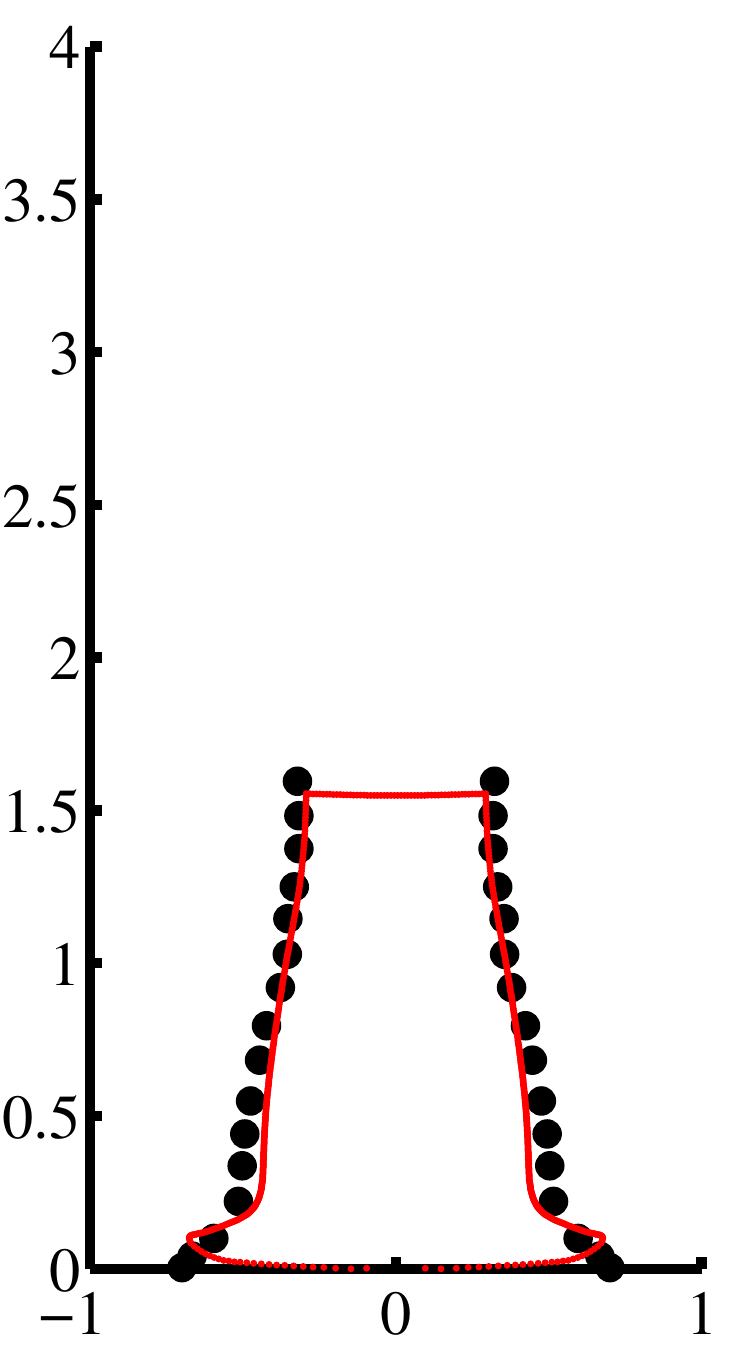}}\\
      (l) SCG (Cu-G) with friction.
    \end{minipage}
    \vspace{12pt}
   \caption{\small Comparison of experimental and computed shapes of ETP copper
            cylinders using the Johnson-Cook (JC), Mechanical Threshold
            Stress (MTS), and Steinberg-Cochran-Guinan (SCG) plasticity models.
            Specimen C-F has an initial temperature of 718 K and Cu-G is initially at
            727 K.  The initial velocities are 188 m/s and 211 m/s, respectively.
            The axes are shown in cm units.}
   \label{fig:CuModelHot}
  \end{figure}

  At higher temperatures, the response of the three plasticity models is 
  quite different.  Comparisons between the computed and experimental 
  profiles of ETP copper specimen Cu-F are shown in 
  Figure~\ref{fig:CuModelHot}(a), (b), and (c).  Those for specimen Cu-G
  are shown in Figure~\ref{fig:CuModelHot}(d), (e), and (f).
  If frictional contact at the impact surface is simulated, the final 
  shapes of the specimens Cu-F and Cu-G are as shown in 
  Figure~\ref{fig:CuModelHot}(g), (h), (i), (j), (k), and (l).

  Notice that though both specimens are nominally at the same temperature and
  has almost identical impact velocities, the final profile is quite different
  even though the final lengths are nearly identical.  It is likely that most of 
  the difference is due the initial conditions with a small contribution from 
  material variability.  This conjecture is partially supported by the fact that
  the profiles predicted by the Johnson-Cook model match the experiments quite well.

  We observe that the Johnson-Cook and Steinberg-Guinan models perform well for 
  specimen Cu-F when friction is not included in the calculation.  In the presence of
  frictional contact, the predicted profiles deviate significantly from the experimental
  profiles in the mushroom region.  This indicates that there is a possibility of 
  inaccurate contact force calculation when friction is included.

  From specimen Cu-G, the slightly higher impact velocity leads to an underestimation 
  of the final length by the Johnson-Cook model, even though the mushroom region is
  predicted accurately.  The MTS model overestimates the length and underestimates the
  mushroom diameter while the Steinberg-Guinan model predicts the final length best but
  fails to predict the mushroom shape.  Once again, frictional contact appears to reduce
  the accuracy of the prediction.

  \subsubsection{Comparisons with FEM}
\begin{figure}[htb!]
    \begin{minipage}[t]{0.24\linewidth}
      \centering
      \scalebox{0.30}{\includegraphics{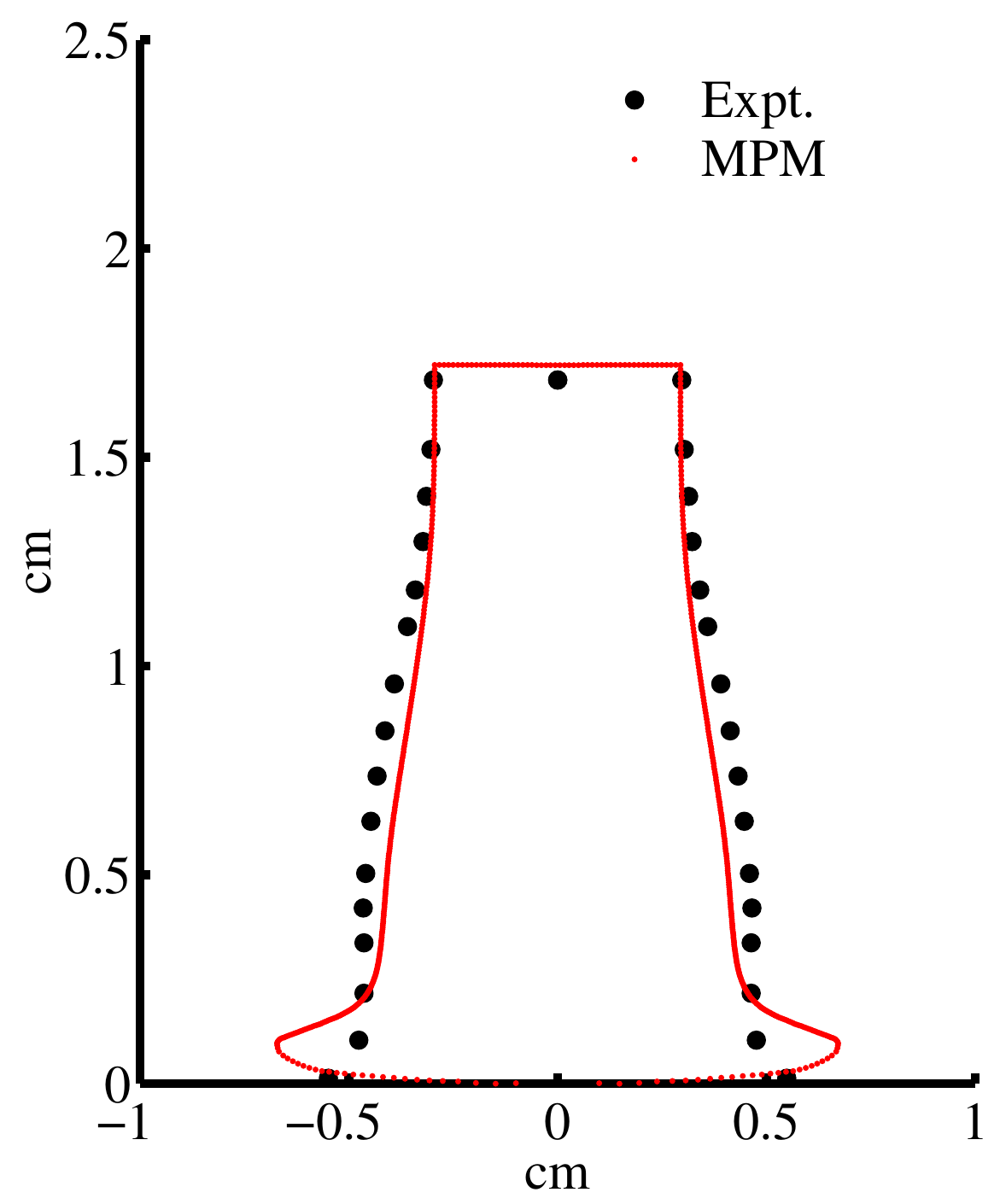}}\\
      (a) MPM (Cu-F)
    \end{minipage}
    \begin{minipage}[t]{0.24\linewidth}
      \centering
      \scalebox{0.30}{\includegraphics{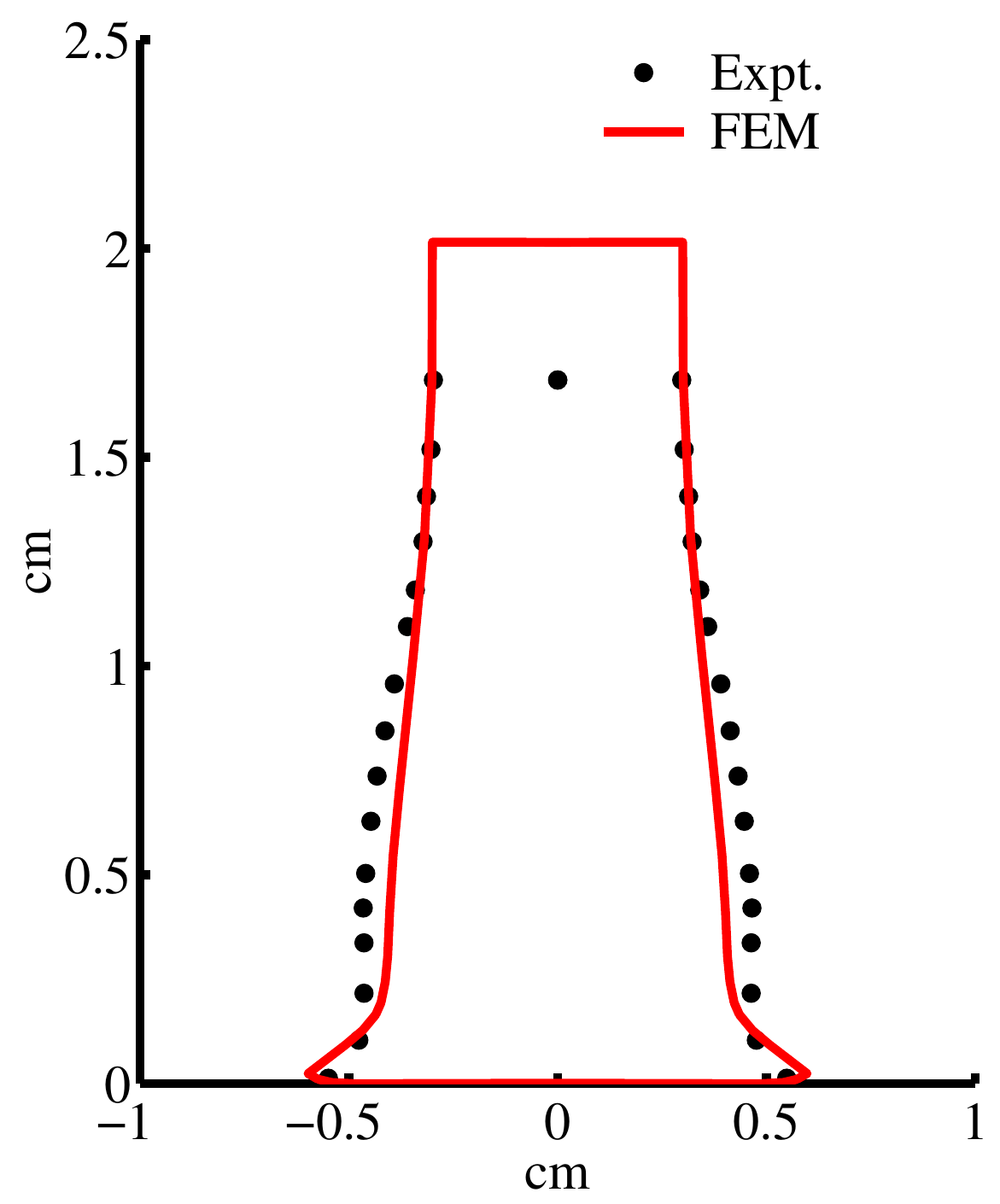}}\\
      (b) FEM (Cu-F)
    \end{minipage}
    \begin{minipage}[t]{0.24\linewidth}
      \centering
      \scalebox{0.30}{\includegraphics{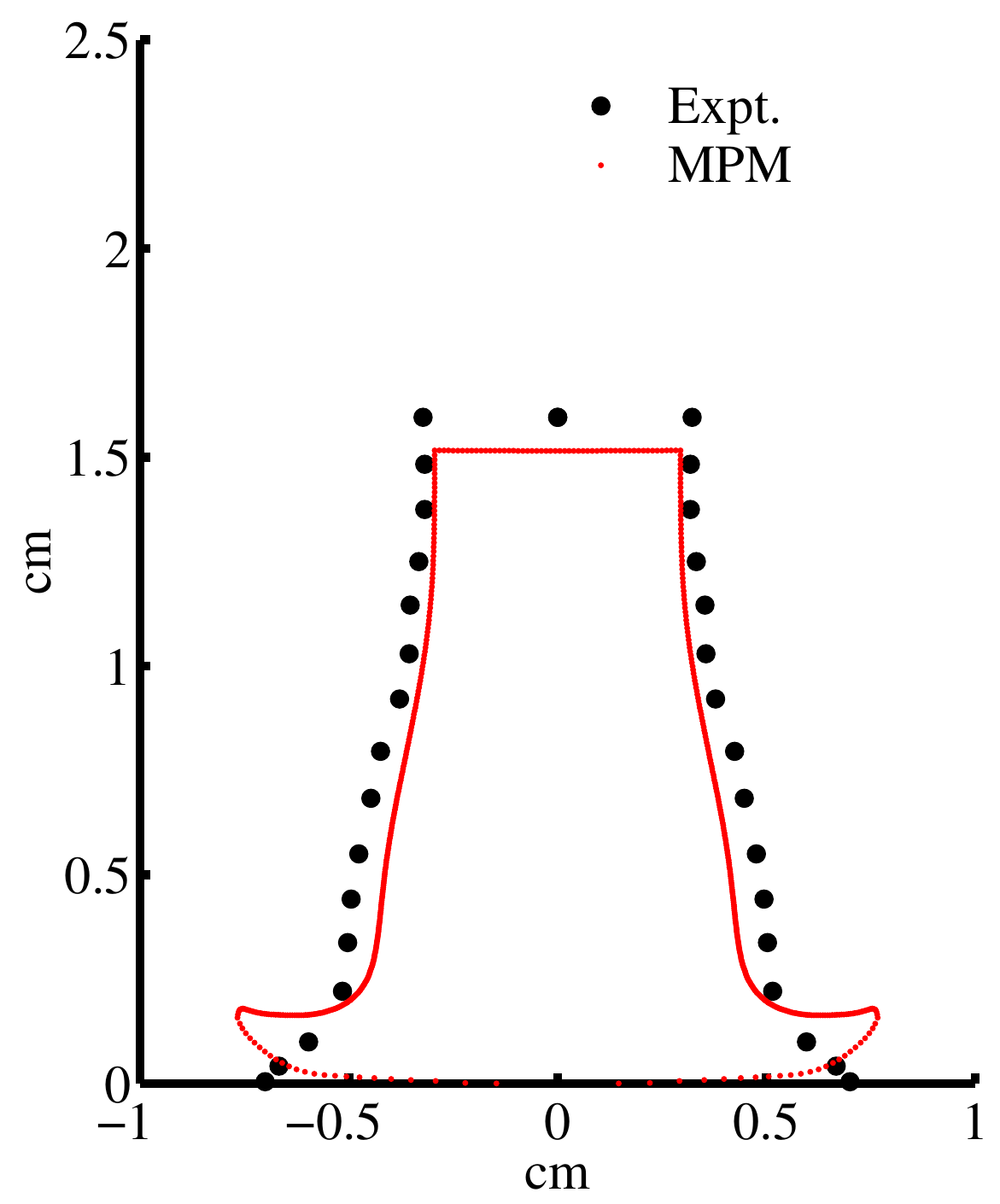}}\\
      (c) MPM (Cu-G)
    \end{minipage}
    \begin{minipage}[t]{0.24\linewidth}
      \centering
      \scalebox{0.30}{\includegraphics{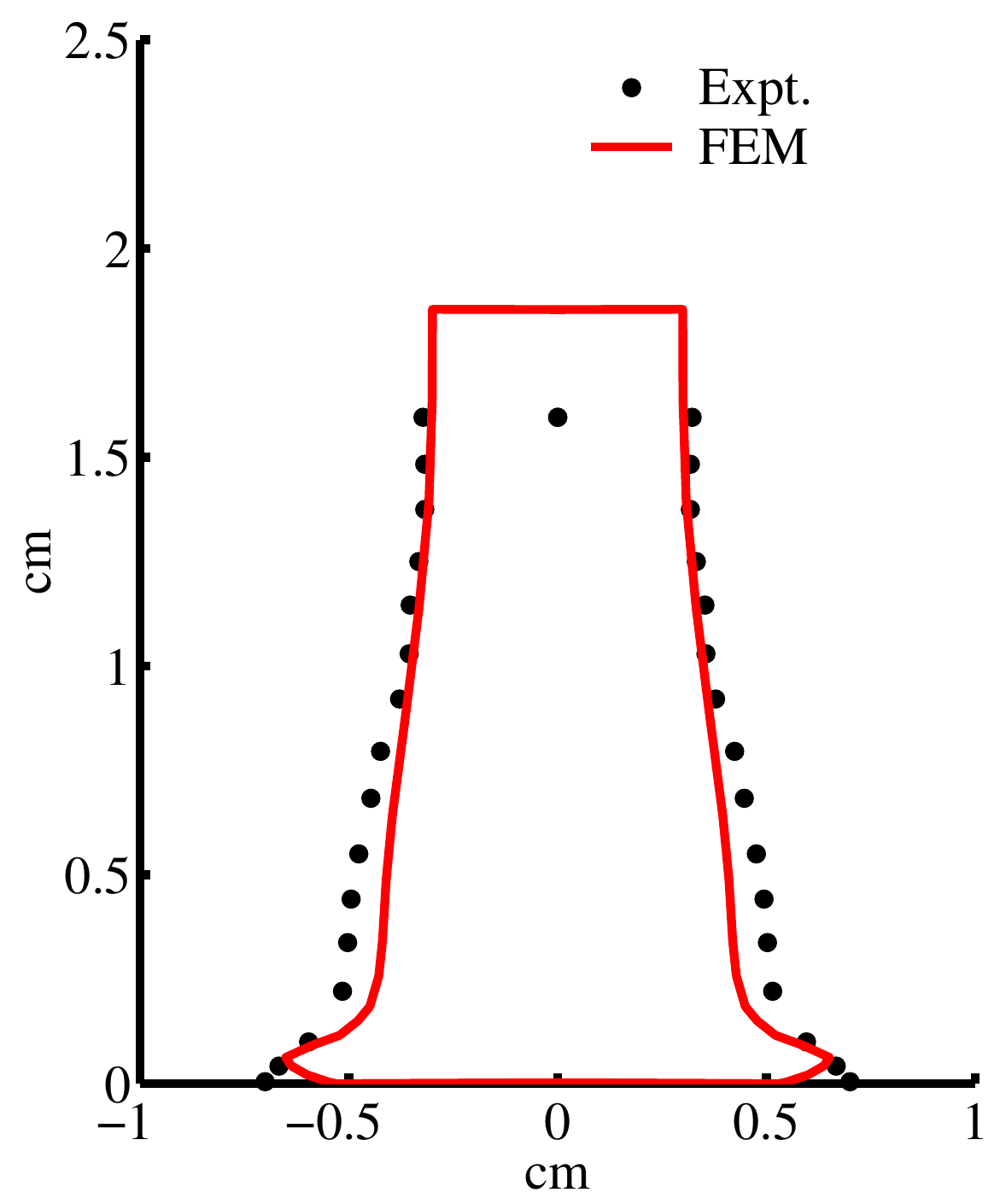}}\\
      (d) FEM (Cu-G)
    \end{minipage}
    \vspace{12pt}
    \caption{\small Comparison of experimental and computed shapes of ETP copper
             cylinders using MPM and FEM. The axes are in cm.}
    \label{fig:CuFE}
  \end{figure}

  To determine how our MPM simulations compare with FEM simulations we have
  run two high temperature ETP copper impact tests using LS-DYNA (with 
  the coupled structural-thermal option).  Figure~\ref{fig:CuFE} shows the
  final deformed shapes for the two cases from the MPM and FEM simulations
  using Johnson-Cook plasticity.  In this case frictional contact has been used.

  The FEM simulations consistently overestimate the final length of the specimen
  though the mushroom diameter is more accurately predicted by FEM.  For the case
  where no contact friction is applied, MPM predictions are consistently better 
  than FEM predictions.
  
  \clearpage
  \subsection{Taylor impact tests on 6061-T6 aluminum alloy}
  In this section we present the results from Taylor tests on 6061-T6
  aluminum specimens for different initial temperatures and impact velocities.
  We have chosen to study this material as it is a well characterized 
  face centered cubic material that has been utilized by \citet{Chhabil99}
  for the validation of high velocity impacts that formed the basis of
  the second stage of our validation simulations.
  Table~\ref{tab:aluminum} shows the initial dimensions, velocity, and 
  temperature of the specimens (along with the type of copper used and
  the source of the data) that we have simulated and compared with 
  experimental data.
  \begin{table}[htb!]
    \caption{\small Initial data for 6061-T6 aluminum simulations. } 
    \begin{tabular}{lllllll}
       \hline
       \hline
       Case & Material
            & Initial & Initial
            & Initial & Initial 
            & Source \\
            & 
            & Length & Diameter 
            & Velocity & Temperature \\
            & 
            & ($L_0$ mm) & ($D_0$ mm)
            & ($V_0$ m/s) & ($T_0$ K)\\
       \hline
       \hline
        Al-A   & 6061-T6 Al & 23.47 & 7.62 & 373 & 298   & \citet{Wilkins73} \\
        Al-B   & 6061-T6 Al & 23.47 & 7.62 & 603 & 298   & \citet{Wilkins73} \\
        Al-C   & 6061-T6 Al & 46.94 & 7.62 & 275 & 298   & \citet{Wilkins73} \\
        Al-D   & 6061-T6 Al & 46.94 & 7.62 & 484 & 298   & \citet{Wilkins73} \\
        Al-E   & 6061-T6 Al & 30    & 6.00 & 200 & 295   & \citet{Gust82} \\
        Al-F   & 6061-T6 Al & 30    & 6.00 & 358 & 295   & \citet{Gust82} \\
        Al-G   & 6061-T6 Al & 30    & 6.00 & 194 & 635   & \citet{Gust82} \\
        Al-H   & 6061-T6 Al & 30    & 6.00 & 354 & 655   & \citet{Gust82} \\
        Al-I   & 6061-T6 Al &       & & & & \citet{Addessio93a}\\
       \hline
       \hline
    \end{tabular}
    \label{tab:aluminum}
  \end{table}

  \subsubsection{Room temperature impact: 6061-T6 Al}
\begin{figure}[htb!]
    \begin{minipage}[t]{0.16\linewidth}
      \centering
      \scalebox{0.30}{\includegraphics{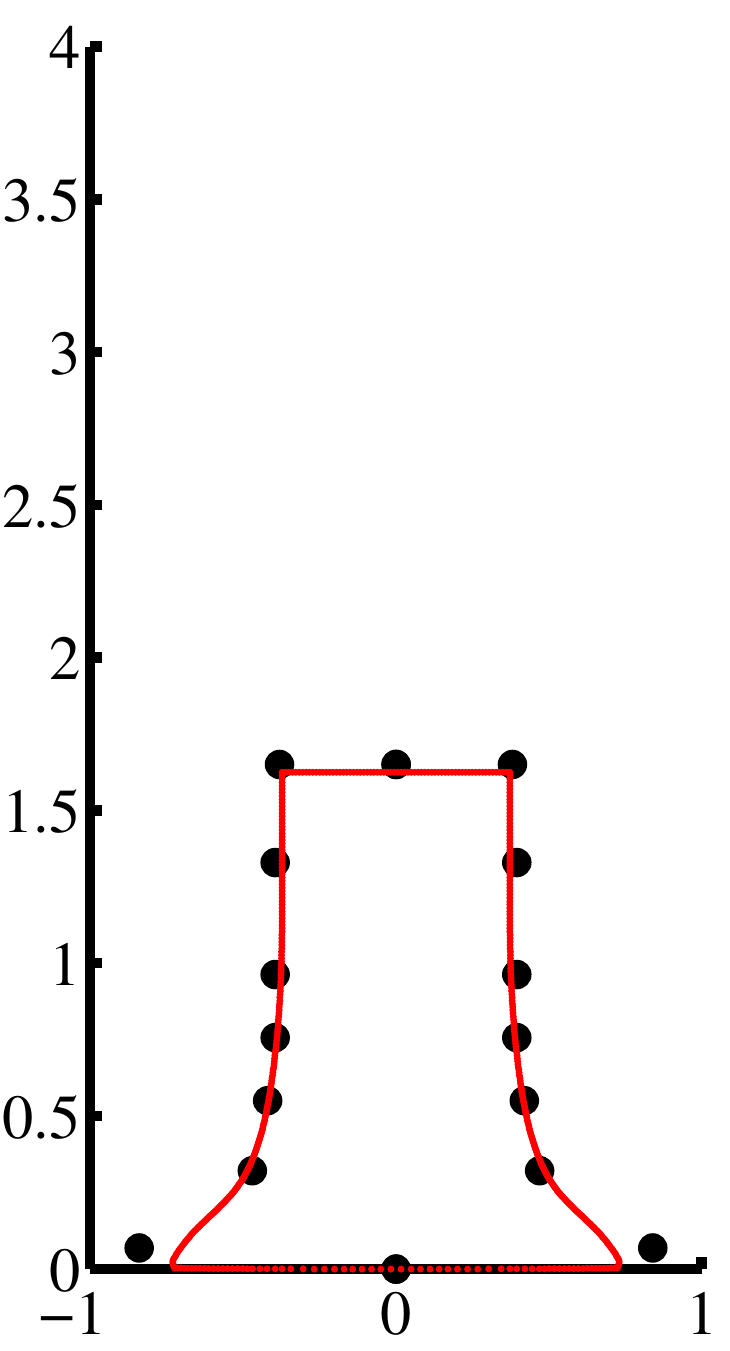}}\\
      (a) JC (Al-A). \\
      \vspace{12pt}
      \scalebox{0.30}{\includegraphics{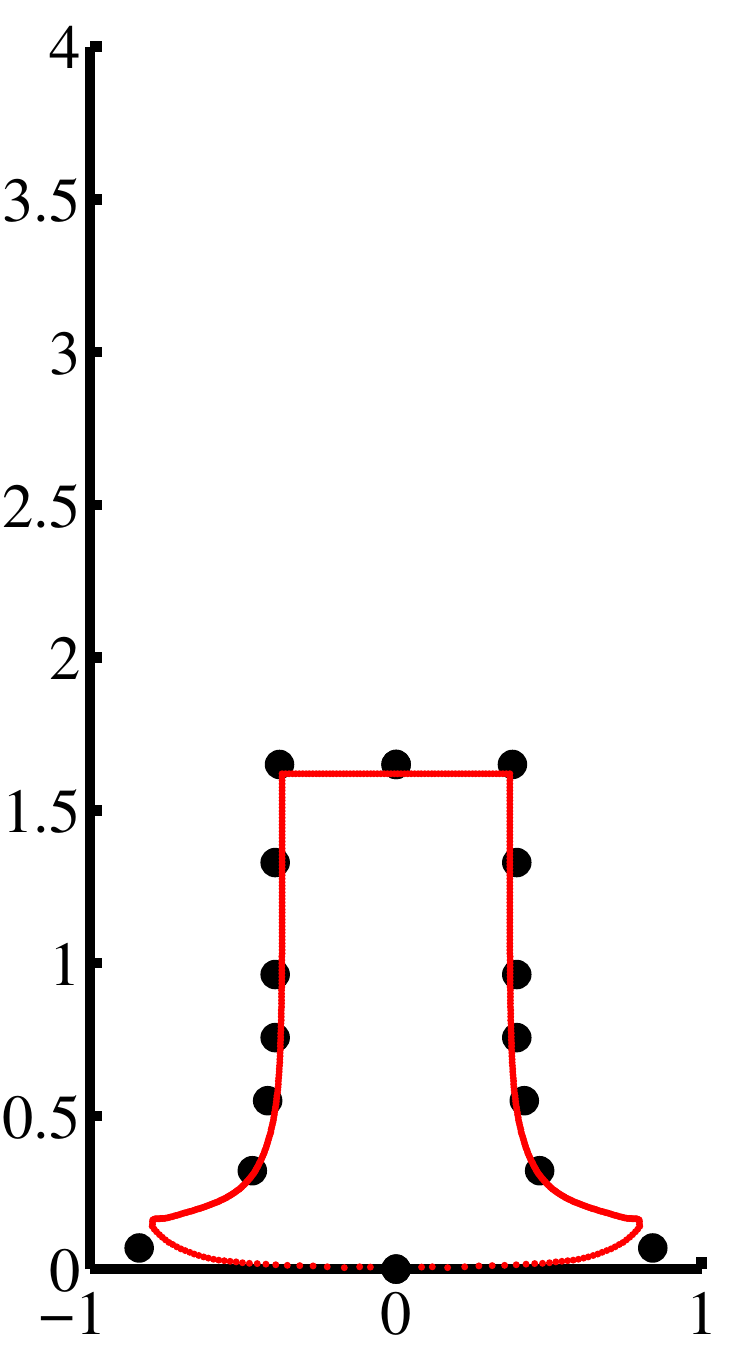}}\\
      (g) JC (Al-A) Friction. 
    \end{minipage}
    \begin{minipage}[t]{0.16\linewidth}
      \centering
      \scalebox{0.30}{\includegraphics{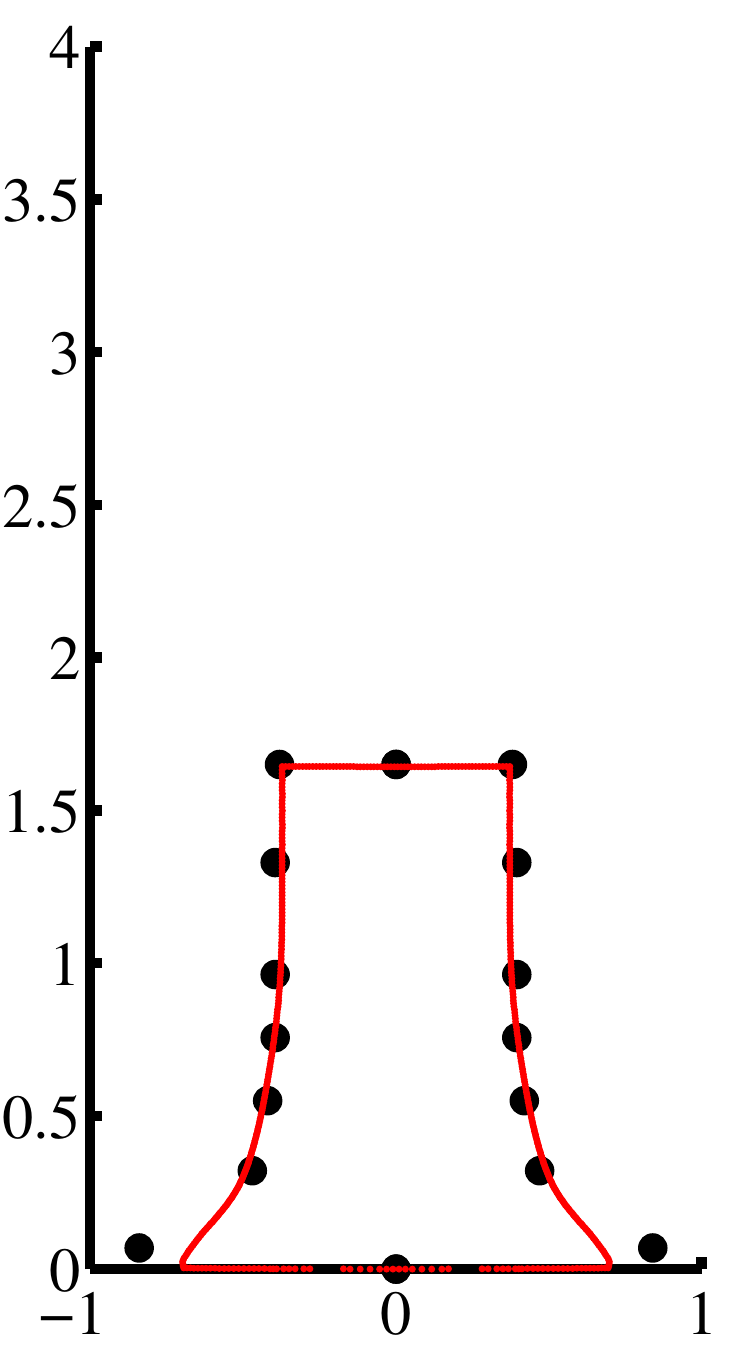}}\\
      (b) MTS (Al-A). \\
      \vspace{12pt}
      \scalebox{0.30}{\includegraphics{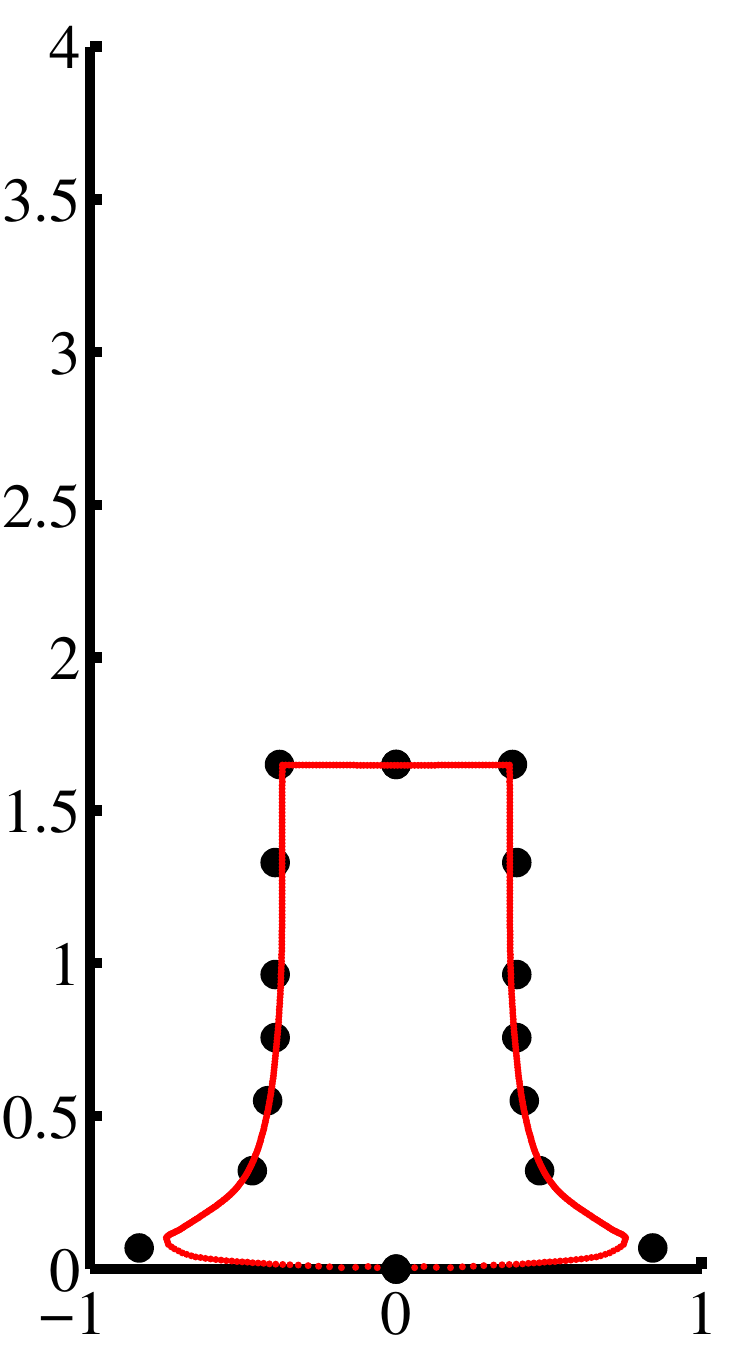}}\\
      (h) MTS (Al-A) Friction.
    \end{minipage}
    \begin{minipage}[t]{0.16\linewidth}
      \centering
      \scalebox{0.30}{\includegraphics{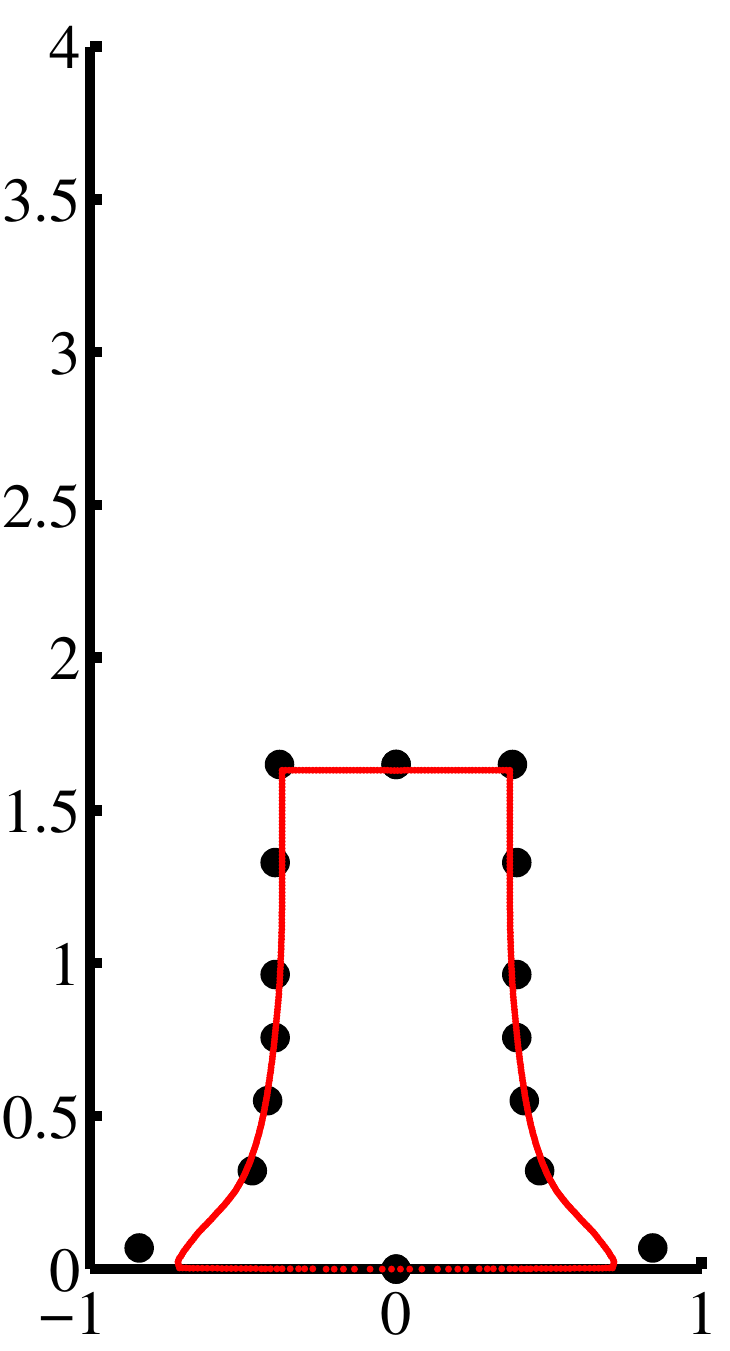}}\\
      (c) SCG (Al-A). \\
      \vspace{12pt}
      \scalebox{0.30}{\includegraphics{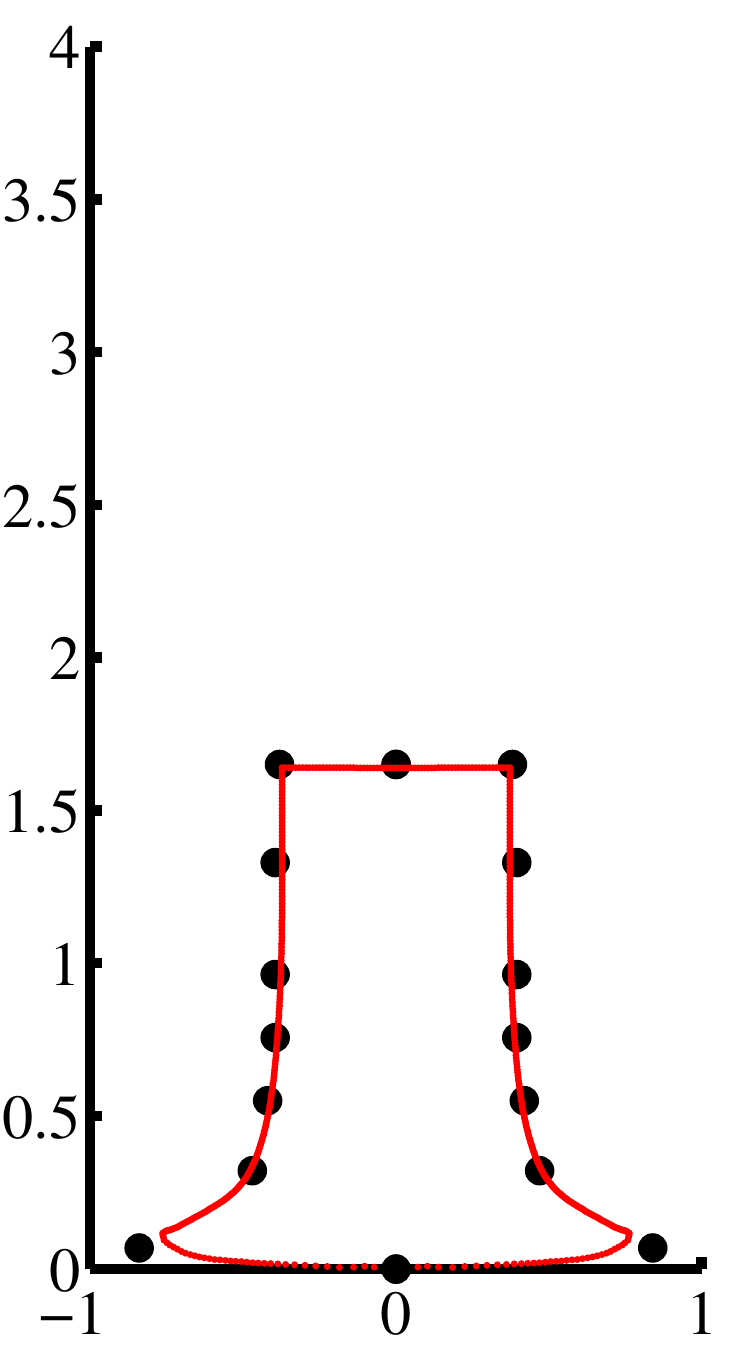}}\\
      (i) SCG (Al-A) Friction.
    \end{minipage}
    \begin{minipage}[t]{0.16\linewidth}
      \centering
      \scalebox{0.30}{\includegraphics{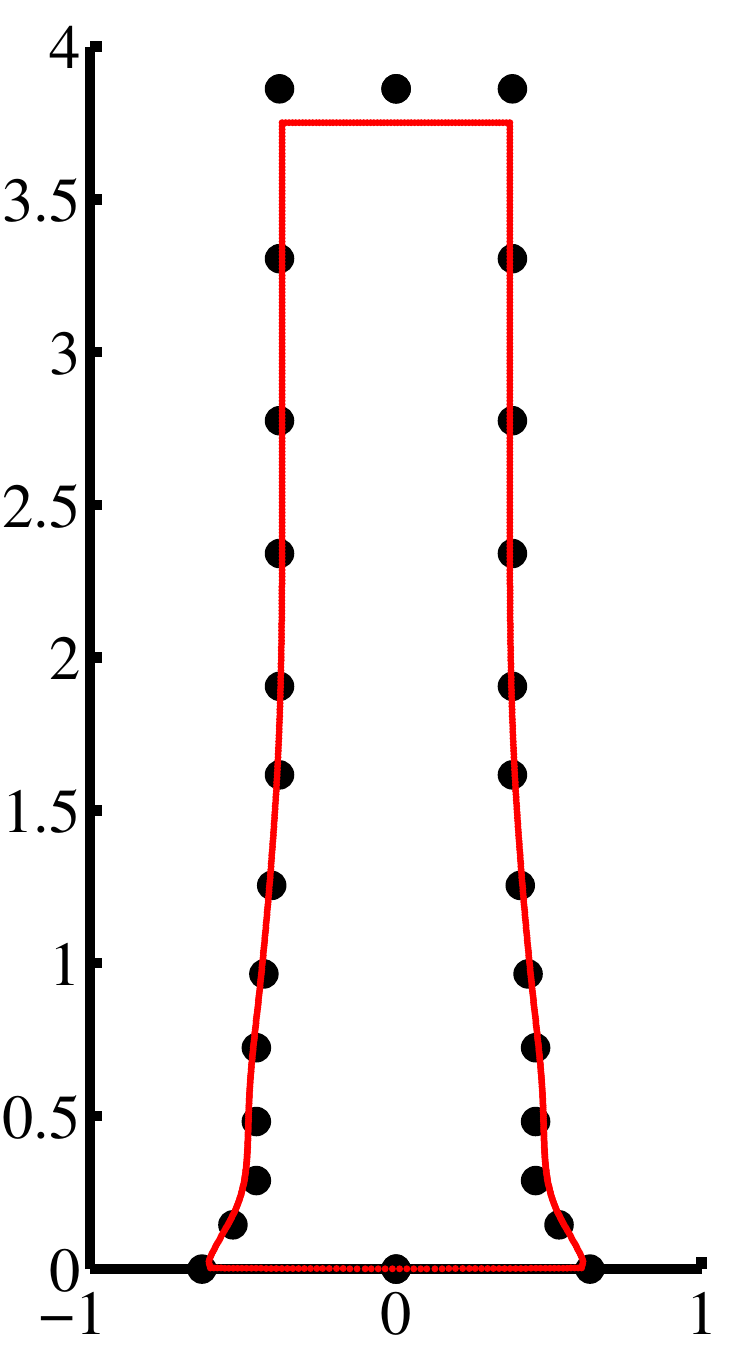}}\\
      (d) JC (Al-C). \\
      \vspace{12pt}
      \scalebox{0.30}{\includegraphics{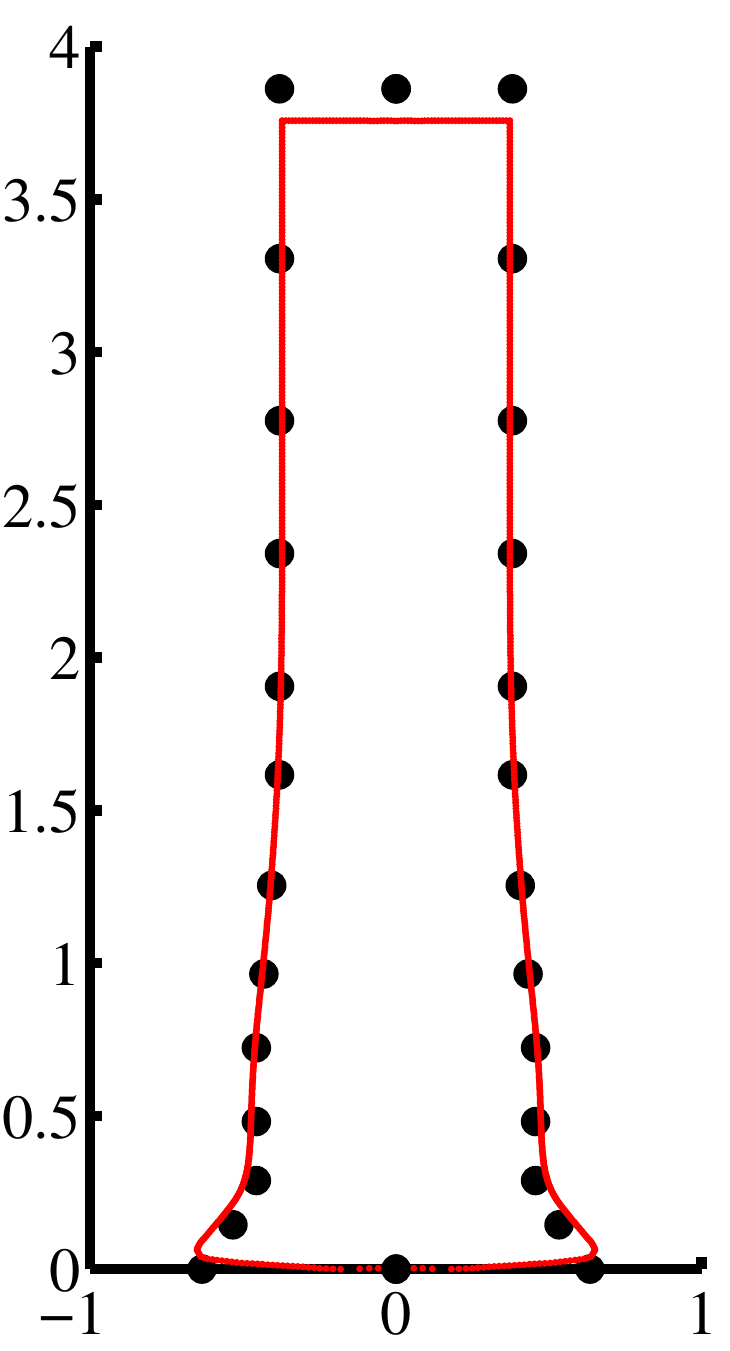}}\\
      (j) JC (Al-C) Friction.
    \end{minipage}
    \begin{minipage}[t]{0.16\linewidth}
      \centering
      \scalebox{0.30}{\includegraphics{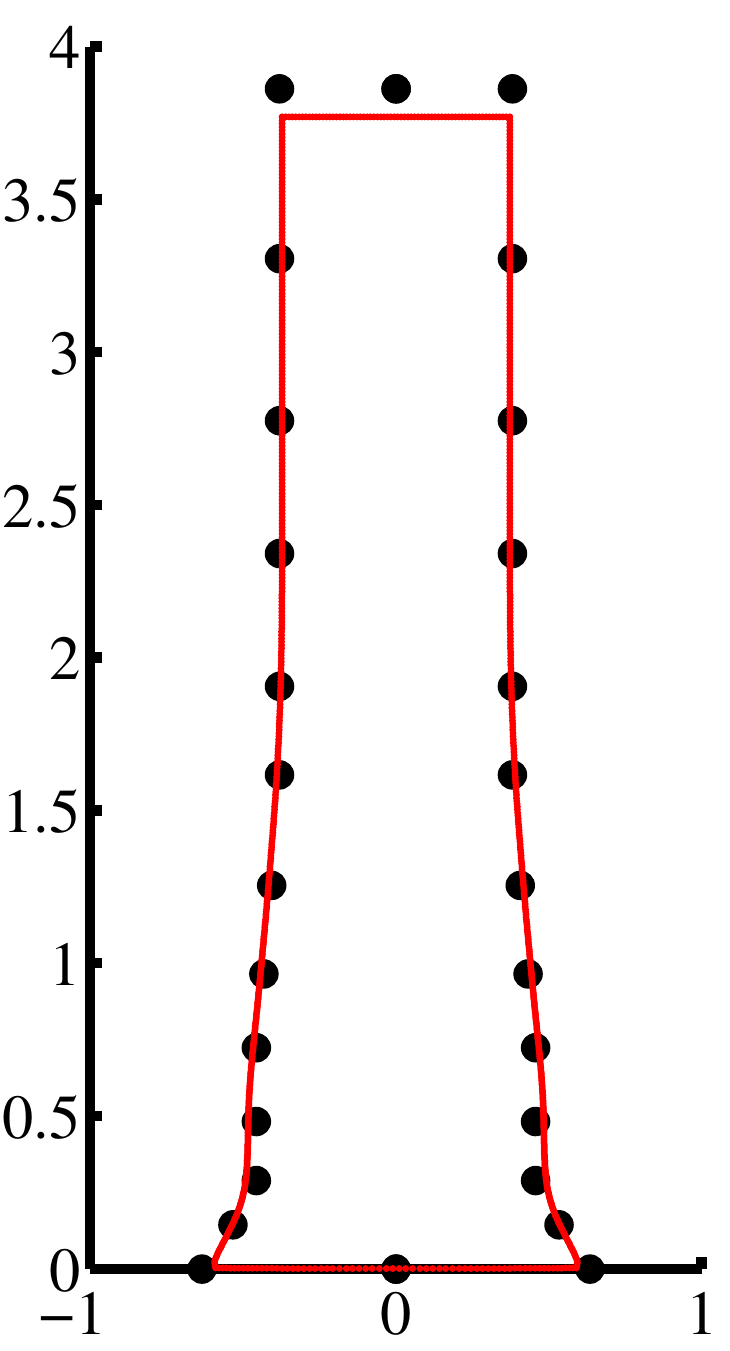}}\\
      (e) MTS (Al-C). \\
      \vspace{12pt}
      \scalebox{0.30}{\includegraphics{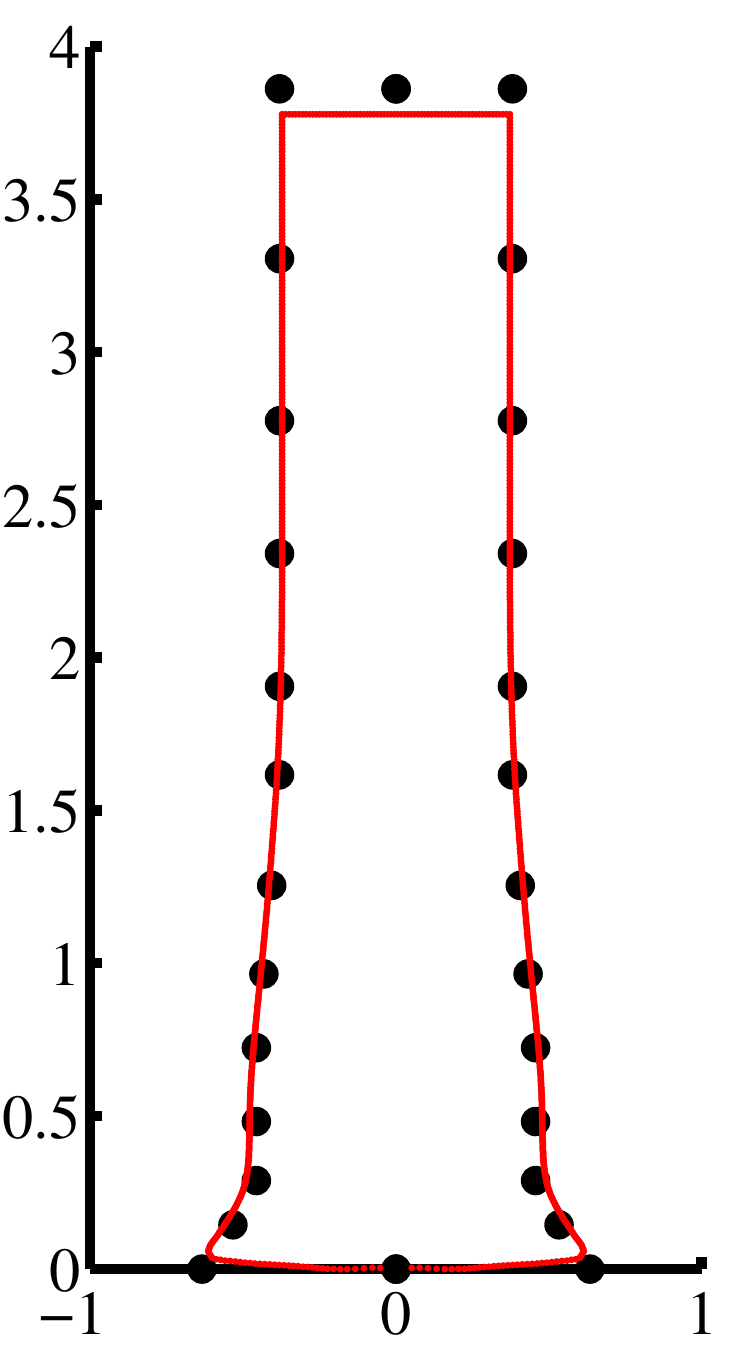}}\\
      (k) MTS (Al-C) Friction.
    \end{minipage}
    \begin{minipage}[t]{0.16\linewidth}
      \centering
      \scalebox{0.30}{\includegraphics{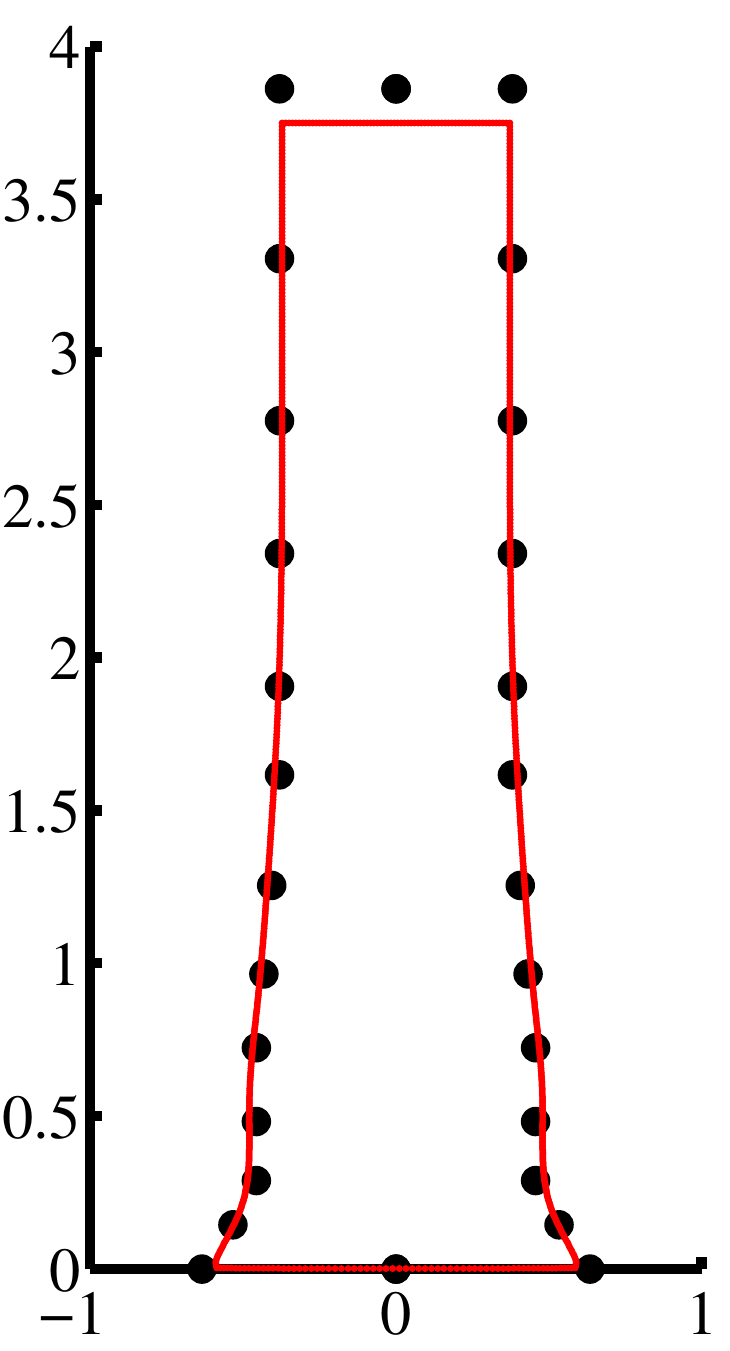}}\\
      (f) SCG (Al-C). \\
      \vspace{12pt}
      \scalebox{0.30}{\includegraphics{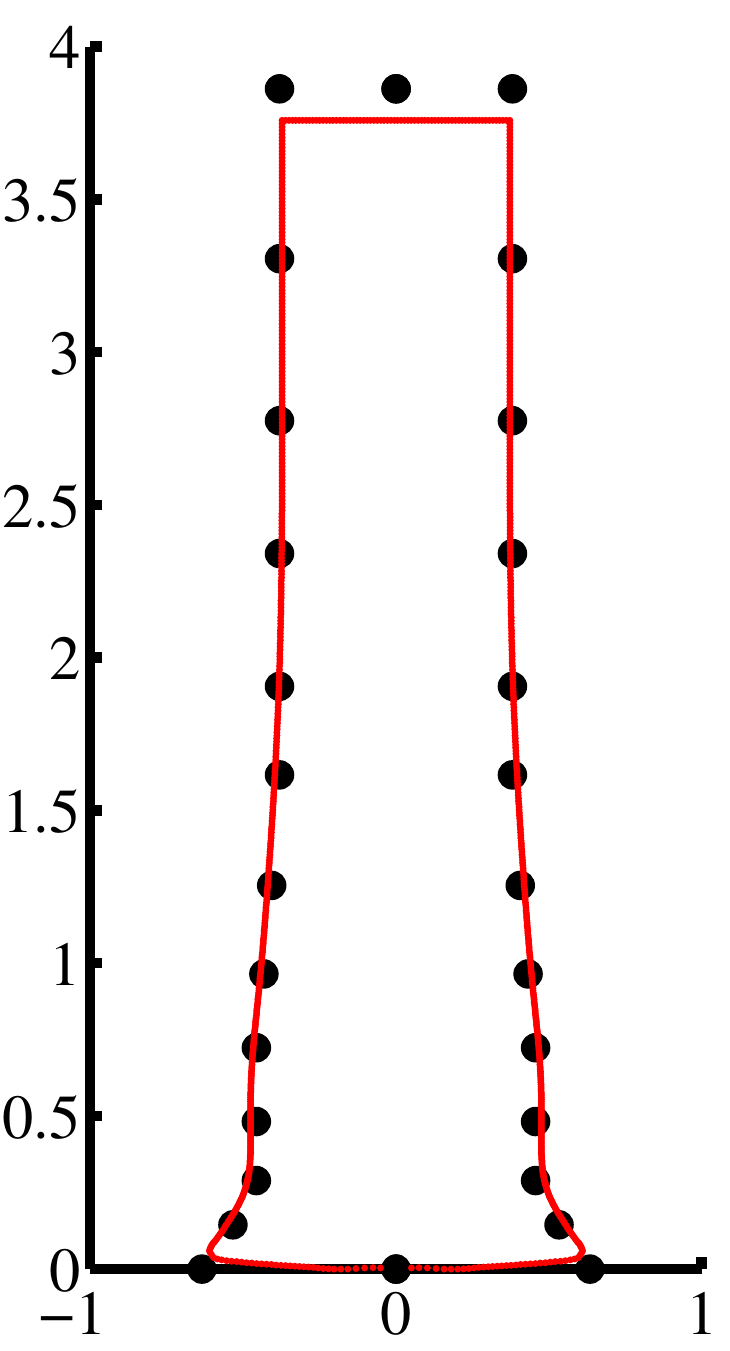}}\\
      (l) SCG (Al-C) Friction.
    \end{minipage}
    \vspace{12pt}
   \caption{\small Comparison of experimental and computed shapes of 6061T6 aluminum
            cylinders using the Johnson-Cook (JC), Mechanical Threshold
            Stress (MTS), and Steinberg-Cochran-Guinan (SCG) plasticity models.
            The figure in the top row are from simulations without friction while 
            those in the bottom row are with friction.
            The axes are shown in cm units.}
   \label{fig:AlModelRoom}
  \end{figure}

  Comparisons between the computed and experimental profiles of 6061T6 
  aluminum alloy specimen Al-A are shown in Figure~\ref{fig:AlModelRoom}(a),
  (b), and (c).  Those for specimen Al-C are shown in 
  Figure~\ref{fig:AlModelRoom}(d), (e), and (f).
  If frictional contact at the impact surface is simulated, the final 
  shapes of the specimens Al-A and Al-C are as shown in 
  Figure~\ref{fig:AlModelRoom} (g), (h), (i), (j), (k), and (l).

  We note that all three models predict essentially identical profiles.  The higher velocity
  impact of the shorter specimen Al-A is best predicted by the MTS model as far as final length 
  is concerned.  The mushroom width is predicted better when some friction is included at the
  anvil-specimen interface.  There is a noticeable amount of curvature under frictional contact.
  We believe that this partly due to the contact algorithm that has been used.

  The longer specimens have lower impact velocities.  However, all three models predict a final
  length that is shorter than that observed in experiment.  We believe that is discrepancy is 
  due to material variability. Note the accuracy with which the profiles are predicted and the 
  noticeably lower curvature of the mushroom under frictional contact compared to specimen Al-A.

  \subsubsection{High temperature impact: 6061-T6 Al}
\begin{figure}[htb!]
    \begin{minipage}[t]{0.16\linewidth}
      \centering
      \scalebox{0.30}{\includegraphics{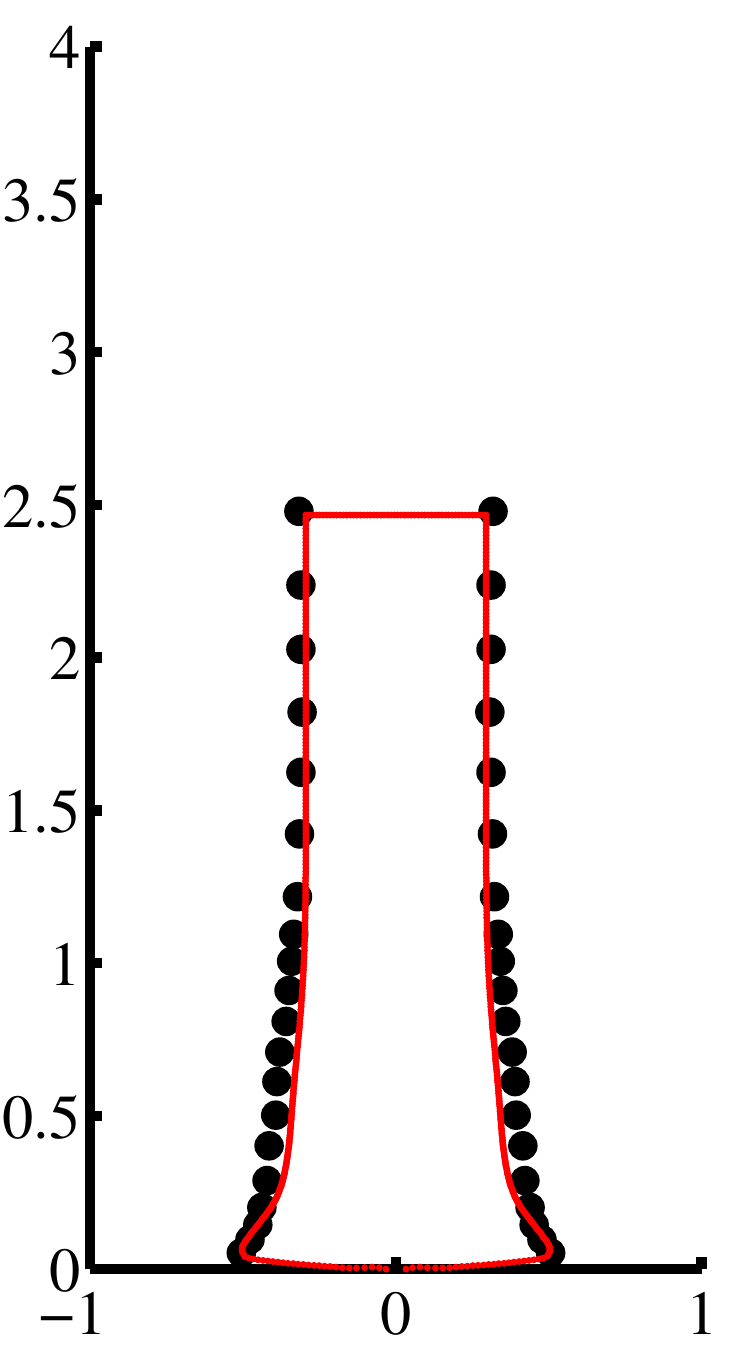}}\\
      (a) JC (Al-G). \\
      \vspace{12pt}
      \scalebox{0.30}{\includegraphics{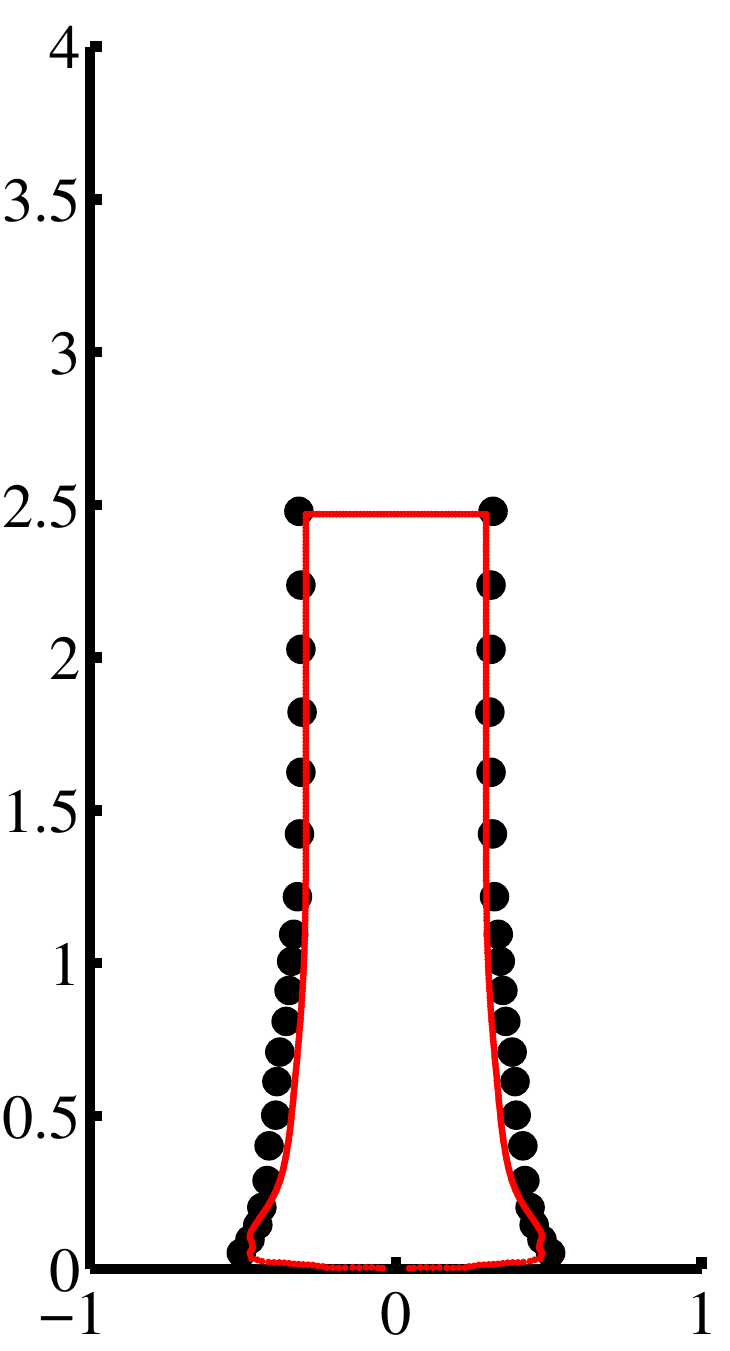}}\\
      (g) JC (Al-G) with erosion. 
    \end{minipage}
    \begin{minipage}[t]{0.16\linewidth}
      \centering
      \scalebox{0.30}{\includegraphics{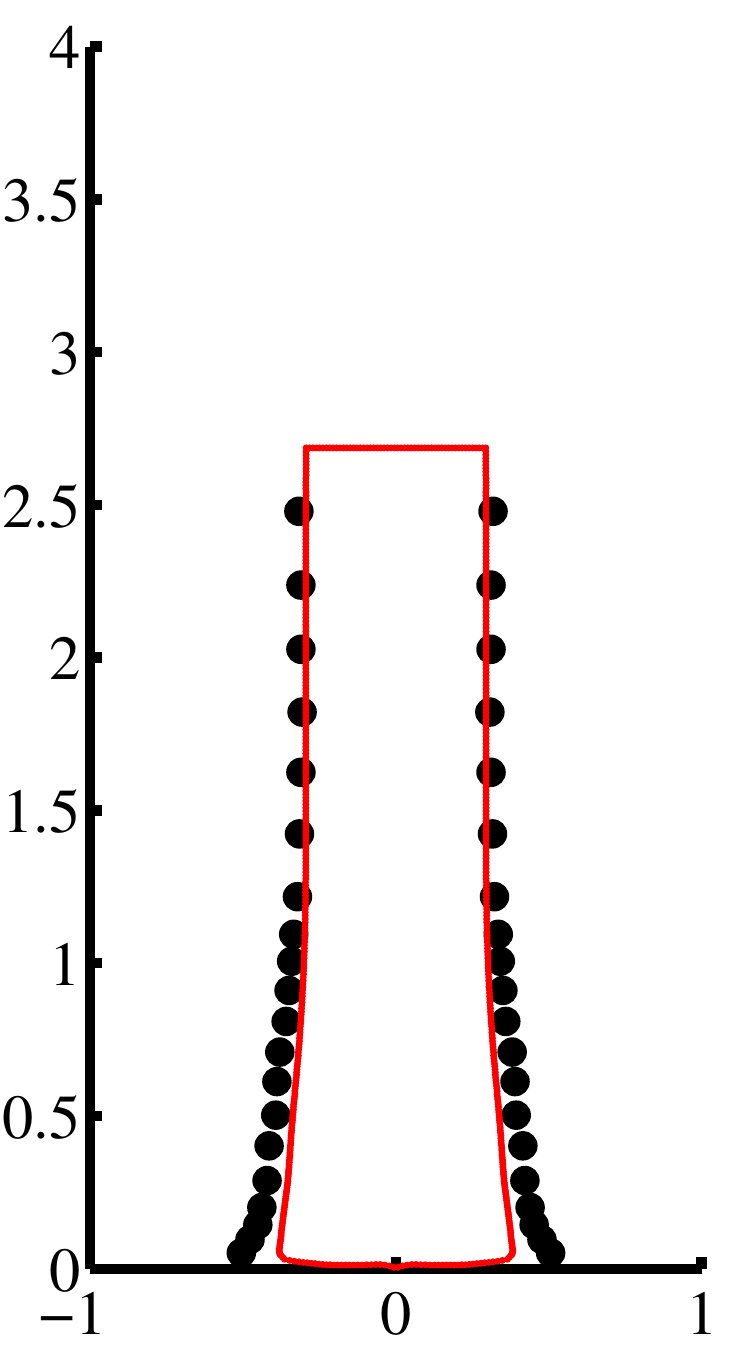}}\\
      (b) MTS (Al-G). \\
      \vspace{12pt}
      \scalebox{0.30}{\includegraphics{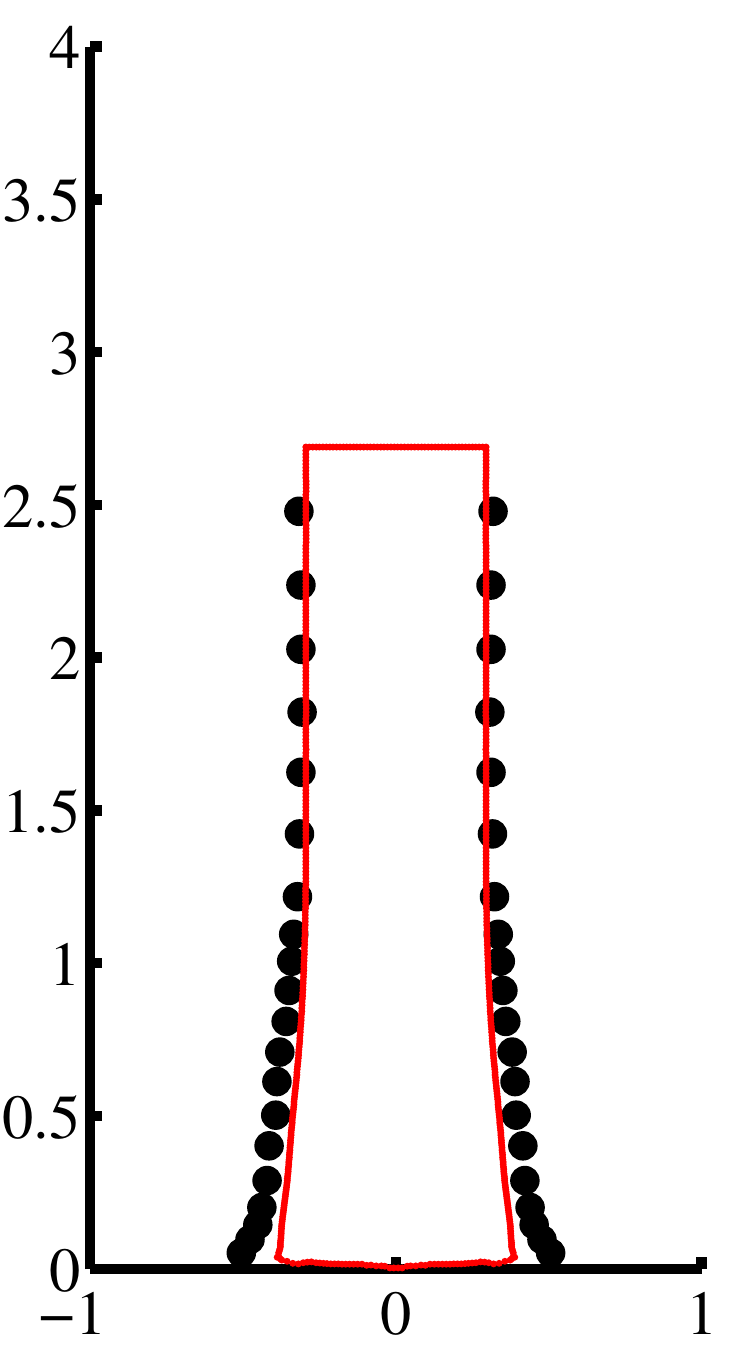}}\\
      (h) MTS (Al-G) with erosion.
    \end{minipage}
    \begin{minipage}[t]{0.16\linewidth}
      \centering
      \scalebox{0.30}{\includegraphics{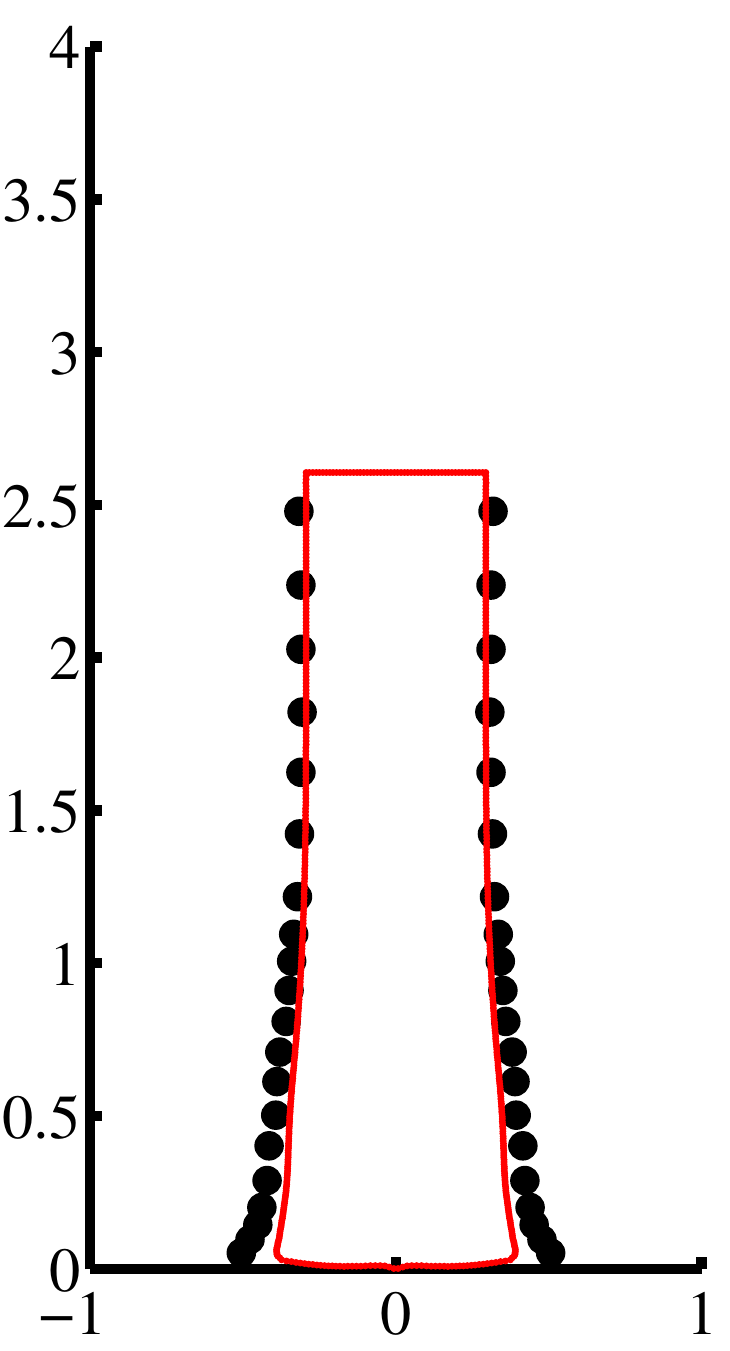}}\\
      (c) SCG (Al-G). \\
      \vspace{12pt}
      \scalebox{0.30}{\includegraphics{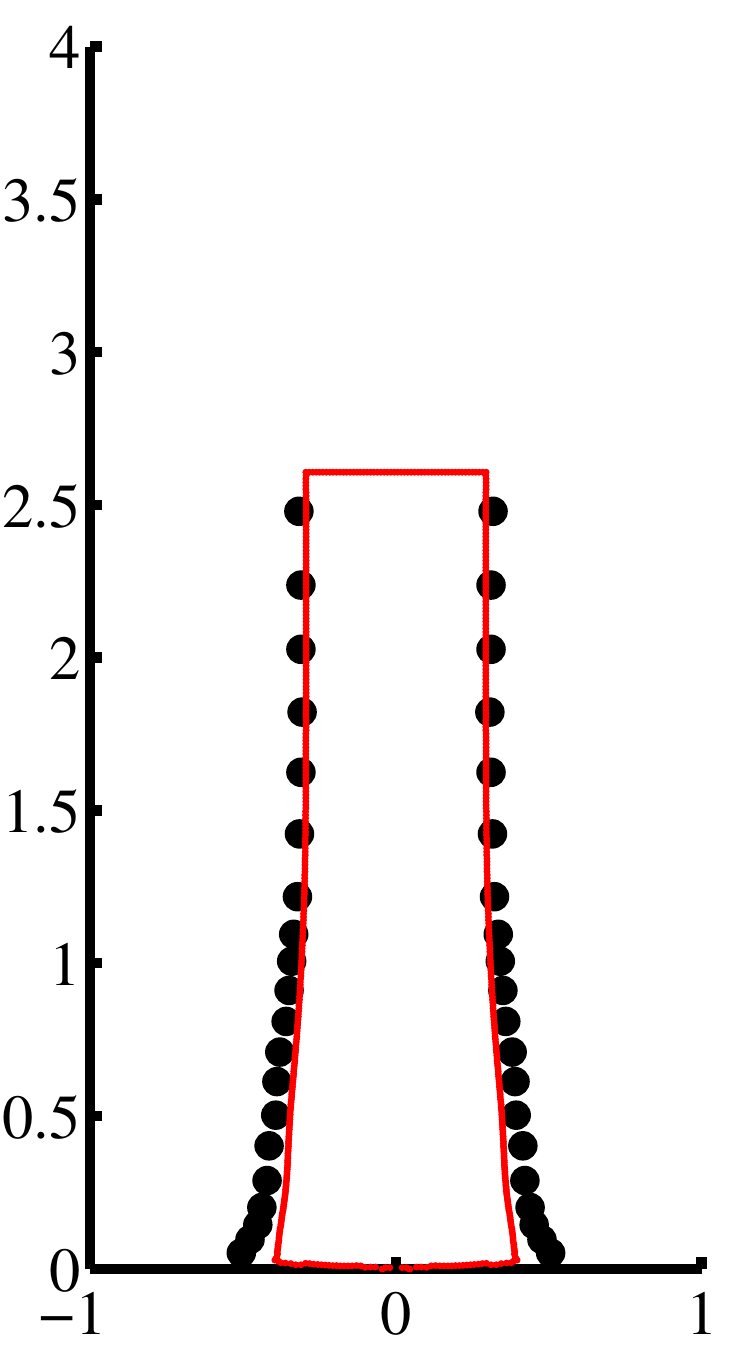}}\\
      (i) SCG (Al-G) with erosion.
    \end{minipage}
    \begin{minipage}[t]{0.16\linewidth}
      \centering
      \scalebox{0.30}{\includegraphics{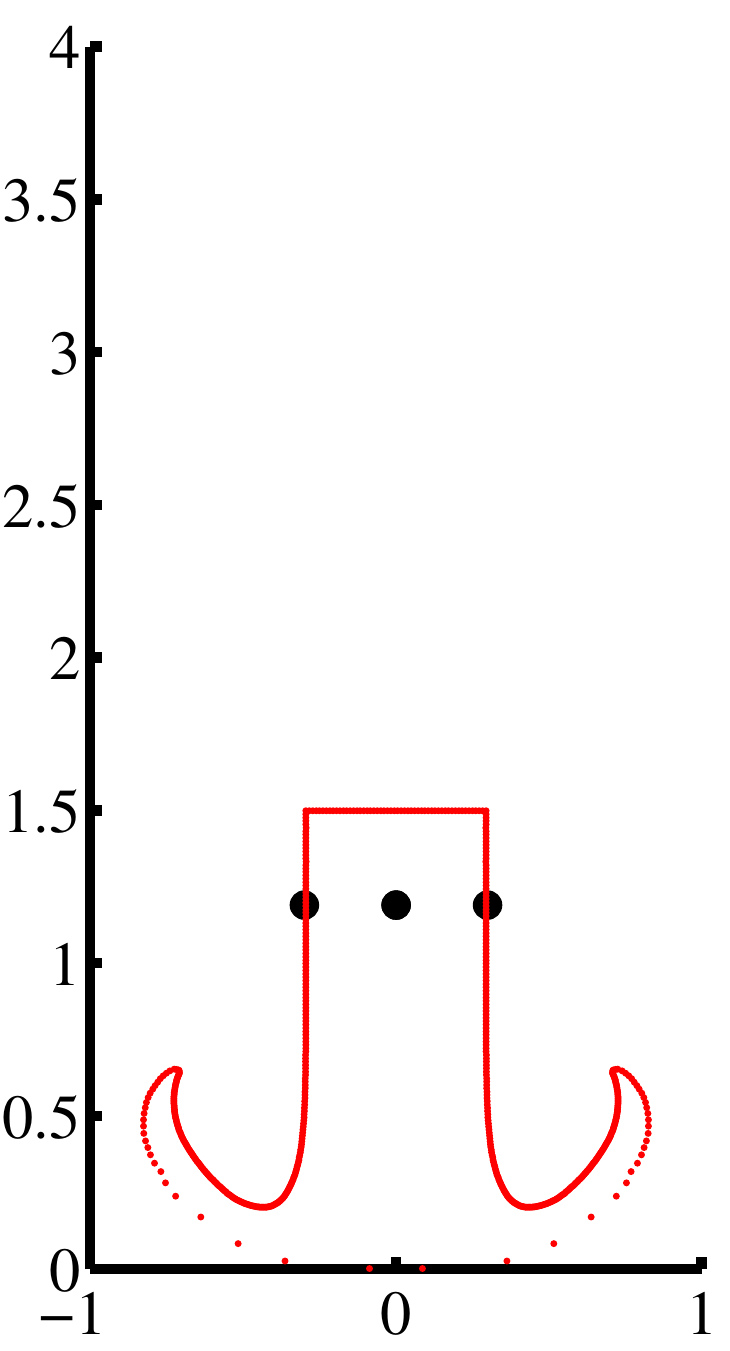}}\\
      (d) JC (Al-H). \\
      \vspace{12pt}
      \scalebox{0.30}{\includegraphics{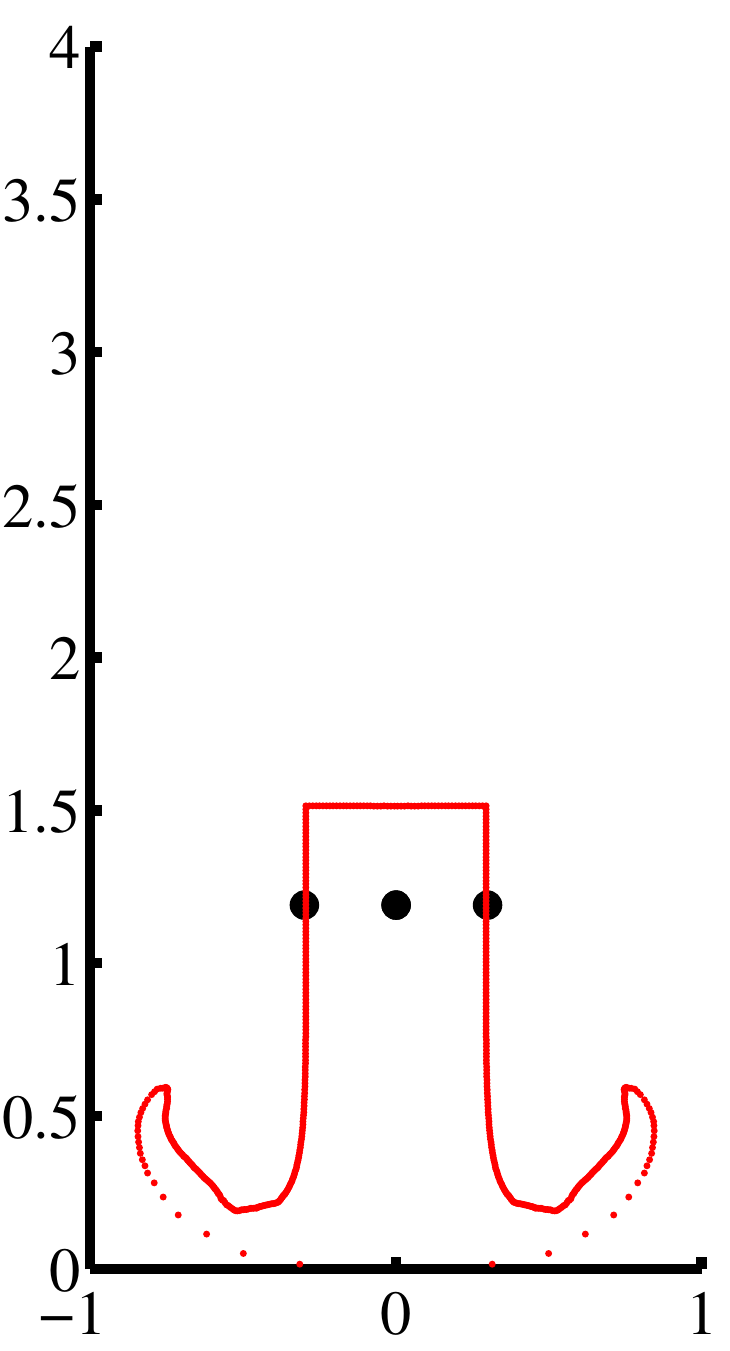}}\\
      (j) JC (Al-H) with erosion.
    \end{minipage}
    \begin{minipage}[t]{0.16\linewidth}
      \centering
      \scalebox{0.30}{\includegraphics{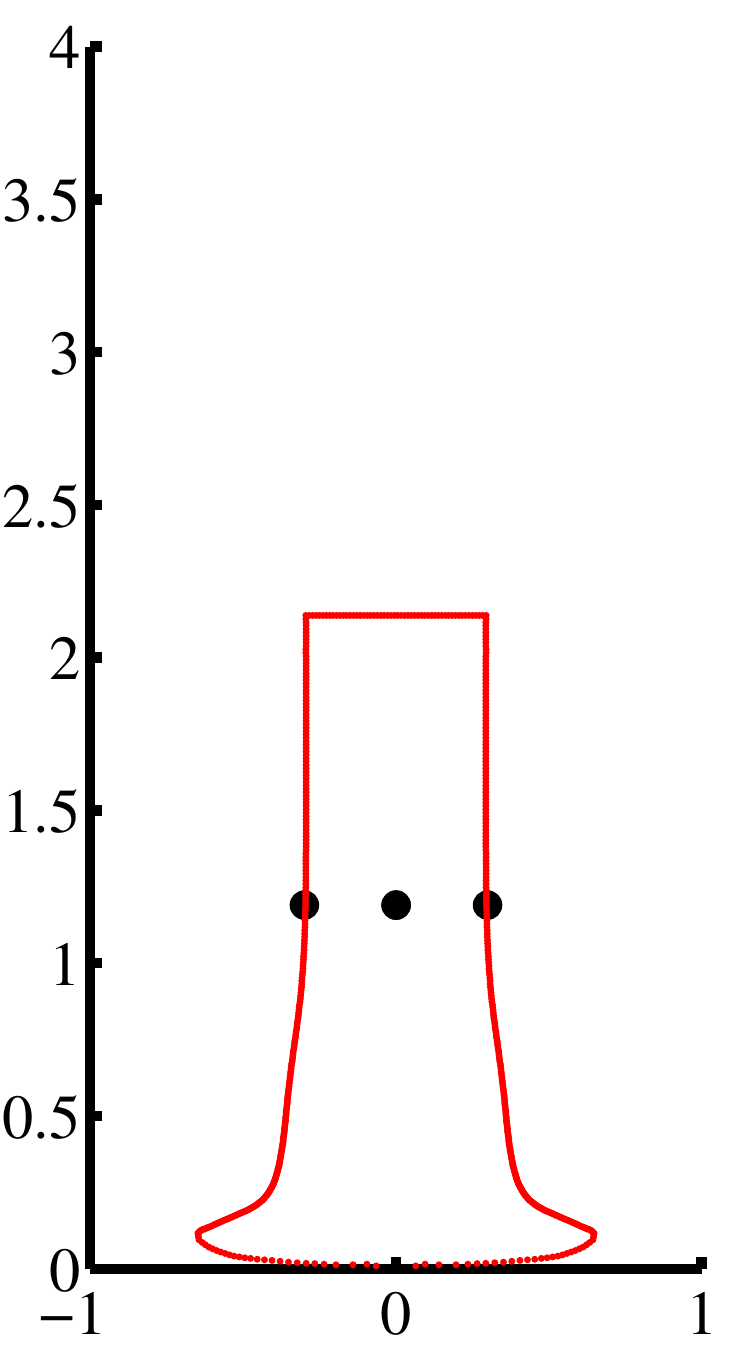}}\\
      (e) MTS (Al-H). \\
      \vspace{12pt}
      \scalebox{0.30}{\includegraphics{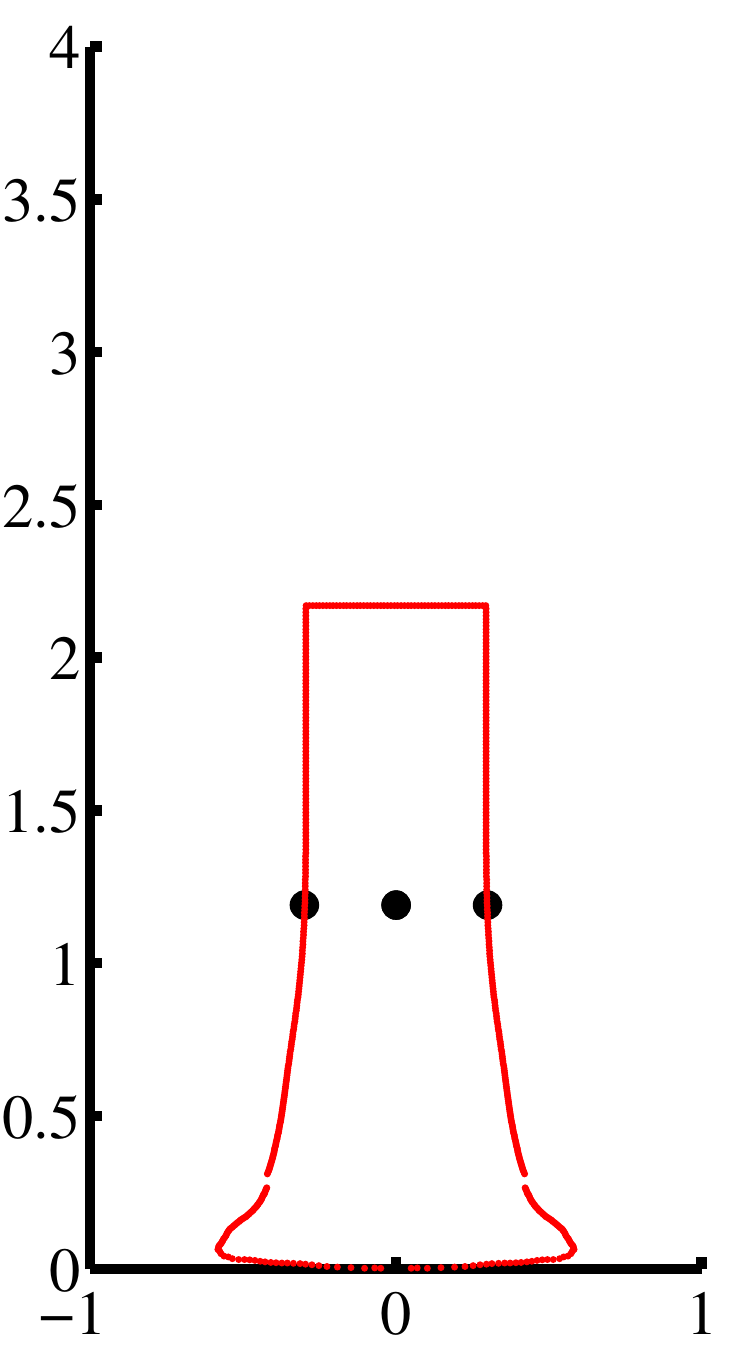}}\\
      (k) MTS (Al-H) with erosion.
    \end{minipage}
    \begin{minipage}[t]{0.16\linewidth}
      \centering
      \scalebox{0.30}{\includegraphics{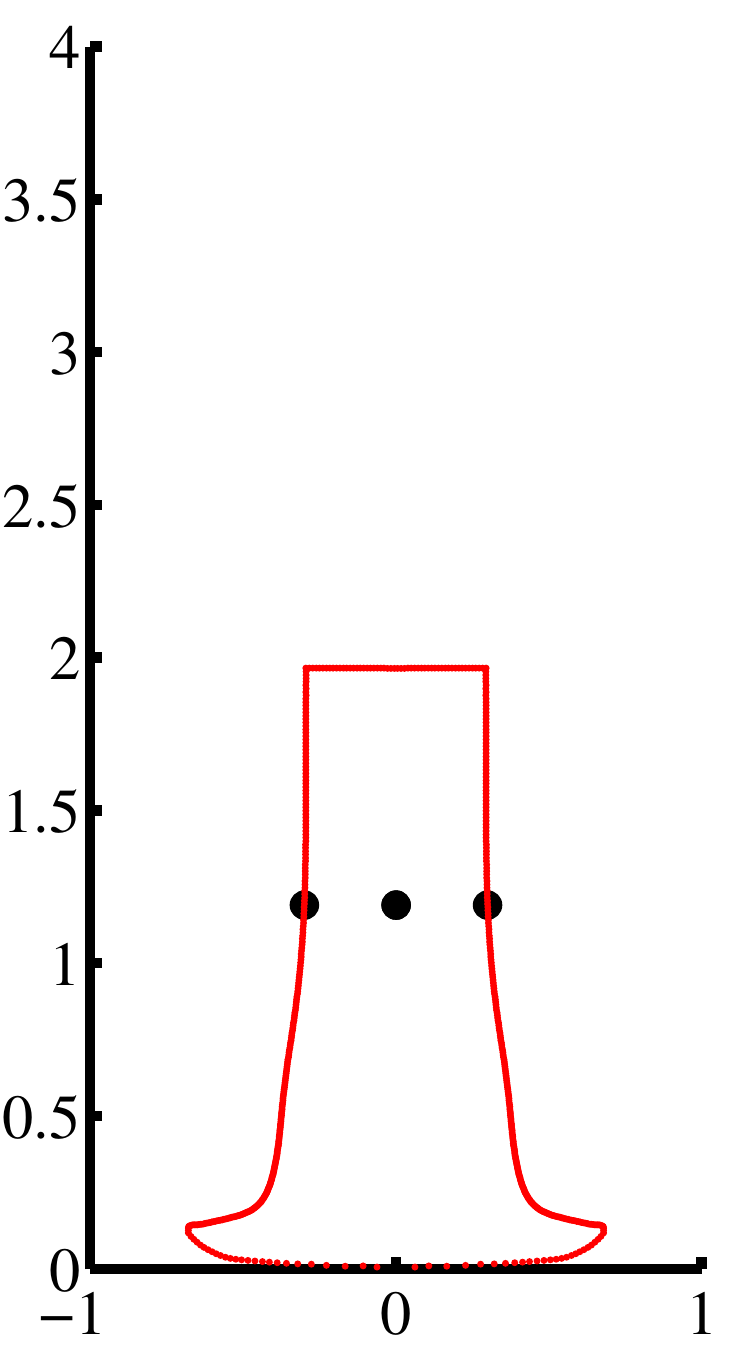}}\\
      (f) SCG (Al-H). \\
      \vspace{12pt}
      \scalebox{0.30}{\includegraphics{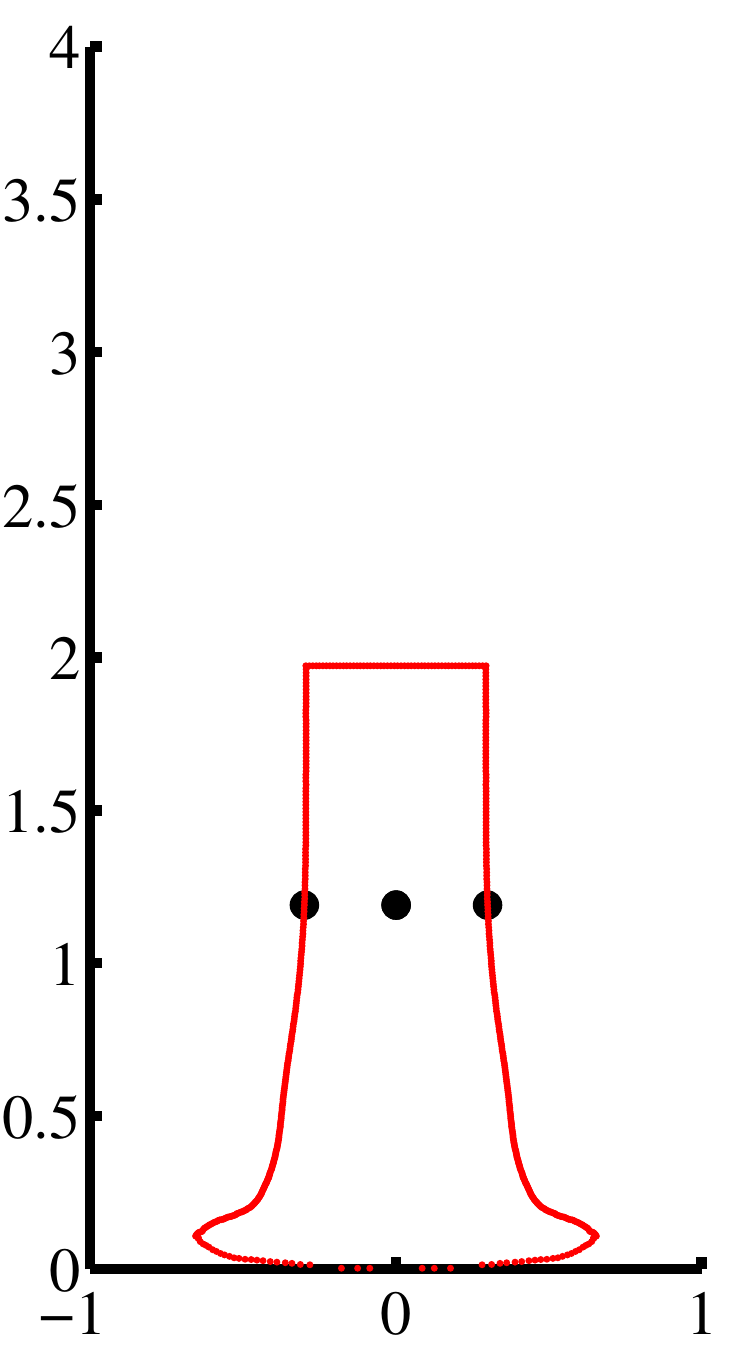}}\\
      (l) SCG (Al-H) with erosion.
    \end{minipage}
    \vspace{12pt}
    \caption{\small Comparison of experimental and computed shapes of 6061T6 aluminum
             cylinders using the Johnson-Cook (JC), Mechanical Threshold
             Stress (MTS), and Steinberg-Cochran-Guinan (SCG) plasticity models.
             Specimens Al-G and Al-H are both initially at 635 K.  Al-G has an impact velocity
             of 194 m/s while Al-H impacts at 354 m/s. The axes are shown in cm units.}
    \label{fig:AlModelHotFric}
  \end{figure}

  At higher temperatures, the response of the three plasticity models is 
  quite different.  Comparisons between the computed and experimental 
  profiles of 6061T6 aluminum alloy specimens have been performed under
  conditions of frictional contact. The final shapes of the specimens Al-G and 
  Al-H are as shown in Figure~\ref{fig:AlModelHotFric}.  If failure simulation
  is included, the profiles are as shown in Figures~\ref{fig:AlModelHotFric}(g),
  (h), (i), (j), (k), and (l).

  For the lower impact velocity of specimen Al-G, the Johnson-Cook model performs the 
  best at predicting both the final length and the mushroom diameter.  Both the MTS 
  and SGC models overestimate the final length and underestimate the mushroom diameter.
  The MTS model fares slightly worse than the SCG model.  However, the differences are
  small enough that they can be attributed to material variability.  Including erosion
  effects in the simulation does not affect the result significantly.

  At the higher impact velocity represented by specimen Al-H, all models fail to predict 
  the final length accurately. The Johnson-Cook model comes closest but overestimates the
  length and has an excessively deformed mushroom region.  The MTS and SCG models have 
  more reasonably shaped mushroom regions but fail to predict the final length by almost 100\%.
  The SCG model is slightly better than the MTS model.

\begin{figure}[htb!]
    \begin{minipage}[t]{0.49\linewidth}
      \centering
      \scalebox{0.48}{\includegraphics{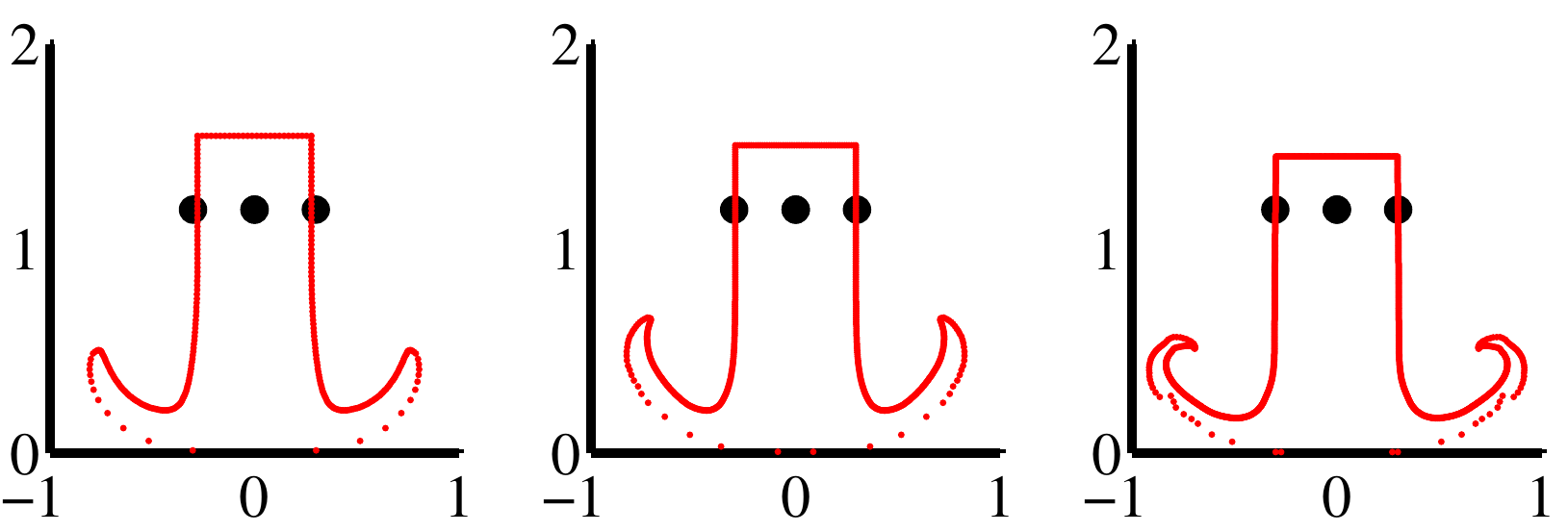}}\\
      (a) Al-H with friction.
    \end{minipage}
    \begin{minipage}[t]{0.49\linewidth}
      \centering
      \scalebox{0.48}{\includegraphics{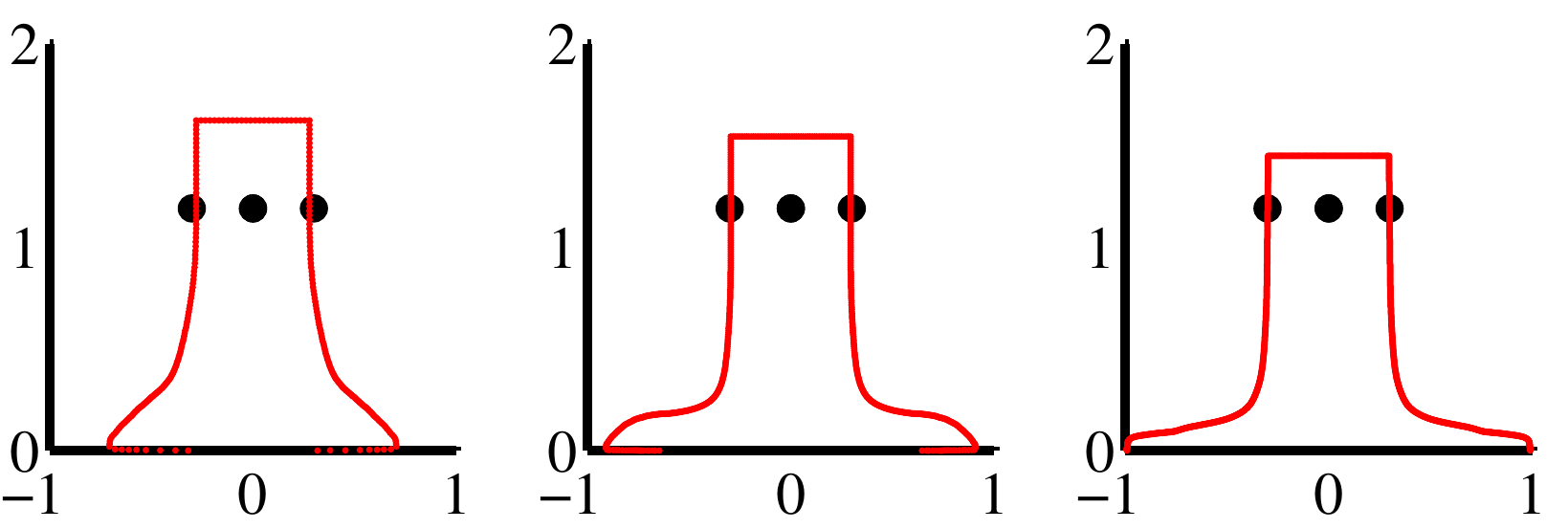}}\\
      (b) Al-H without friction.
    \end{minipage}
    \vspace{12pt}
    \caption{\small Comparison of experimental and computed shapes of 6061T6 aluminum
             cylinders (Al-H) using the Johnson-Cook (JC) with increasing mesh
             refinement. The axes are in cm.}
    \label{fig:AlRefine}
  \end{figure}

  It is possible that the discrepancy that we observe for specimen Al-H is due to inadequate 
  discretization.  Simulations of impact specimen Al-H with increasing mesh refinement are shown
  in Figure~\ref{fig:AlRefine}.  The number of grid cells in the plane of the specimen profile
  has been doubled with each refinement.

  If we examine the profiles shown in Figure~\ref{fig:AlRefine}(a), we observe that the cylinder
  does appear to shorten with increasing refinement.  However, there is unphysical curling of the
  end of the specimen.  On the other hand, if we eliminate friction from the calculation, the
  mushroom appears to increase with increased refinement while the length decreases.  This indicates
  that there is some amount of mesh dependence of the solution that is probably due to the
  softening behavior of the material.

  \subsubsection{Comparisons with FEM}
\begin{figure}[htb!]
    \begin{minipage}[t]{0.23\linewidth}
      \centering
      \scalebox{0.30}{\includegraphics{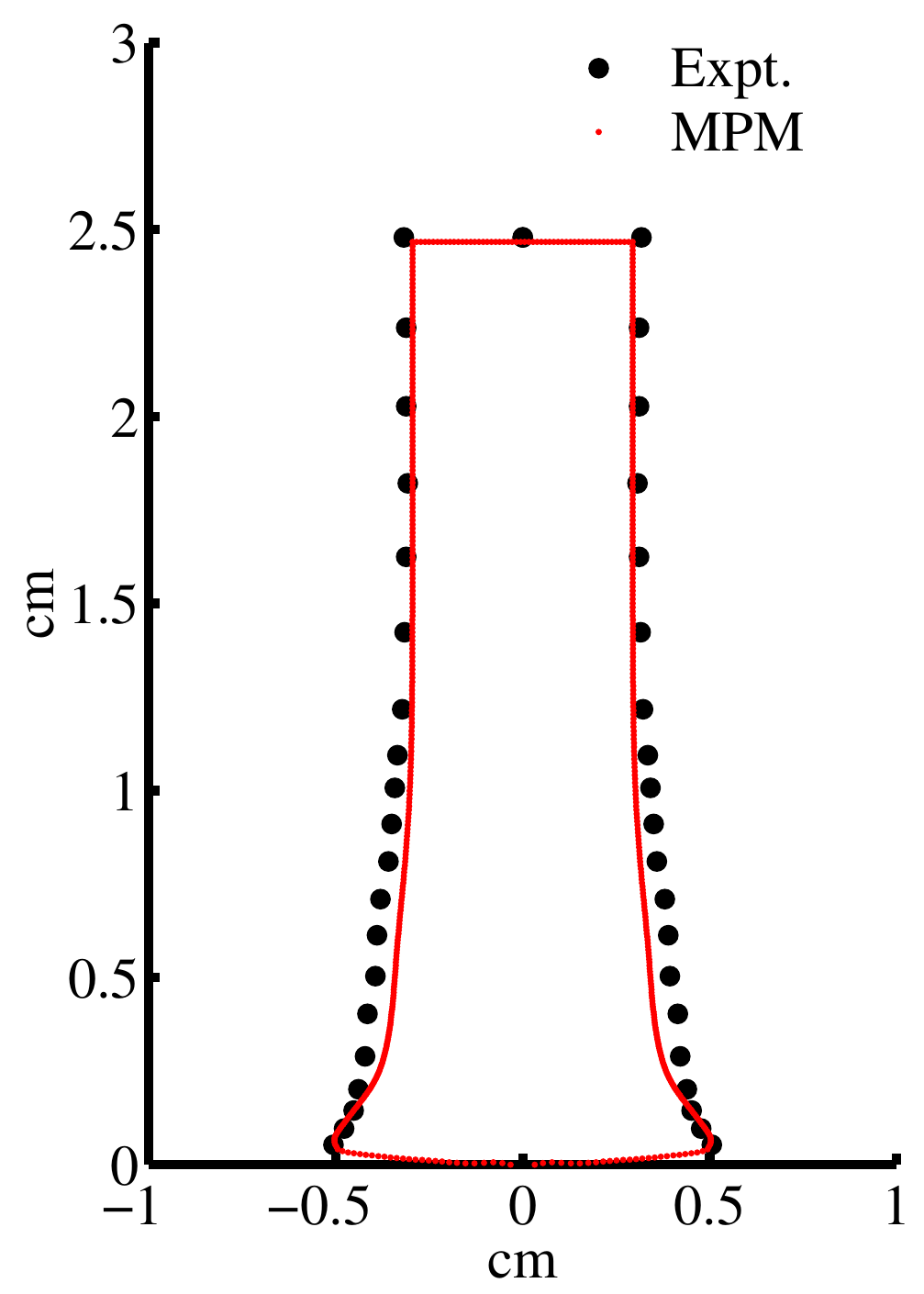}}\\
      (a) MPM (Al-G)
    \end{minipage}
    \begin{minipage}[t]{0.23\linewidth}
      \centering
      \scalebox{0.30}{\includegraphics{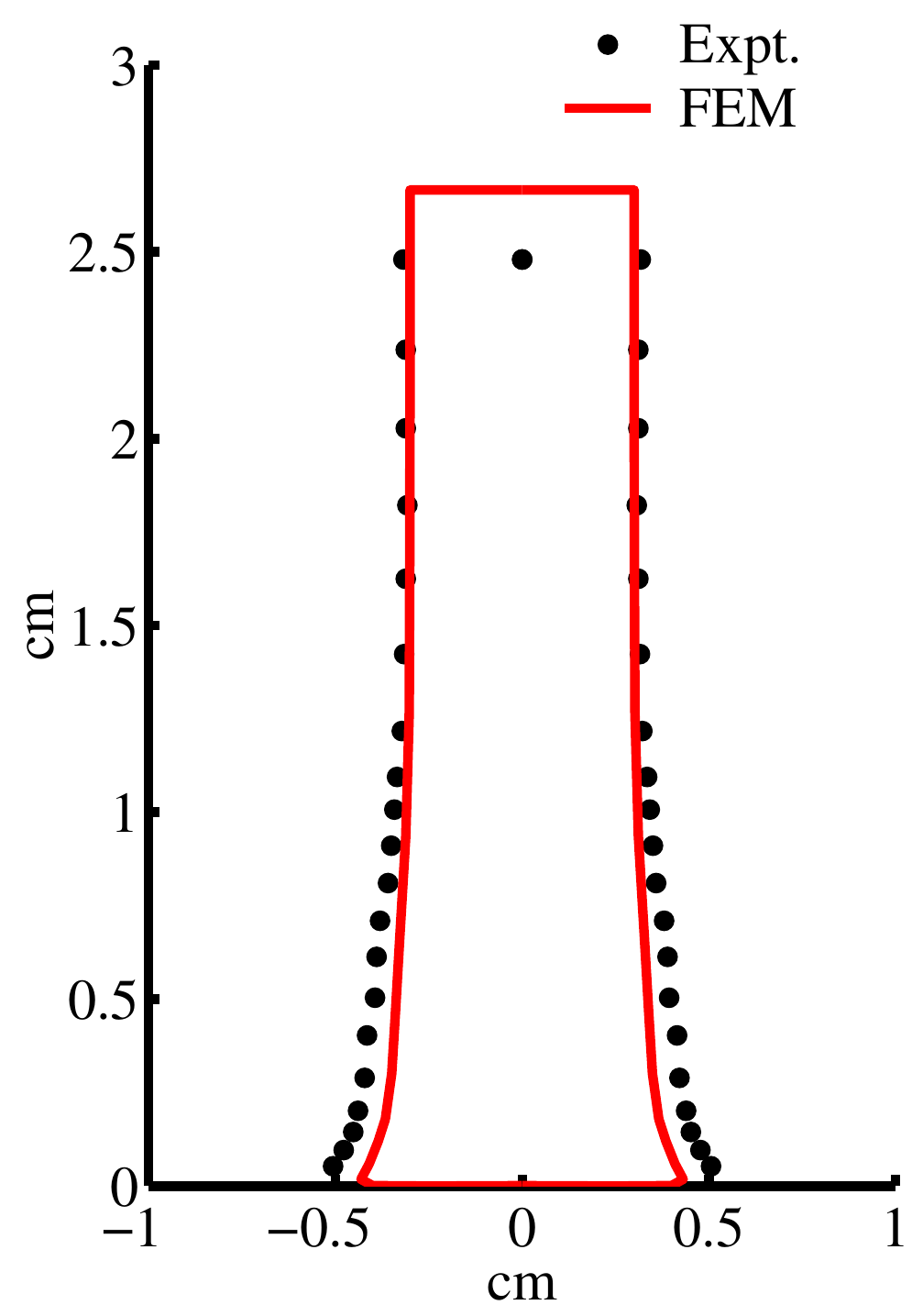}}\\
      (b) FEM (Al-G)
    \end{minipage}
    \begin{minipage}[t]{0.23\linewidth}
      \centering
      \scalebox{0.30}{\includegraphics{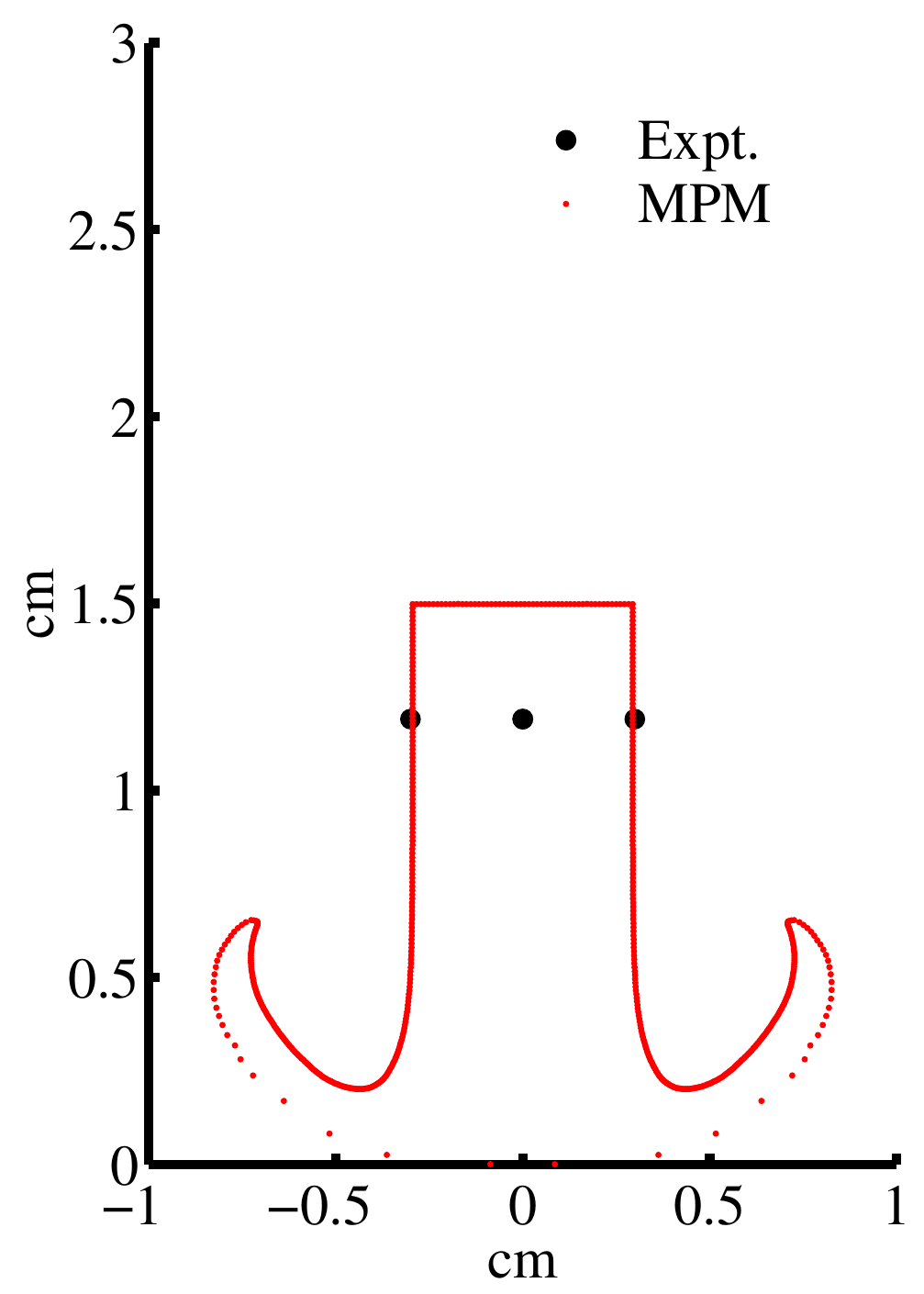}}\\
      (c) MPM (Al-H)
    \end{minipage}
    \begin{minipage}[t]{0.23\linewidth}
      \centering
      \scalebox{0.30}{\includegraphics{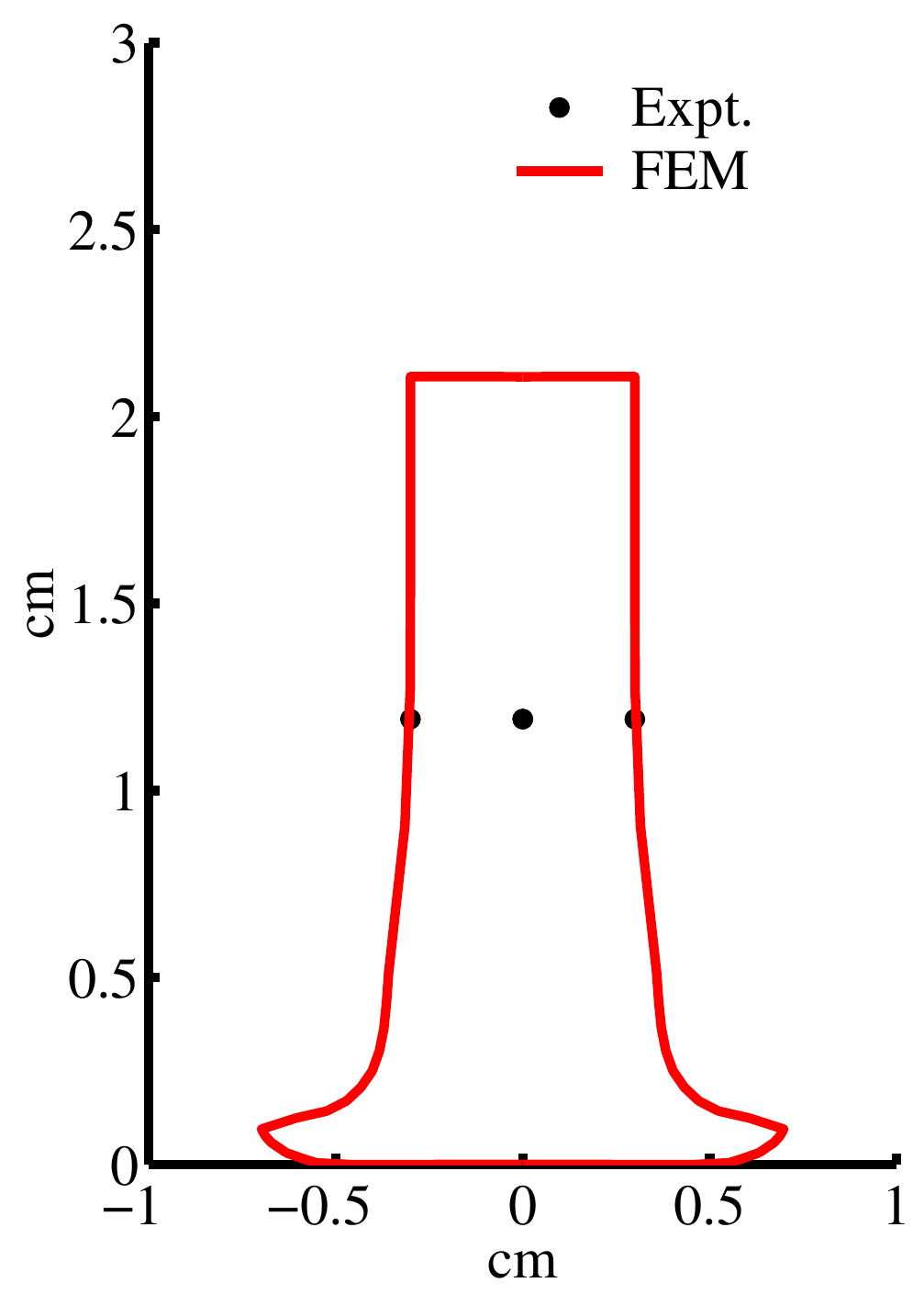}}\\
      (d) FEM (Al-H)
    \end{minipage}
    \vspace{12pt}
    \caption{\small Comparison of experimental and computed shapes of 6061T6 aluminum
             cylinders using MPM and FEM. The axes are in cm.}
    \label{fig:AlFE}
  \end{figure}

  To determine how our MPM simulations compare with FEM simulations we have
  run two high temperature aluminum impact tests using LS-DYNA (with 
  the coupled structural-thermal option).  Figure~\ref{fig:AlFE} shows the
  final deformed shapes for the two cases from the MPM and FEM simulations
  using Johnson-Cook plasticity.  The FEM simulations consistently overestimate the
  final length and underestimate the mushroom diameter at high temperatures.

  \clearpage

  \subsection{Taylor impact tests on 4340 steel}
  In this section we present the results from Taylor tests on 4340 steel
  specimens for different initial temperatures and impact velocities.
  Table~\ref{tab:steel} shows the initial dimensions, velocity, and 
  temperature of the specimens (along with the type of copper used and
  the source of the data) that we have simulated and compared with 
  experimental data.  Note that only a few representative results are shown in
  this report.
  \begin{table}[htb!]
    \caption{\small Initial data for 4340 steel simulations. } 
    \begin{tabular}{lllllll}
       \hline
       \hline
       Case & Hardness
            & Initial & Initial
            & Initial & Initial 
            & Source \\
            & 
            & Length & Diameter 
            & Velocity & Temperature \\
            & 
            & ($L_0$ mm) & ($D_0$ mm)
            & ($V_0$ m/s) & ($T_0$ K)\\
       \hline
       \hline
        St-A   & $R_c = 40$ & 30 & 6.00 & 158 & 295   & \citet{Gust82} \\
        St-B   & $R_c = 40$ & 30 & 6.00 & 232 & 295   & \citet{Gust82} \\
        St-C   & $R_c = 40$ & 30 & 6.00 & 183 & 715   & \citet{Gust82} \\
        St-D   & $R_c = 40$ & 30 & 6.00 & 312 & 725   & \citet{Gust82} \\
        St-E   & $R_c = 40$ & 30 & 6.00 & 136 & 1285   & \citet{Gust82} \\
        St-F   & $R_c = 40$ & 30 & 6.00 & 160 & 1285   & \citet{Gust82} \\
        St-G   & $R_c = 30$ & 25.4  & 7.62 & 208 & 298   & \citet{Johnson83} \\
        St-H   & $R_c = 30$ & 12.7  & 7.62 & 282 & 298   & \citet{Johnson83} \\
        St-I   & $R_c = 30$ & 8.1   & 7.62 & 343 & 298   & \citet{Johnson83} \\
        St-J   &  & & & & &\citet{Addessio93a}\\
       \hline
       \hline
    \end{tabular}
    \label{tab:steel}
  \end{table}

  \subsubsection{Room temperature impact: steel}
\begin{figure}[htb!]
    \begin{minipage}[t]{\linewidth}
      \centering
      \scalebox{0.35}{\includegraphics{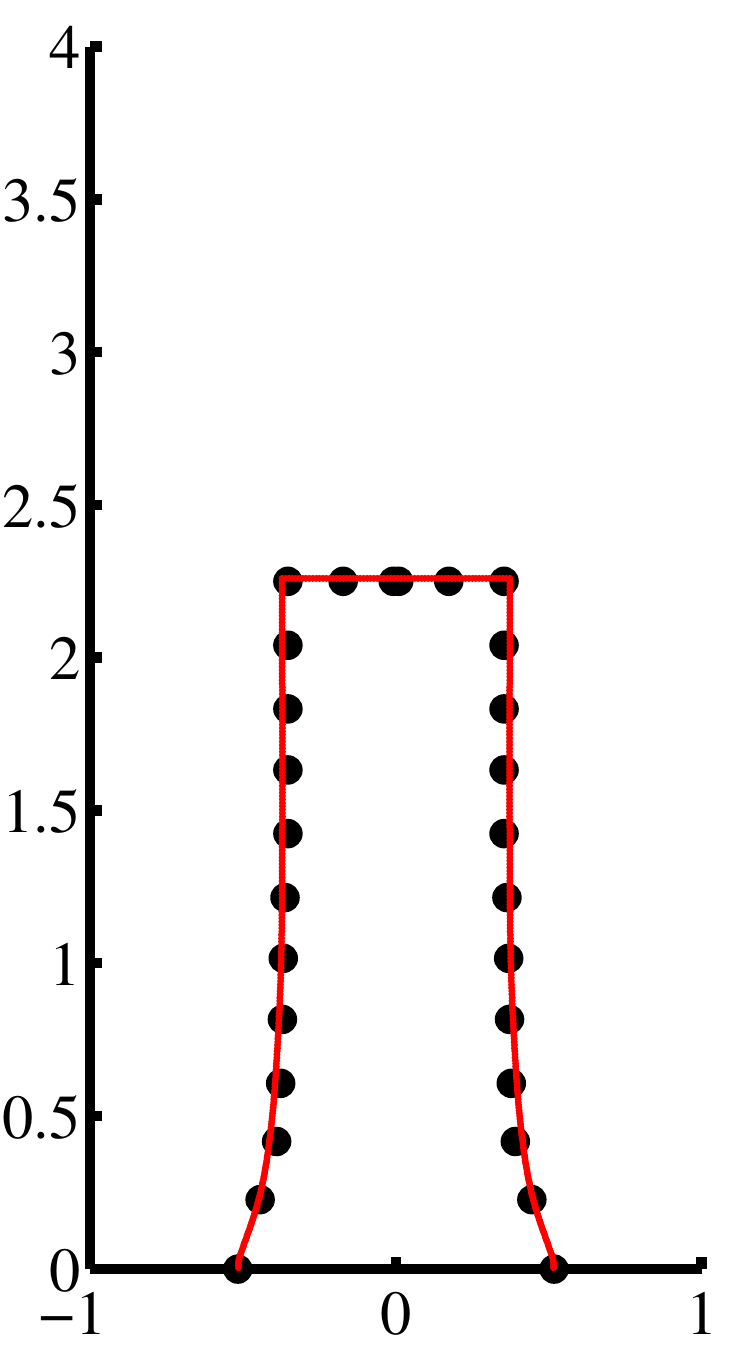}}\\
    \end{minipage}
    \caption{\small Comparison of experimental and computed shape of 4340 steel
             cylinder (St-G) without friction. The axes are in cm.}
    \label{fig:4340RoomNoFric}
  \end{figure}

  Figure~\ref{fig:4340RoomNoFric} shows the simulated profile of case St-G 
  without friction.  The Johnson-Cook model performs quite well in predicting the
  deformed profile of the specimen.  An almost identical profile is obtained
  if we incorporate friction at the impact surface.  Similar results are obtained 
  for the other room temperature specimens.

  \subsubsection{High temperature impact: steel}
\begin{figure}[htb!]
    \begin{minipage}[t]{0.23\linewidth}
      \centering
      \scalebox{0.35}{\includegraphics{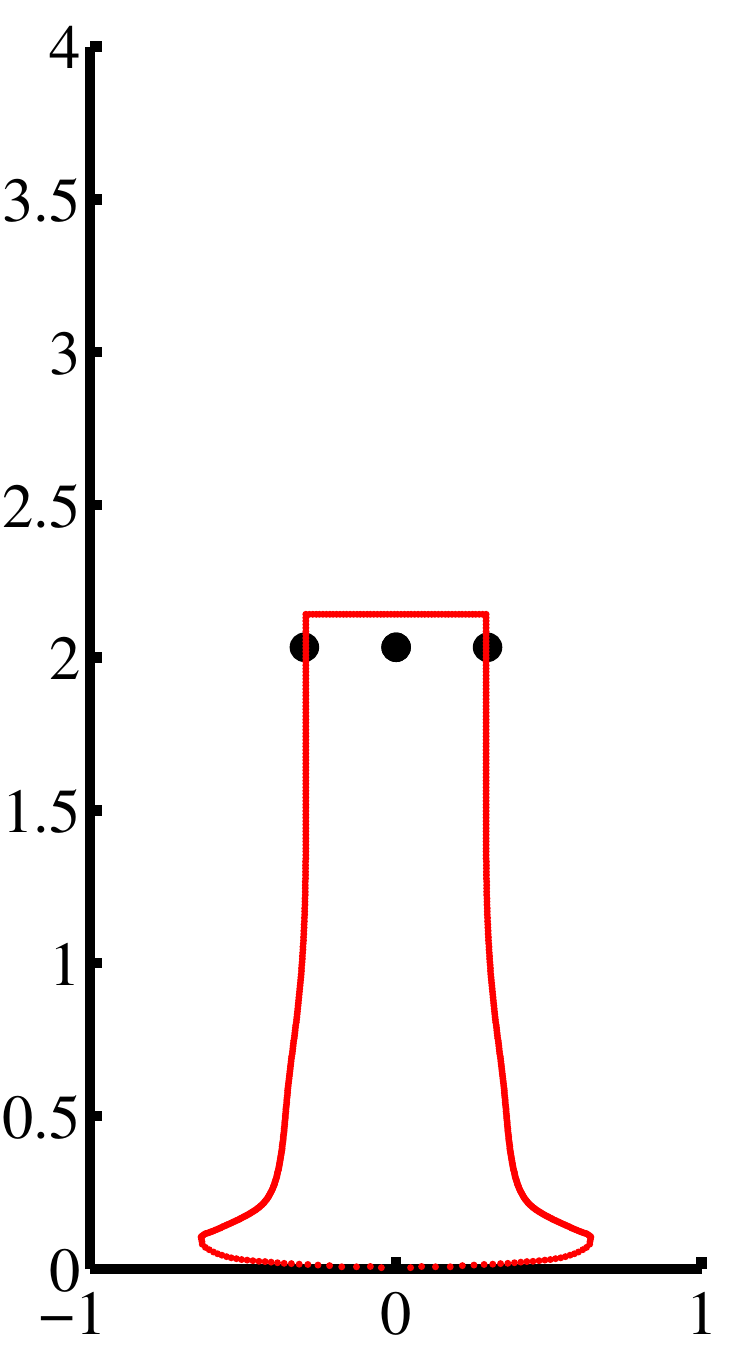}}\\
      (a) JC (St-D). \\
      \scalebox{0.35}{\includegraphics{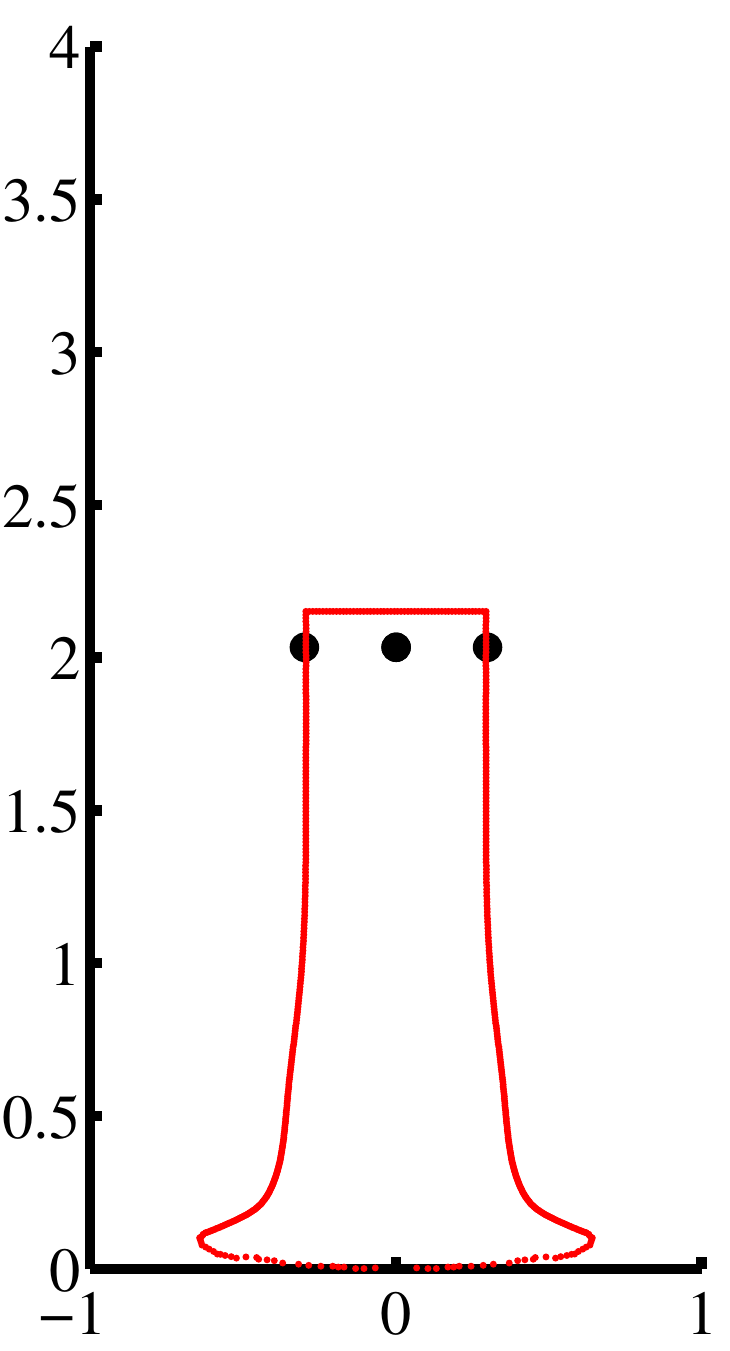}}\\
      (e) JC (St-D) with erosion.
    \end{minipage}
    \begin{minipage}[t]{0.23\linewidth}
      \centering
      \scalebox{0.35}{\includegraphics{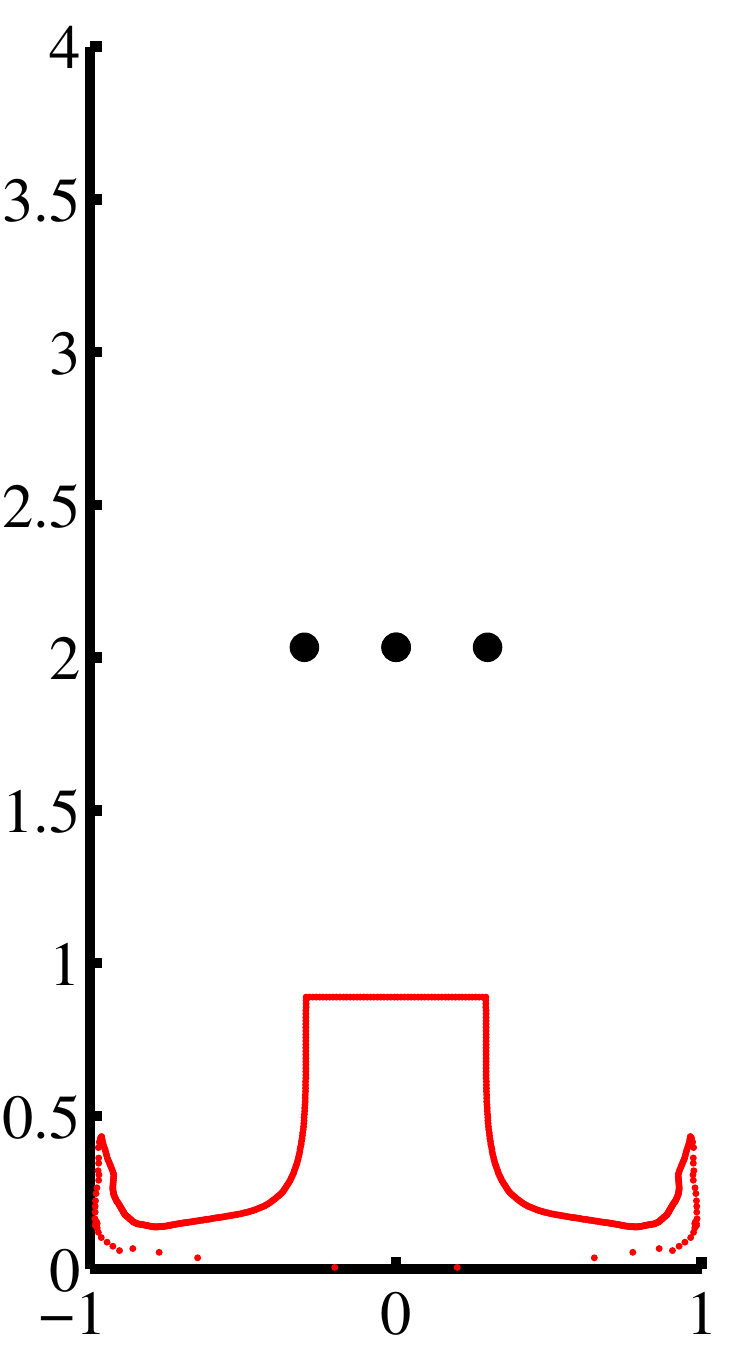}}\\
      (b) SCG (St-D). \\
      \scalebox{0.35}{\includegraphics{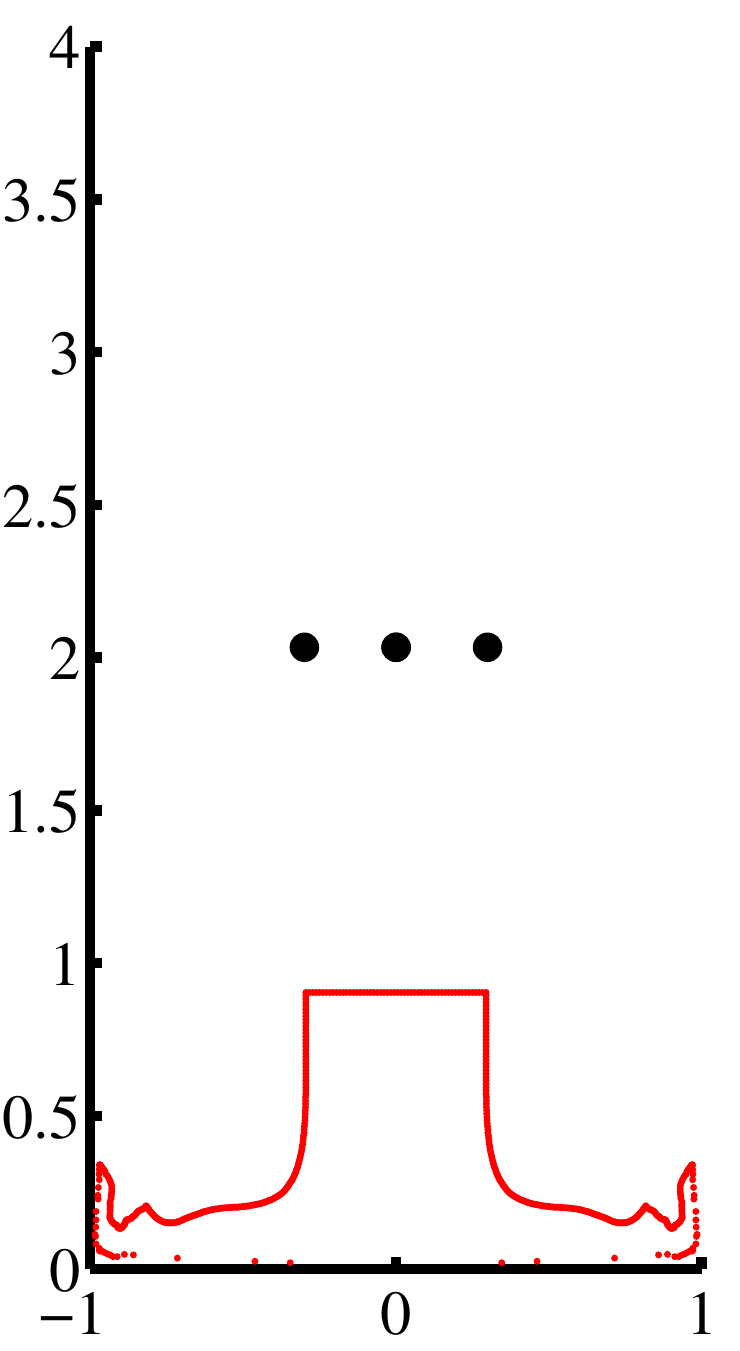}}\\
      (f) SCG (St-D) with erosion. 
    \end{minipage}
    \begin{minipage}[t]{0.23\linewidth}
      \centering
      \scalebox{0.35}{\includegraphics{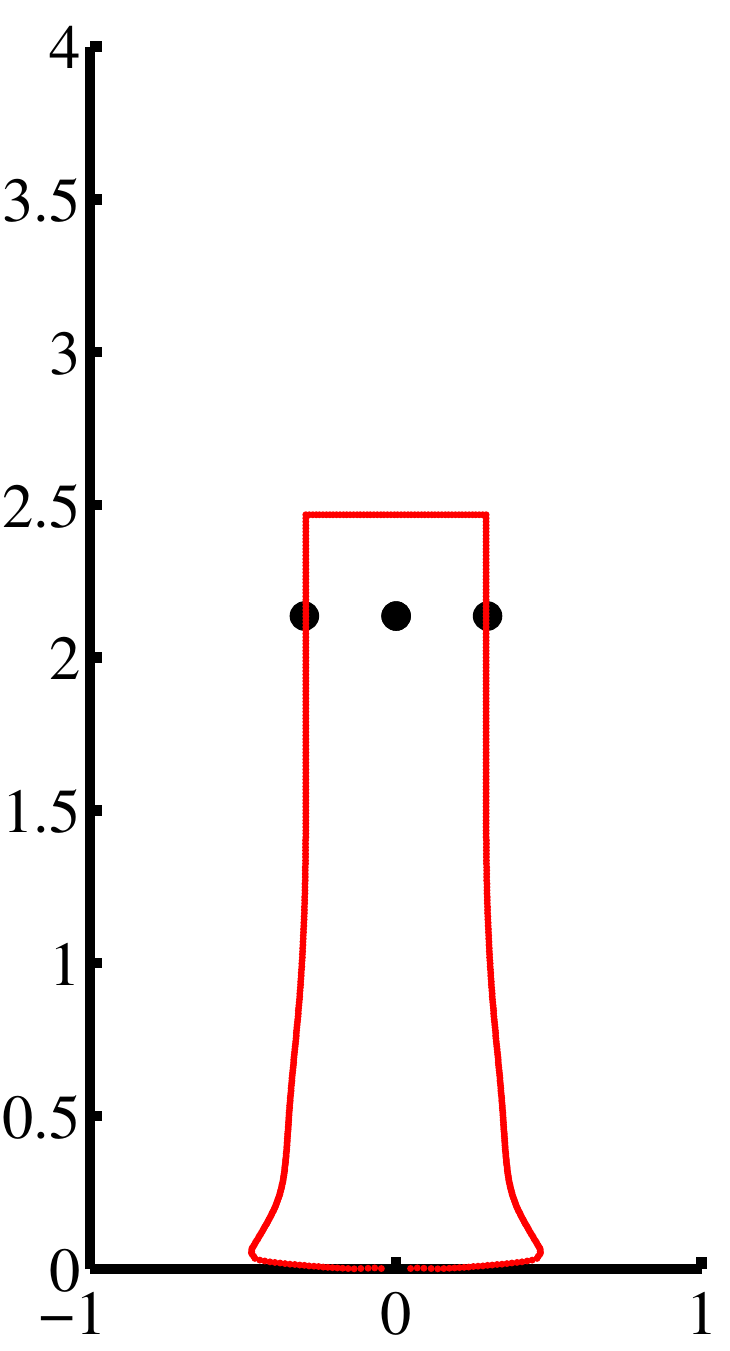}}\\
      (c) JC (St-F). \\
      \scalebox{0.35}{\includegraphics{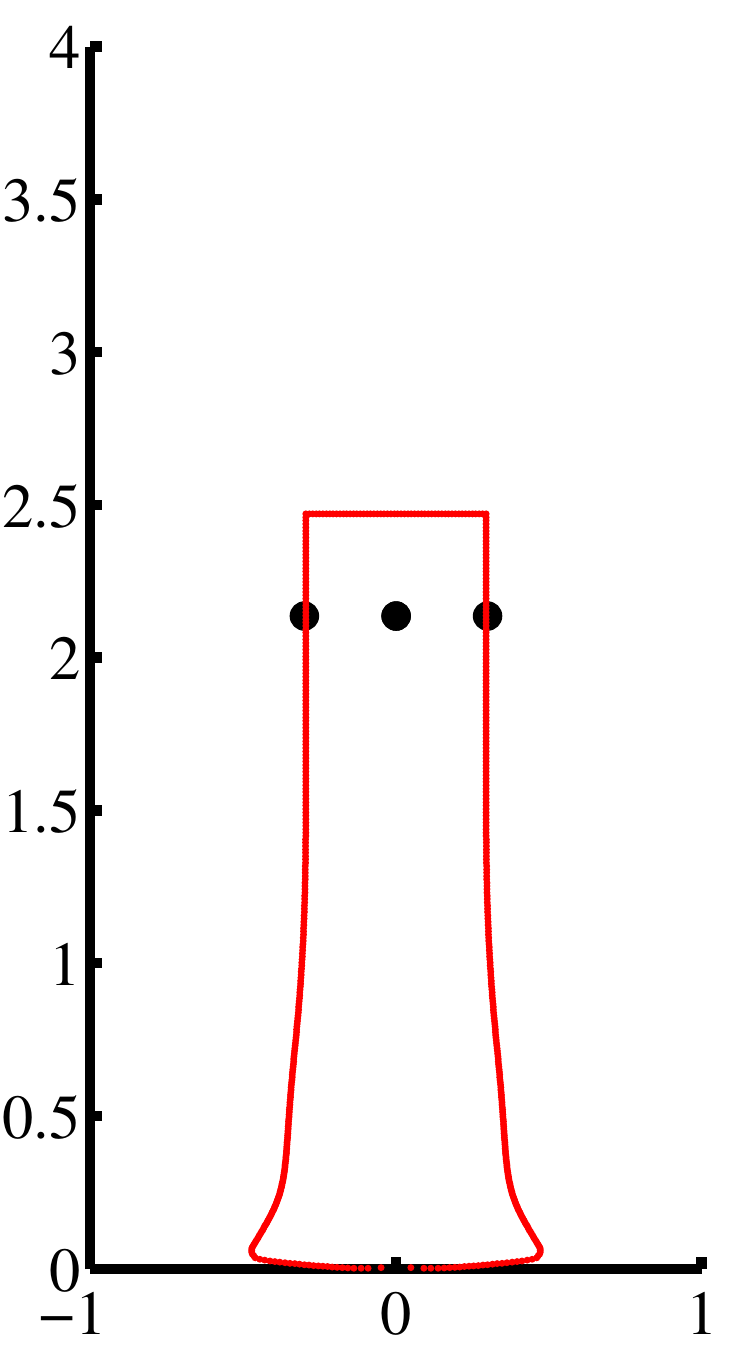}}\\
      (g) JC (St-F) with erosion.
    \end{minipage}
    \begin{minipage}[t]{0.23\linewidth}
      \centering
      \scalebox{0.35}{\includegraphics{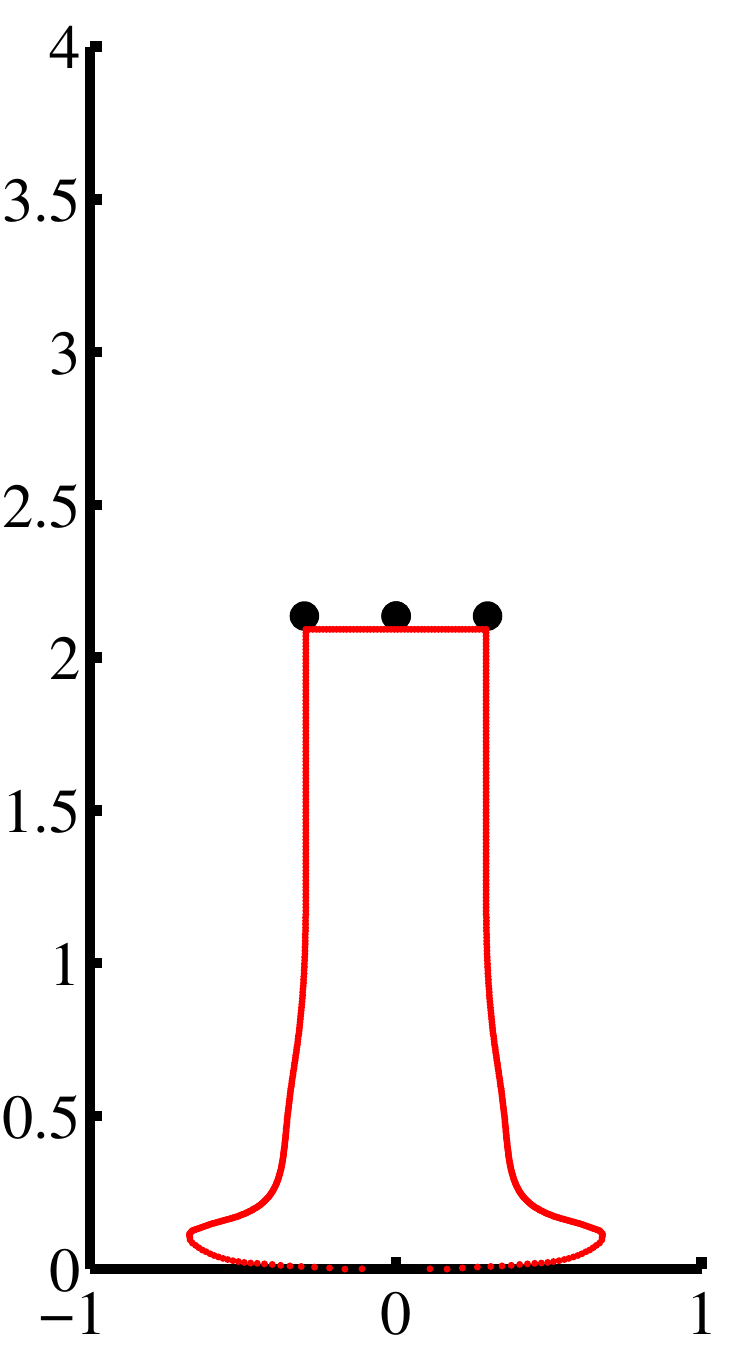}}\\
      (d) SCG (St-F). \\
      \scalebox{0.35}{\includegraphics{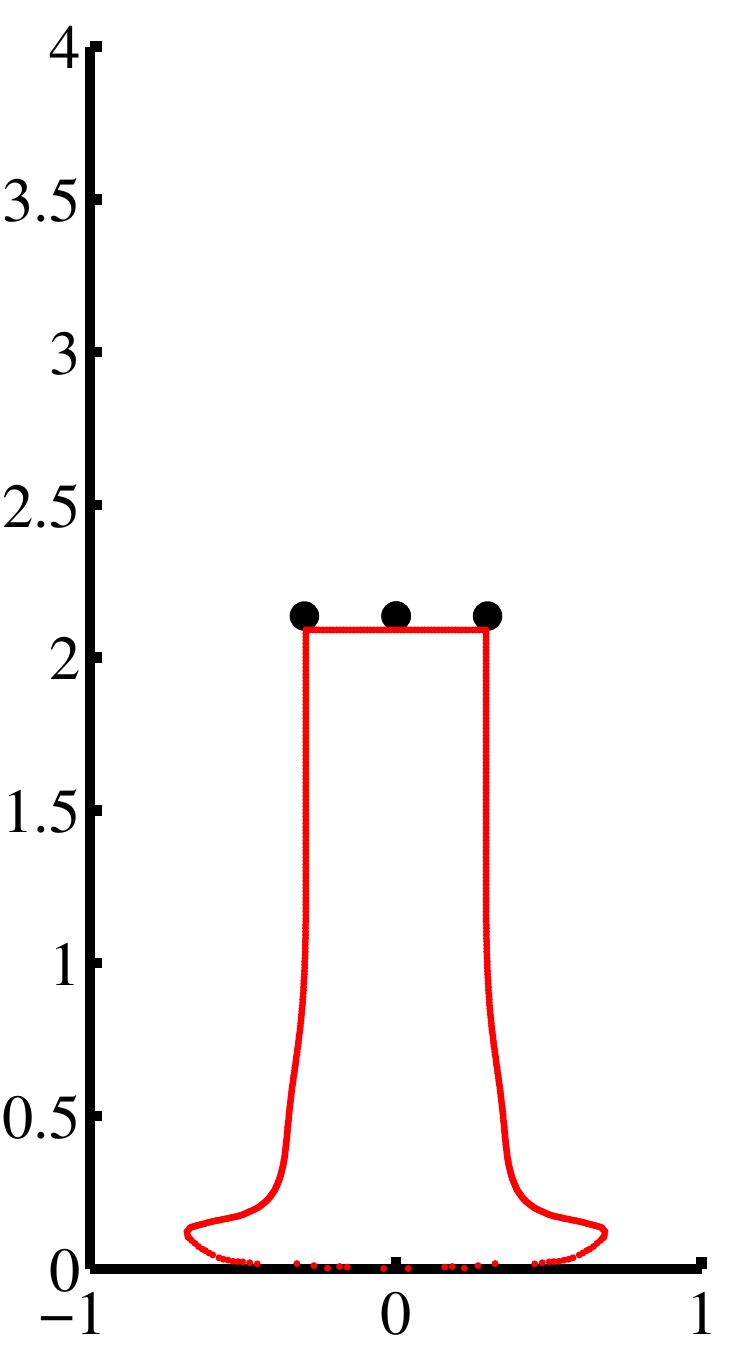}}\\
      (h) SCG (St-F) with erosion.
    \end{minipage}
    \vspace{12pt}
    \caption{\small Comparison of experimental and computed shapes of 4340 steel
             cylinder with friction. St-D is at 725 K and 312 m/s.  St-F is at 1285 K and
             160 m/s. The axes are in cm.}
    \label{fig:4340HotFric}
  \end{figure}

  For high temperature impacts tests, the effect of friction is more obvious in that
  there is a curling of the edges.  Figures~\ref{fig:4340HotFric}(a),(b),(c),(d) show the simulated 
  profiles of cases St-D and St-F with friction.  Specimen D is at a lower temperature
  than specimen F but the impact velocity of the form is almost double that of the 
  latter.  The Johnson-Cook model predicts the final length of the St-D accurately but
  underestimates the final length of St-G.  This indicates that the high temperature 
  behavior of the model is not quite correct even though the rate dependence is captured
  well.  On the other hand, the Steinberg-Guinan model fails miserably at predicting the
  high velocity response but does well for the low velocity/high temperature response.

  Figures~\ref{fig:4340HotFric}(e), (f), (g), and (h) show Taylor impact simulations for cases St-D and
  St-F with particle erosion.  No significant difference can be seen in the computed
  profiles when we compare these to the plots in Figure~\ref{fig:4340HotFric}, except
  for the SCG model for the St-D sample.  Erosion and fracture of the mushroom end does 
  not appear to have a first-order effect on the final length of the impact specimen.

\section{CONCLUSION}\label{sec:conclude}

  Lower temperature simulations lead to predicted profiles that are close to those observed 
  in experiment.  This is true over a range of impact velocities.  However, high temperature
  impacts do not fare so well.

  For the copper specimens, the Johnson-Cook and Steinberg-Guinan models perform better than
  the MTS model both at low and high temperatures.  In future work we show that this is partially 
  due to the stress integration algorithm used in these calculations.  Also, FEM simulations
  consistently overestimate the final length of specimens at high temperatures.

  For the aluminum specimens, the three models, Johnson-Cook, MTS and Steinberg-Guinan, predict 
  accurate final lengths and mushroom diameters at room temperature.  However, at high temperatures
  all three models deviate from experiment, especially when the strain rate is increased.  This
  indicates a coupling of strain rate and temperature that is either not captured by these models or
  requires further calibration.  Mesh dependence due to softening appears to be an issue in the 
  MPM simulations.  FEM simulations with LS-DYNA predict profiles (at high temperatures) that 
  are less deformed than those predicted by MPM suggesting that the constitutive model evaluation
  is not as accurate in FEM calculations.

  For the steel specimens, both Johnson-Cook and Steinberg-Guinan perform well at room temperature.
  However, the Steinberg-Guinan model fails under a combination of high impact velocities and
  high temperatures.  The Johnson-Cook model is not very accurate at high temperatures but captures
  rate-dependent effects quite well.  Failure of the material at the mushroom end does not appear
  to affect the final length of the specimen significantly.

\section*{Acknowledgments}
  This work was supported by the the U.S. Department of Energy through the 
  Center for the Simulation of Accidental Fires and Explosions, under grant 
  W-7405-ENG-48.
\bibliographystyle{unsrtnat}
{\footnotesize
\bibliography{mybiblio}
}

\end{document}

%% file: mymacros.tex
\newcommand{\Bnabla}{\ensuremath{\boldsymbol{\nabla}}}

\newcommand{\Bsig}{\ensuremath{\boldsymbol{\sigma}}}

\newcommand{\Ba}{\ensuremath{\mathbf{a}}}

\newcommand{\Bf}{\ensuremath{\mathbf{f}}}
\newcommand{\Bg}{\ensuremath{\mathbf{g}}}
\newcommand{\Bm}{\ensuremath{\mathbf{m}}}
\newcommand{\Bn}{\ensuremath{\mathbf{n}}}

\newcommand{\Bv}{\ensuremath{\mathbf{v}}}

\newcommand{\Bx}{\ensuremath{\mathbf{x}}}

\newcommand{\BD}{\ensuremath{\mathbf{D}}}

\newcommand{\BG}{\ensuremath{\mathbf{G}}}

\newcommand{\Tint}{\ensuremath{\text{int}}}
\newcommand{\Text}{\ensuremath{\text{ext}}}

\newcommand{\Half}{\ensuremath{\frac{1}{2}}}

\newcommand{\Grad}[1]{\ensuremath{\Bnabla #1}}

%% file: taylorImpact.bbl
\begin{thebibliography}{45}
\providecommand{\natexlab}[1]{#1}
\providecommand{\url}[1]{\texttt{#1}}
\expandafter\ifx\csname urlstyle\endcsname\relax
  \providecommand{\doi}[1]{doi: #1}\else
  \providecommand{\doi}{doi: \begingroup \urlstyle{rm}\Url}\fi

\bibitem[Taylor(1948)]{Taylor48}
G.~I. Taylor.
\newblock The use of flat-ended projectiles for determining dynamic yield
  stress {I.} {T}heoretical considerations.
\newblock \emph{Proc. Royal Soc. London A}, 194\penalty0 (1038):\penalty0
  289--299, 1948.

\bibitem[Whiffin(1948)]{Whiffin48}
A.~C. Whiffin.
\newblock The use of flat-ended projectiles for determining dynamic yield
  stress {II.} {T}ests on various metallic materials.
\newblock \emph{Proc. Royal Soc. London A}, 194\penalty0 (1038):\penalty0
  300--322, 1948.

\bibitem[Johnson and Holmquist(1988)]{Johnson88a}
G.~R. Johnson and T.~J. Holmquist.
\newblock Evaluation of cylinder-impact test data for constitutive models.
\newblock \emph{J. Appl. Phys.}, 64\penalty0 (8):\penalty0 3901--3910, 1988.

\bibitem[Zerilli and Armstrong(1987)]{Zerilli87}
F.~J. Zerilli and R.~W. Armstrong.
\newblock Dislocation-mechanics-based constitutive relations for material
  dynamics calculations.
\newblock \emph{J. Appl. Phys.}, 61\penalty0 (5):\penalty0 1816--1825, 1987.

\bibitem[Sulsky et~al.(1994)Sulsky, Chen, and Schreyer]{Sulsky94}
D.~Sulsky, Z.~Chen, and H.~L. Schreyer.
\newblock A particle method for history dependent materials.
\newblock \emph{Comput. Methods Appl. Mech. Engrg.}, 118:\penalty0 179--196,
  1994.

\bibitem[Sulsky et~al.(1995)Sulsky, Zhou, and Schreyer]{Sulsky95}
D.~Sulsky, S.~Zhou, and H.~L. Schreyer.
\newblock Application of a particle-in-cell method to solid mechanics.
\newblock \emph{Computer Physics Communications}, 87:\penalty0 236--252, 1995.

\bibitem[Armstrong et~al.(1999)Armstrong, Gammon, Geist, Keahey, Kohn, McInnes,
  Parker, and Smolinski]{Armstrong99}
R.~Armstrong, D.~Gammon, A.~Geist, K.~Keahey, S.~Kohn, L.~McInnes, S.~Parker,
  and B.~Smolinski.
\newblock Toward a {C}ommon {C}omponent {A}rchitecture for high-performance
  scientific computing.
\newblock In \emph{Proc. 1999 Conference on High Performance Distributed
  Computing}, 1999.

\bibitem[de~St.~Germain et~al.(2000)de~St.~Germain, McCorquodale, Parker, and
  Johnson]{Dav2000}
J.~D. de~St.~Germain, J.~McCorquodale, S.~G. Parker, and C.~R. Johnson.
\newblock Uintah: a massively parallel problem solving environment.
\newblock In \emph{Ninth IEEE International Symposium on High Performance and
  Distributed Computing}, pages 33--41. IEEE, Piscataway, NJ, Nov 2000.

\bibitem[Long and Wight(2002)]{Long02}
G.~T. Long and C.~A. Wight.
\newblock Thermal decomposition of a melt-castable high explosive:
  isoconversional analysis of {TNAZ}.
\newblock \emph{J. Phys. Chem. B}, 106:\penalty0 2791--2795, 2002.

\bibitem[Guilkey et~al.(2004)Guilkey, Harman, Kashiwa, and McMurtry]{Guilkey04}
J.~E. Guilkey, T.~B. Harman, B.~A. Kashiwa, and P.~A. McMurtry.
\newblock An {E}ulerian-{L}agrangian approach to large deformation
  fluid-structure interaction problems.
\newblock Submitted, 2004.

\bibitem[Bennett et~al.(1998)Bennett, Haberman, Johnson, Asay, and
  Henson]{Bennet98}
J.~G. Bennett, K.~S. Haberman, J.~N. Johnson, B.~W. Asay, and B.~F. Henson.
\newblock A constitutive model for non-shock ignition and mechanical response
  of high explosives.
\newblock \emph{J. Mech. Phys. Solids}, 46\penalty0 (12):\penalty0 2303--2322,
  1998.

\bibitem[Bardenhagen and Kober(2004)]{Bard04}
S.~G. Bardenhagen and E.~M. Kober.
\newblock The generalized interpolation material point method.
\newblock \emph{Comp. Model. Eng. Sci.}, 2004.
\newblock to appear.

\bibitem[Bardenhagen et~al.(2001)Bardenhagen, Guilkey, Roessig, BrackBill,
  Witzel, and Foster]{Bard01}
S.~G. Bardenhagen, J.~E. Guilkey, K.~M Roessig, J.~U. BrackBill, W.~M. Witzel,
  and J.~C. Foster.
\newblock An improved contact algorithm for the material point method and
  application to stress propagation in granular material.
\newblock \emph{Computer Methods in the Engineering Sciences}, 2\penalty0
  (4):\penalty0 509--522, 2001.

\bibitem[Zocher et~al.(2000)Zocher, Maudlin, Chen, and
  Flower-Maudlin]{Zocher00}
M.~A. Zocher, P.~J. Maudlin, S.~R. Chen, and E.~C. Flower-Maudlin.
\newblock An evaluation of several hardening models using {T}aylor cylinder
  impact data.
\newblock In \emph{Proc. , European Congress on Computational Methods in
  Applied Sciences and Engineering}, Barcelona, Spain, 2000. ECCOMAS.

\bibitem[Hill and Hutchinson(1975)]{Hill75}
R.~Hill and J.~W. Hutchinson.
\newblock Bifurcation phenomena in the plane tension test.
\newblock \emph{J. Mech. Phys. Solids}, 23:\penalty0 239--264, 1975.

\bibitem[Bazant and Belytschko(1985)]{Bazant85}
Z.~P. Bazant and T.~Belytschko.
\newblock Wave propagation in a strain-softening bar: {E}xact solution.
\newblock \emph{ASCE J. Engg. Mech}, 111\penalty0 (3):\penalty0 381--389, 1985.

\bibitem[Tvergaard and Needleman(1990)]{Tver90}
V.~Tvergaard and A.~Needleman.
\newblock Ductile failure modes in dynamically loaded notched bars.
\newblock In J.~W. Ju, D.~Krajcinovic, and H.~L. Schreyer, editors,
  \emph{Damage Mechanics in Engineering Materials: AMD 109/MD 24}, pages
  117--128. American Society of Mechanical Engineers, New York, NY, 1990.

\bibitem[Ramaswamy and Aravas(1998{\natexlab{a}})]{Ramaswamy98}
S.~Ramaswamy and N.~Aravas.
\newblock Finite element implementation of gradient plasticity models {P}art
  {I}: {G}radient-dependent yield functions.
\newblock \emph{Comput. Methods Appl. Mech. Engrg.}, 163:\penalty0 11--32,
  1998{\natexlab{a}}.

\bibitem[Hao et~al.(2000)Hao, Liu, and Qian]{Hao00}
S.~Hao, W.~K. Liu, and D.~Qian.
\newblock Localization-induced band and cohesive model.
\newblock \emph{J. Appl. Mech.}, 67:\penalty0 803--812, 2000.

\bibitem[Coplien(1992)]{Coplien92}
J.~O. Coplien.
\newblock \emph{Advanced C++ Programming Styles and Idioms}.
\newblock Addison-Wesley, Reading, MA, 1992.

\bibitem[Steinberg et~al.(1980)Steinberg, Cochran, and Guinan]{Steinberg80}
D.~J. Steinberg, S.~G. Cochran, and M.~W. Guinan.
\newblock A constitutive model for metals applicable at high-strain rate.
\newblock \emph{J. Appl. Phys.}, 51\penalty0 (3):\penalty0 1498--1504, 1980.

\bibitem[Johnson and Cook(1983)]{Johnson83}
G.~R. Johnson and W.~H. Cook.
\newblock A constitutive model and data for metals subjected to large strains,
  high strain rates and high temperatures.
\newblock In \emph{Proc. 7th International Symposium on Ballistics}, pages
  541--547, 1983.

\bibitem[Follansbee and Kocks(1988)]{Follans88}
P.~S. Follansbee and U.~F. Kocks.
\newblock A constitutive description of the deformation of copper based on the
  use of the mechanical threshold stress as an internal state variable.
\newblock \emph{Acta Metall.}, 36:\penalty0 82--93, 1988.

\bibitem[Goto et~al.(2000{\natexlab{a}})Goto, Bingert, Reed, and
  Garrett]{Goto00a}
D.~M. Goto, J.~F. Bingert, W.~R. Reed, and R.~K. Garrett.
\newblock Anisotropy-corrected {MTS} constitutive strength modeling in {HY}-100
  steel.
\newblock \emph{Scripta Mater.}, 42:\penalty0 1125--1131, 2000{\natexlab{a}}.

\bibitem[Gurson(1977)]{Gurson77}
A.~L. Gurson.
\newblock Continuum theory of ductile rupture by void nucleation and growth:
  {P}art 1. {Y}ield criteria and flow rules for porous ductile media.
\newblock \emph{ASME J. Engg. Mater. Tech.}, 99:\penalty0 2--15, 1977.

\bibitem[Tvergaard and Needleman(1984)]{Tver84}
V.~Tvergaard and A.~Needleman.
\newblock Analysis of the cup-cone fracture in a round tensile bar.
\newblock \emph{Acta Metall.}, 32\penalty0 (1):\penalty0 157--169, 1984.

\bibitem[Ramaswamy and Aravas(1998{\natexlab{b}})]{Ramaswamy98a}
S.~Ramaswamy and N.~Aravas.
\newblock Finite element implementation of gradient plasticity models {P}art
  {II}: {G}radient-dependent evolution equations.
\newblock \emph{Comput. Methods Appl. Mech. Engrg.}, 163:\penalty0 33--53,
  1998{\natexlab{b}}.

\bibitem[Chu and Needleman(1980)]{Chu80}
C.~C. Chu and A.~Needleman.
\newblock Void nucleation effects in biaxially stretched sheets.
\newblock \emph{ASME J. Engg. Mater. Tech.}, 102:\penalty0 249--256, 1980.

\bibitem[Borvik et~al.(2001)Borvik, Hopperstad, Berstad, and
  Langseth]{Borvik01}
T.~Borvik, O.~S. Hopperstad, T.~Berstad, and M.~Langseth.
\newblock A computational model of viscoplastcity and ductile damage for impact
  and penetration.
\newblock \emph{Eur. J. Mech. A/Solids}, 20:\penalty0 685--712, 2001.

\bibitem[Goto et~al.(2000{\natexlab{b}})Goto, Bingert, Chen, Gray, and
  Garrett]{Goto00}
D.~M. Goto, J.~F. Bingert, S.~R. Chen, G.~T. Gray, and R.~K. Garrett.
\newblock The mechanical threshold stress constitutive-strength model
  description of {HY}-100 steel.
\newblock \emph{Metallurgical and Materials Transactions A}, 31A:\penalty0
  1985--1996, 2000{\natexlab{b}}.

\bibitem[Johnson and Cook(1985)]{Johnson85}
G.~R. Johnson and W.~H. Cook.
\newblock Fracture characteristics of three metals subjected to various
  strains, strain rates, temperatures and pressures.
\newblock \emph{Int. J. Eng. Fract. Mech.}, 21:\penalty0 31--48, 1985.

\bibitem[Hancock and MacKenzie(1976)]{Hancock76}
J.~W. Hancock and A.~C. MacKenzie.
\newblock On the mechanisms of ductile failure in high-strength steels
  subjected to multi-axial stress-states.
\newblock \emph{J. Mech. Phys. Solids}, 24:\penalty0 147--167, 1976.

\bibitem[Johnson and Addessio(1988)]{Johnson88}
J.~N. Johnson and F.~L. Addessio.
\newblock Tensile plasticity and ductile fracture.
\newblock \emph{J. Appl. Phys.}, 64\penalty0 (12):\penalty0 6699--6712, 1988.

\bibitem[Drucker(1959)]{Drucker59}
D.~C. Drucker.
\newblock A definition of stable inelastic material.
\newblock \emph{J. Appl. Mech.}, 26:\penalty0 101--106, 1959.

\bibitem[Rudnicki and Rice(1975)]{Rudnicki75}
J.~W. Rudnicki and J.~R. Rice.
\newblock Conditions for the localization of deformation in pressure-sensitive
  dilatant materials.
\newblock \emph{J. Mech. Phys. Solids}, 23:\penalty0 371--394, 1975.

\bibitem[Perzyna(1998)]{Perzyna98}
P.~Perzyna.
\newblock Constitutive modelling of dissipative solids for localization and
  fracture.
\newblock In Perzyna P., editor, \emph{Localization and Fracture Phenomena in
  Inelastic Solids: CISM Courses and Lectures No. 386}, pages 99--241.
  SpringerWien, New York, 1998.

\bibitem[Becker(2002)]{Becker02}
R.~Becker.
\newblock Ring fragmentation predictions using the gurson model with material
  stability conditions as failure criteria.
\newblock \emph{Int. J. Solids Struct.}, 39:\penalty0 3555--3580, 2002.

\bibitem[Oberkampf et~al.(2002)Oberkampf, Trucano, and Hirsch]{Oberkampf02}
W.~L. Oberkampf, T.~G. Trucano, and C.~Hirsch.
\newblock Verification, validation, and predictive capability in computational
  engineering and physics.
\newblock In \emph{Verfification and Validation for Modeling and Simulation in
  Computational Science and Engineering Applications}, Johns Hopkins
  University, Laurel, Maryland, 2002. Foundations for Verfification and
  Validation in the 21st Century Workshop.

\bibitem[Babuska and Oden(2004)]{Babuska04}
I.~Babuska and J.~T. Oden.
\newblock Verification and validation in computational engineering and science:
  basic concepts.
\newblock \emph{Comput. Methods Appl. Mech. Engrg.}, 193:\penalty0 4057--4066,
  2004.

\bibitem[Sutanthavibul et~al.(2002)]{XFig04}
S.~Sutanthavibul et~al.
\newblock \emph{Xfig User Manual Version 3.2.4}.
\newblock http://www.xfig.org, 2002.

\bibitem[Carrington and Gayler(1948)]{Carrington48}
W.~E. Carrington and M.~L.~V Gayler.
\newblock The use of flat-ended projectiles for determining dynamic yield
  stress {III.} {C}hanges in microstructure caused by deformation under impact
  at high-striking velocities.
\newblock \emph{Proc. Royal Soc. London A}, 194\penalty0 (1038):\penalty0
  323--331, 1948.

\bibitem[Wilkins and Guinan(1973)]{Wilkins73}
M.~L. Wilkins and M.~W. Guinan.
\newblock Impact of cylinders on a rigid boundary.
\newblock \emph{J. Appl. Phys.}, 44\penalty0 (3):\penalty0 1200--1206, 1973.

\bibitem[Gust(1982)]{Gust82}
W.~H. Gust.
\newblock High impact deformation of metal cylinders at elevated temperatures.
\newblock \emph{J. Appl. Phys.}, 53\penalty0 (5):\penalty0 3566--3575, 1982.

\bibitem[Addessio et~al.(1993)Addessio, Johnson, and Maudlin]{Addessio93a}
F.~L. Addessio, J.~N. Johnson, and P.~J. Maudlin.
\newblock The effect of void growth on {T}aylor cylinder impact experiments.
\newblock \emph{J. Appl. Phys.}, 73\penalty0 (11):\penalty0 7288--7297, 1993.

\bibitem[Chhabildas et~al.(1998)Chhabildas, Konrad, Mosher, Reinhart, Duggins,
  Trucano, Summers, and Peery]{Chhabil99}
L.~C. Chhabildas, C.~H. Konrad, D.~A. Mosher, W.~D. Reinhart, B.~D. Duggins,
  T.~G. Trucano, R.~M. Summers, and J.~S. Peery.
\newblock A methodology to validated {3D} arbitrary {L}agrangian {E}ulerian
  codes with applications to {ALEGRA}.
\newblock \emph{Int. J. Impact Engrg.}, 23:\penalty0 101--112, 1998.

\end{thebibliography}
